\documentclass[onecolumn]{ametsocV6.1}
\usepackage{amsmath,amsfonts,amssymb,bm}
\usepackage{mathptmx}
\usepackage{newtxtext}
\usepackage{newtxmath}
\usepackage{gensymb}

\title{On the archetypal `flavours', indices and teleconnections of ENSO revealed by global sea surface temperatures}
\authors{Didier~P.~Monselesan~\aff{a}\correspondingauthor{Didier P. Monselesan, didier.monselesan@csiro.au},
James~S.~Risbey~\aff{a},
Benoit~Legresy~\aff{a},
Sophie Cravatte~\aff{b},
Bastien~Pagli~\aff{c},
Takeshi Izumo~\aff{c},
Christopher~C.~Chapman~\aff{a},
Mandy~Freund~\aff{a},
Abdelwaheb~Hannachi~\aff{d},
Damien~Irving~\aff{a},
P.~Jyoteeshkumar~Reddy~\aff{a}, 
Doug~Richardson~\aff{e},
Dougal~T.~Squire~\aff{a},
Carly~R.~Tozer~\aff{a}}

\affiliation{
\aff{a}{CSIRO Environment, Hobart, Tasmania, Australia}\\
\aff{b}{LEGOS université de Toulouse, IRD, CNES, CNRS, UPS, Toulouse, France}\\
\aff{c}{Institut de Recherche pour le Développement (IRD), 
Ecosystemes Insulaires Océaniens, Université de Polynésie Française}\\
\aff{d}{Department of Meteorology and Bolin Centre for Climate Research, Stockholm University, Stockholm, Sweden}\\
\aff{e}{ARC Centre of Excellence for Climate Extremes, University of New South Wales, Sydney, Australia}
}
\abstract{El Ni\~no-Southern Oscillation global (ENSO) imprint on sea surface temperature comes in many guises. To identify its tropical fingerprints and impacts on the rest of the climate system, we propose a global approach based on archetypal analysis (AA), a pattern recognition method based on the identification of extreme configurations in the dataset under investigation. Relying on detrended sea surface temperature monthly anomalies over the 1982 to 2022 period, the technique recovers central and eastern Pacific ENSO types identified by more traditional methods and allows one to hierarchically add extra flavours and nuances to both persistent and transient phases of the phenomenon. Archetypal patterns found compare favorably to phase identification from K-means, fuzzy C-means and recently published network-based machine-learning algorithms. The AA implementation is modified for the identification of ENSO phases in sub-seasonal-to-seasonal prediction systems and complements current alert systems in characterising the diversity of ENSO and its teleconnections. Tropical and extra-tropical teleconnection composites from various oceanic and atmospheric fields derived from the analysis are shown to be robust and physically relevant. Extending AA to sub-surface ocean fields improves the discrimination between phases when the characterisation of ENSO based on sea surface temperature is uncertain. We show that AA on detrended sea-level monthly anomalies provides a clearer expression of ENSO types.}

\begin{document}
\nolinenumbers
\maketitle

%%%%%%%%%%%%%%%%%%%%%%%%%%%%%%%%%%%%%%%%%%%%%%%%%%%%%%%%%%%%%%%%%%%%%
% MAIN BODY OF PAPER
%%%%%%%%%%%%%%%%%%%%%%%%%%%%%%%%%%%%%%%%%%%%%%%%%%%%%%%%%%%%%%%%%%%%%
\section{Significance statement}

ENSO comes in many guises and is challenging to classify in a changing climate. This issue is becoming critical as operational centers recognize the shortcomings of simple two category ENSO systems. Archetypal analysis, a pattern recognition technique, is applied to sea surface temperature and sea-level anomalies to provide a robust characterisation of more flavours of ENSO. The additional flavours are related to the state of the Pacific Ocean. The inclusion of sea-level anomalies resolves issues in the classification of ENSO when sea surface temperature alone is inconclusive. The ability of the method to recover previously detected ENSO flavours and to extend its application to the classification of nuances in the onset, main phase, and decay of the phenomenon, is assessed.

%%%%%%%%%%%%%%%%%%%%%%%%%%%%%%%%%%%
%   Section 1
%%%%%%%%%%%%%%%%%%%%%%%%%%%%%%%%%%%
\section{Introduction}\label{introduction}
Operational centers have long realised that their climate driver alert systems in general, and ENSO alert protocols in particular, may suffer from non-stationary processes, driven, not only by anthropogenic climate change, but also by long-period internal climate modes and their likely interactions \citep{bom22:brr72}. The advice given by operational centers mainly relies on simple indices derived from various oceanic and atmospheric anomaly fields reaching and exceeding critical thresholds combined with a persistence criterion. This advice may also be conditional on multivariate alert mechanisms derived from additional atmospheric and oceanic diagnostics reviewed by a panel of experts\footnote{For example, \url{https://www.cpc.ncep.noaa.gov/products/analysis_monitoring/enso_advisory/ensodisc.pdf} and \url{http://www.bom.gov.au/climate/enso/outlook/#tabs=About-ENSO-and-the-Outlook}}. For ENSO alert and prediction systems, operational centers typically issue an all-encompassing type of advice for \textsl{La Ni\~{n}a}, \textsl{La Ni\~{n}a watch}, \textsl{Neutral}, \textsl{El Ni\~{n}o watch}, and \textsl{El Ni\~{n}o} phases.  A detailed inspection of ENSO global imprint on sea surface temperature (SST) alone over the 1982-2022 period reveals a diversity of expressions of ENSO and its teleconnections with the rest of the climate system and their environmental impact, unlikely to be adequately described by a single index. The non-stationarity of SST at long-time scales also interferes with alert protocols rendering them prone to mischaracterisation \citep{lhe13:cd, tim18:nat, tur19:ijc, old21:erl}.

Hereafter, we propose an approach based on archetypal analysis (AA) adding more flavours\footnote{Here, we, the authors, understand \textsl{flavour} in the ENSO context as a range of clearly differentiated (arche)typical patterns manifest in the tropical SST anomalies with well defined centers of action or loci of absolute maximum amplitudes, which expression may have a distinct dynamical origin.} to the reclassification of historical ENSO events \citep{han17:jcli,bla22:aies} and readily applicable to the identification of future ENSO phases in subseasonal-to-seasonal (S2S) predictions. The method allows the unsupervised detection of Eastern and Central Pacific (EP and CP) ENSO phases amongst others, and can easily be extended not only to capture the diversity in the onset and decay of the phenomenon, but also to other climate drivers such as the North Atlantic Oscillation (NAO), the Pacific–North American pattern (PNA), the Pacific-South American pattern (PSA), the Southern Annular Mode (SAM) or the Indian Ocean Dipole (IOD) for example when applied to specific region of interests \citep[for a review of Sir Gilbert Walker contributions to the identification of large-scale climate teleconnections such as the Southern Oscillation or the NAO]{rog82:mwr, lau94:jas, saj99:nat, tho00:jcli, kat02:ss}. To further frame the issue, a short and far from exhaustive history of ENSO follows with a focus on the seminal works leading to the identification and naming of the different flavours of ENSO.

Following \cite{bje66:tel, bje69:mwr} early work on El Ni\~no atmospheric teleconnections linking it to the Sir Gilbert Walker `Southern Oscillation', \cite{wyr75:jpo} proposed that El Ni\~no episodes were due to anomalously high southeast trades present in the central Pacific, over the two years preceding El Ni\~no, intensifying the subtropical gyre and associated South Equatorial Current of the South Pacific, leading up to a build-up of water in the western equatorial Pacific and increasing the sea level east-west slope across the basin. Wyrtki postulated that the relaxation of these enhanced and persistent trades would then trigger the opposite reaction whereby the water accumulated in the West of the basin, over two years prior, would flow back eastwards initiating El Ni\~no, leading to a warming of the eastern equatorial SST and a deepening of the equatorial thermocline. It is interesting to note that, at the time, the La Ni\~na phase of ENSO had not been coined neither by Bjerknes nor Wyrtki in their seminal papers. However comparing and contrasting the onset, persistence and decay of the 1957-1958, 1965-1966 and 1972-1973 El Ni\~no events, \cite{wyr75:jpo} was led to conclude 
\begin{quote}
    \sl El Ni\~no also appears in varying intensity and, like any other large-scale event in the ocean-atmosphere system, El Ni\~no certainly does not have a single cause; no two El Ni\~no events are quite alike
\end{quote}
intimating at this early stage the diversity or flavours of El Ni\~no. As far as La Ni\~na is concerned, the realisation that the condition corresponded to a distinct phase of the ENSO phenomenon came some fifteen years later, although Wyrtki may have alluded to it when mentioning the anomalously high southeast trades present in the central Pacific as a precursor to El Ni\~no, as the enhancement of the trades would likely cool the Eastern equatorial Pacific below its climatological values and lead to a further shoaling of the thermocline there. According to the Oxford English Dictionary search tool, Guill\'en et al.\footnote{\url{https://www.oed.com/view/Entry/252722?isAdvanced=true&result=2&rskey=RRdzwh&}. Interestingly, it has been rather challenging to find the peer-reviewed publication where La Ni\~na is first mentioned as the opposite phase of El Ni\~no. \cite{izu23:misc} notes that the term La Ni\~na may have emanated from George Philander work through his Spanish speaking South-American wife.} reported on SST anomalies having a negative effects on primary production in \cite{gla81:book} and, it appears, mentioned for the first time the La Ni\~na moniker for the El Ni\~no colder counterpart. The weather and climate community had to wait for the aftermath of the intense 1997-1998 Ni\~no, which was followed by a rapid swing to a strong Ni\~na condition, for directing its attention to La Ni\~na and its impacts \citep{gla02:book}. The peculiar characteristics and teleconnections of the 2002-2003 El Ni\~no, when compared to the conventional El Ni\~nos \citep[for an early review]{ras82:mwr}, prompted \cite{ash07:jgr} to undertake a comprehensive review of the event and led to the definition of a distinct type of El Ni\~no named El Ni\~no Modoki. The colder El Ni\~no Modoki counterpart, La Ni\~na Modoki, was not mentioned in \cite{ash07:jgr}, but was introduced by \cite{cai09a:grl} two years later when discussing the asymmetric impact on Australian rainfall of what appears to be four distinct phases of ENSO. By 2009, four possibly distinct ENSO conditions have been identified in the literature: El Ni\~no, La Ni\~na, El Ni\~no Modoki and La Ni\~na Modoki. 

Multiple approaches have been applied to the characterisation of the flavours and naming of ENSO types. The specificity of the ENSO types discovered by each of these methods is inherent to the methodological choices made, as it is with our approach, but is not necessarily inherent to the phenomenon these methods intend to capture. Hereafter, we will opt for the CP and EP naming convention based on resemblance, not equivalence. We could have used Central/Eastern Pacific \citep{kug09:jcli,tak11:grl,sul16:sr}, Canonical/Modoki \citep{can83:sc, ash07:jgr, cai09a:grl} or Canonical/Dateline \citep{tre01:jcli} ENSO terminology interchangeably.

With the reliance on both supervised and unsupervised machine learning (ML) algorithms applied to weather and climate research, the number of new ENSO classes is increasing, leading some to question if the inflationary nature of the ML discovery of these classes corresponds to truly different climatic processes \citep{chr07:jcli, joh13:jcli, kar13:grl, kas21:ptrs}. Although there is still some debate whether or not the different flavours of ENSO correspond to dynamically distinct phenomena \citep{ash07:jgr, kao09:jcli, kar13:grl, joh13:jcli, cap15:bams,  lig19:grl, ama19:cccr, cap20:ch4, feng20:cd}, various classification methods \citep{lel07:cd, tak11:grl, sul16:sr, fre19:ng, lem19:cd, wie21:fc, bla22:aies, sch23:arX} allow to discriminate between the manifestation and evolution of either a singular mode, plural modes or even a continuum of modes \citep{tre01:jcli, joh13:jcli, wil18a:grl, feng20:cd, sch21:jcli, die21:cec} of tropical variability based on sea surface temperature anomalies (SSTAs).

In this work, we build on the AA results reported by \cite{han17:jcli}, \cite{bla22:aies} and \cite{cha22:ncom} for SSTA, who showed AA unique ability in identifying extreme SST configurations. Section \ref{data} introduces all the datasets considered in this study and justifies some of the AA data pre-processing steps. Section \ref{aa} summarises the AA method, together with Section S1 of the supplementary material, provides an overview of AA in layperson's terms. It also compares and contrasts it to two commonly used clustering algorithms: K-mean and fuzzy C-means. Section \ref{sec:flavours} presents AA results applied to global SSTAs and discusses the resulting AA ENSO patterns and indices, which are compared to more traditional methods of classification in Section \ref{sec:tradition}. Section \ref{sec:online} demonstrates that AA can be applied to S2S forecasts and could complement current ENSO alert systems in characterising ENSO diversity. Section \ref{sec:tele} illustrates the reliability and relevance of the AA classification of ENSO types by showing that extreme configurations detected in detrended SSTAs (dSSTAs) correspond to robust and explainable teleconnection patterns of various atmospheric and oceanic fields. We also present AA results performed on detrended sea level anomalies showing that they provide a clearer expression of ENSO types. ENSO phase transitions characterised by AA are illustrated in Section \ref{sec:transition}. Section \ref{conclusion} concludes the paper by summarising our findings and discusses potential applications of AA to possibly advance weather and climate research.

We would like to conclude this introduction with an important caveat. Although we will borrow from the ENSO nomenclature to label the AA discovered phases with the CP and EP qualifiers, we do not claim that these phases correspond exactly to those mentioned in the cited literature, but we do show that they resemble these types as much as these types resemble one another. As mentioned previously, we characterise the flavour or type in the ENSO context as a range of clearly differentiated patterns manifest in the SST anomalies in the Pacific with well defined centers of action or loci of absolute maximum amplitudes and broad spatial extent.

%%%%%%%%%%%%%%%%%%%%%%%%%%%%%%%%%%%
%   Section 2
%%%%%%%%%%%%%%%%%%%%%%%%%%%%%%%%%%%
\section{Data}\label{data}
We apply AA to linearly detrended Optimum Interpolation Sea Surface Temperature (OISST) v2.1 high resolution dataset \citep{rey07:jcli} global SST anomalies (SSTAs) over the 1982-2022 period. We mainly focus here on monthly anomalies represented as the departure from monthly climatological values defined as the time-mean of SST values for each month-of-year across all years considered in the 1982-2022 interval with the local linear trend computed over the same period removed at every grid point. Only complete calendar years are considered.

Our justification for going global and not reducing the geographical domain in the analysis is manifold. Firstly, limiting the domain to a given latitude-longitude box is subjective, and may incur the risk of the box boundaries cutting through important teleconnection pathways potentially useful to the classification of ENSO events and their impacts.

Secondly and often without being explicitly stated, reducing the domain is tantamount to infer \textsl{a priori} that the geophysical processes under consideration within it are the most important drivers of interactions with the rest of the climate system outside the initially chosen domain. As causal links are particularly difficult to ascertain, these choices are at time problematic. For high temporal resolution, from 1 to 1/5 days, \cite{bac19:jcli} (Fig.~8) show that the local atmosphere-ocean predictability, based on a spectral causality argument, is mainly driven by atmospheric processes apart from a narrow tropical band, where the ocean drives the atmosphere. However, for frequencies lower than 1/month, the regions where the ocean drives predictability expand to the sub-tropics and higher latitudes. These findings are based on local interactions between atmosphere and ocean only. The picture is likely to change when remote teleconnections are taken into account. A multivariate approach to assess non-local interactions between ocean and atmosphere is challenging. Lagged correlation and linear response theoretical arguments have been put forward to explain this relationship \citep{cap17:GRL, peg20:cd, cap21:scirep}. Lagged-regression, -correlation or -covariance analyses are often used to assess causality in weather and climate studies to investigate relationships between variables. These methods often suffer from drawbacks where variables' auto-correlation and their non-linear interactions have a confounding effect as discussed recently by \cite{mcg18:jcli}, for example. They advocate for the implementation of dedicated causality-probing methods \citep{gra69:eco, bar14:jnm} as in \cite{bac19:jcli}. However, these approaches are based on time-mean (expectation) statistics and are difficult to compare directly with AA results, which are not. As consequences of the monthly temporal resolution used in this study and the potential causal and non-local relationships between tropical, extra-tropical and high-latitude SSTs, enlarging our domain of study can be justified. This choice can help to link the ENSO tropical signature - its location, extent and intensity - to concurrent extra-tropical conditions and to inform on its potential remote impacts.

Thirdly, the choice of fixed geographical boxes, apart from being subjective\footnote{The choice of climate indices, based on restricted geographical domains and commonly used to benchmark climate driver strength and phases, may suffer from 1) the non-stationarity of the underlying field under study and 2) a lack of sample diversity when initially established as the history of ENSO diversity exposed in Section \ref{introduction} attests.}, corresponds to a crude way to capture dynamical processes and is probably not suitable for a direct comparison with ocean-atmosphere global climate model (OAGCM) simulations due to unresolved model biases in the model representation of geophysical processes in general, and in the spatiotemporal representation of ENSO in particular \citep[for a recent expos\'e]{gui20:book}. As a case in point, \cite{feng20:cd} recently embarked in a thorough examination of the implication of these choices for the detection and their power of discrimination when applied to the ENSO diversity based on SSTAs. We refer the reader to it as an illustration on how arbitrary and critical these choices can be in the definition of various ENSO indices and associated patterns.

As we are interested in detecting the different flavours of ENSO, removing a linear trend to SSTA prior to AA implementation helps to approximately separate the anthropogenic greenhouse forcing and natural secular trends from sub-seasonal-to-decadal internal climate mode variability, \textsl{de facto} assuming that the internal variability of the climate system can be separated from these processes operating at longer timescales and ignoring their potential interaction. The assumption of separability is difficult to test given the short high-quality global SST satellite records of 41 years at our disposal and would be probably untenable for centennial to millennial timescales \citep[for ENSO variability over the last millenium]{bun09:jcli, lev17:grl, han20:qi, emi20:book, jia21:pp, pow21:sci}.

\begin{figure}[htbp]
    \includegraphics[width=\textwidth]{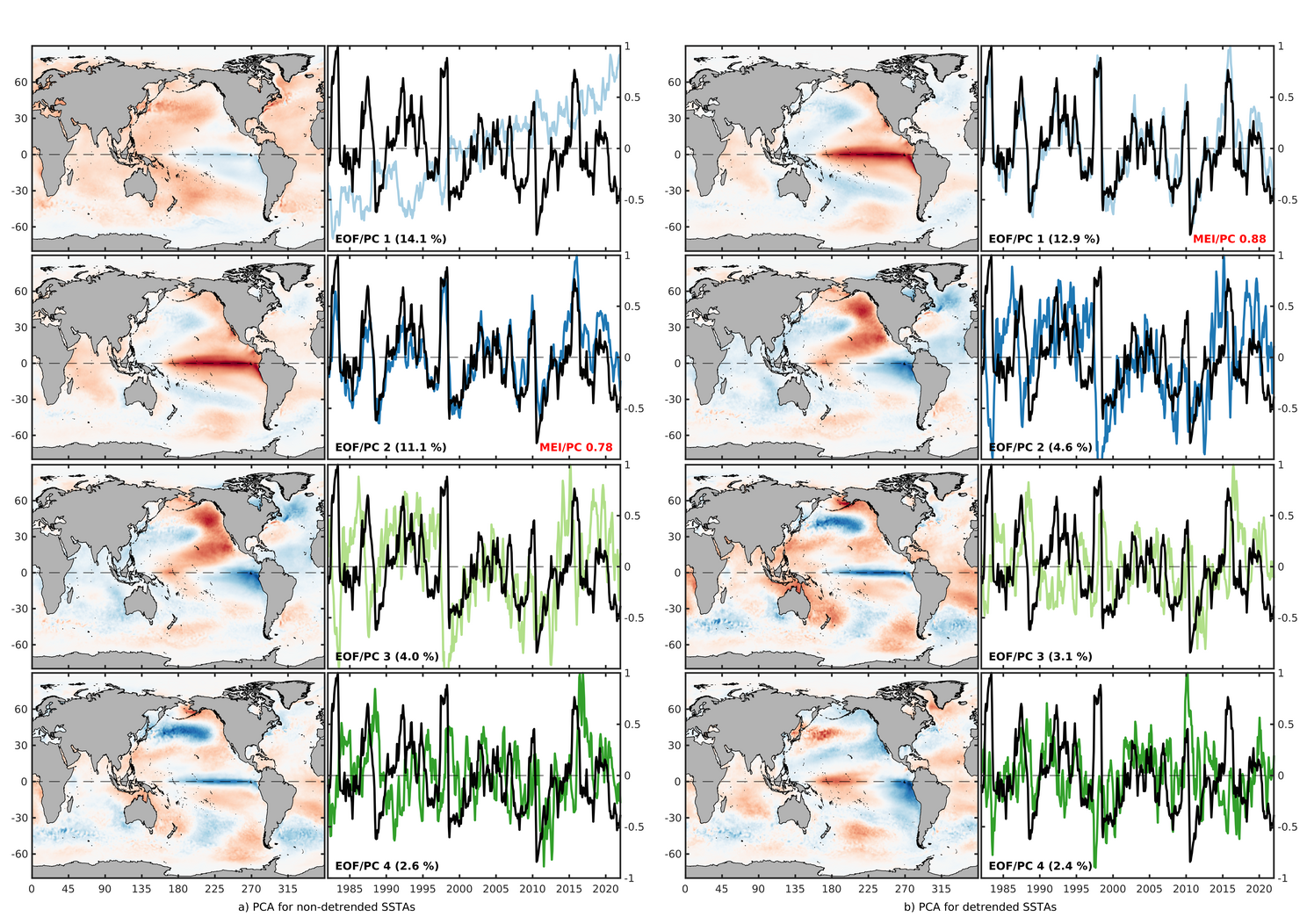}
    \caption{First four (rows) empirical orthogonal function (EOF, left column) and principal component (PC, right column) modes of global a) non-detrended and b) detrended SSTAs over the 1982 to 2022 period. The MEI index (black line) is overlaid on the PCs. For each mode, the percentage of variance explained is given on the bottom left corner of each subplot on the right and, on the left. Reported in red, is the correlation coefficient computed at zero lag over the 1982 to 2022 period between MEI and $PC_{2}$ and $PC_{1}$ for a) and b), respectively.}
    \label{fig:ssta_pca}
\end{figure}

In the following, we justify the removal of the 41-year long linear trend from SSTAs by referring to the recent publications  of \cite{wil18b:grl}, \cite{wil20:jcli} and \cite{jeb20:jcli}. \cite{wil18b:grl} propose supervised ML pattern recognition methods to separate the anthropogenic forced response from internal variability in both OAGCMs and observations using a combination of low-pass filtering to separate long-timescale signal -- the forced response -- from short-timescale noise -- the internal variability -- and a fingerprinting technique to maximize the associated signal-to-noise variance. As observational records, 100 years of \cite{cow14:qjrms} infilled Hadley Center Climate Research Unit version 4 surface temperature (HadCRUT4) is used. Not only the global mean HadCRUT4 temperature, but also the basin mean SST difference between eastern and western equatorial Pacific \citep[Figures 15a and 15b]{wil20:jcli} for example, displays strong non-linear trends when considered over the entire 1920 to 2020 period. However from 1980 onward, these trends are approximately linear in \cite{wil20:jcli}. In comparison, Figures \ref{fig:ssta_pca}a and  \ref{fig:ssta_pca}b display the first four empirical orthogonal function (EOF) and principal component (PC) modes for both non-detrended and linearly detrended SSTAs for OISST v2.1 \citep{rey07:jcli}. The first EOF/PC pair in Figure \ref{fig:ssta_pca}a, corresponding to 14.1\% of the variance, captures the global warming pattern as indicated by the upward trending PC combined to the overall positive EOF pattern, although the South-Eastern Pacific shows significant cooling in a wedge-shaped domain, which has been linked to atmospherically driven enhanced up-welling in response to both increasing greenhouse gas and ozone depletion since the 1980s by \cite{jeb20:jcli} through dedicated detection and attribution experiments. The first EOF/PC pair is followed by three EOF/PC pairs highly ($> 0.9$) and significantly (p-value $< 10^{-6}$ at 0.05 significance level) correlated to the first three EOF/PC pairs of Figure \ref{fig:ssta_pca}b with a slight increase in variance for the individual modes of the detrended case.

As we strive for simplicity in the data preparation, we posit that the forced response to anthropogenic climate change and secular natural trends can be approximated by a linear trend and removed from SSTAs over the 1982-2022 period. Furthermore, the choice of OISST v2.1 dataset over longer gridded SST records such as NOAA Extended Reconstructed Sea Surface Temperature (ERSST) Version 5\footnote{Boyin Huang, Peter W. Thorne, Viva F. Banzon, Tim Boyer, Gennady Chepurin, Jay H. Lawrimore, Matthew J. Menne, Thomas M. Smith, Russell S. Vose, and Huai-Min Zhang (2017): NOAA Extended Reconstructed Sea Surface Temperature (ERSST), Version 5. NOAA National Centers for Environmental Information. doi:10.7289/V5T72FNM [2023-07-07].} or COBE-SST Version 2 \citep{hir14:jcli} is motivated by concerns that reconstructed SST products prior to the satellite area may suffer from spatiotemporal sampling inhomogeneities due to the disparity in the number of available observations between the Southern and Northern hemispheres for example, potentially leading to the mischaracterisation of ENSO phases or even the detection of spurious ones, as illustrated and discussed in Sections S2 and S3 of the supplementary material.

%%  M-file
%   JCLI_fig1.m
%

The archetypes spatial and temporal imprints, and their linkages to extra-tropical atmospheric circulation are revealed by compositing Japanese 55-year Reanalysis (JRA-55) monthly averaged atmospheric fields \citep{kob15:jmsj} over overlapping time periods as described in \cite{bla22:aies}. The imprint of ENSO on global precipitation is further ascertained using Global Precipitation Climatology Project (GPCP) Version 2.3 \citep{adl18:atmos} gridded data for the entire globe. 
Sea level anomalies (SLA) from AVISO$^{+}$ Satellite altimetry data\footnote{The Ssalto/Duacs altimeter products were produced and distributed by the Copernicus Marine and Environment Monitoring Service (CMEMS) (\url{http://www.marine.copernicus.eu})} and Simple Ocean Data Assimilation (SODA) Version 3.15.2\footnote{\url{https://www2.atmos.umd.edu/~ocean/index_files/soda3.15.2_mn_download_b.htm}} \citep{car08:mwr} ocean temperature composites will provide additional evidence of the relevance of AA to the study of ENSO diversity for sub-surface ocean fields. The Multivariate ENSO \citep{wol11:ijc} and the Oceanic ENSO Indices (MEI\footnote{\url{https://psl.noaa.gov/enso/mei}} and ONI\footnote{\url{https://origin.cpc.ncep.noaa.gov/products/analysis_monitoring/ensostuff/ONI_v5.php}}) will be compared to indices derived by AA.

%%%%%%%%%%%%%%%%%%%%%%%%%%%%%%%%%%%
%   Section Archetypal analysis
%%%%%%%%%%%%%%%%%%%%%%%%%%%%%%%%%%%
\section{Archetypal analysis}\label{aa}
Archetypal analysis is an unsupervised pattern recognition method. Since its development by \cite{cut94:tec}, AA has only recently been adopted by the weather and climate community to identify extreme geophysical configurations and to investigate their impacts at the regional and global scales \citep{ste15:jcli, han17:jcli, ric21:jhm, ris21:mwr, bla22:aies, cha22:ncom}.

The AA representation involves stochastic or probability matrices, the matrices \textbf{C} and \textbf{S} defined hereafter. These matrices correspond to the probability of expression of dSSTA records in archetypes (cluster definition), themselves extreme geophysical configurations of the dataset as a whole, and the probability of expression of archetypes in each and every dSSTA record (cluster membership). This unique property is crucial and leads to a probabilistic expression of the identified phases definition and membership. It can be also applied to composites of any geophysical fields associated with ENSO. 

For a detailed review of the method and its various implementations, the reader is referred to the following monographs, \cite{han21a:sas} and \cite{tre21:mfa}, and specifically to \cite{han17:jcli} and \cite{bla22:aies} when applied to global SST. Hereafter, AA is briefly introduced and compared to the commonly used principal component analysis (PCA) in Table \ref{tab:aa}, PCA being also used hereafter as a dimension reduction first step to speed-up computation as in \cite{ric21:jhm}, \cite{ris21:mwr}, \cite{bla22:aies} and \cite{cha22:ncom}. 
\begin{table}[htbp]
    \centering
    \begin{tabular}{p{0.45\textwidth}p{0.45\textwidth}}
    \hline
    Singular value decomposition (SVD) & Archetypal analysis (AA)\\
    \hline\hline
    $\textbf{X}=\textbf{U}\Lambda\textbf{V}^{T}$ & $\textbf{X}\approx\textbf{XCS}$ \\
    (Deterministic) solution of & (Non-deterministic) solution of \\
    \quad$\underset{\textbf{U},\textbf{S},\textbf{V}}{\arg\min}\|\textbf{X} - \textbf{U}\Lambda\textbf{V}^{T}\|_{F}^{2}$ & \quad$\underset{\textbf{C},\textbf{S}}{\arg\min}\|\textbf{X} - \textbf{XCS}\|_{F}^{2}$ \\
    with (\textbf{quadratic}) constraints & with (\textbf{convexity}) constraints \\
    $\sum_{s} U_{sr} U_{st} =\delta_{rt}$ & $\sum_{p} S_{pt}=\textbf{1}_{t},\, S_{pt}\ge 0,\, \forall p,t$ \\
    $\sum_{t} V_{rt} V_{st} =\delta_{rs}$ & $\sum_{t} C_{tp}=\textbf{1}_{p},\, C_{tp}\ge 0,\, \forall p,t$ \\
    $\text{Trace}(\Lambda\Lambda^{T}) = \text{Trace}(\textbf{XX}^T) = \| \textbf{X}\textbf{X}^{T} \|_{F}^{2}$ & \\
    \hline
    \end{tabular}
    \caption{Compare and contrast SVD and AA matrix factorisation methods. The symbol $\|\textbf{X}\|_{F} = \sqrt{\sum_{i}\sum_{j} |X_{ij}|^2}$ denotes the Froebenius norm of the matrix $\textbf{X}$. The superscript $T$ denotes the matrix transpose. }
    \label{tab:aa}
\end{table}	

AA is not only a matrix factorisation method \citep{cic09:wiley,ela10:spr,eld19:siam,gan20:siam,gil21:siam,han21a:sas,tre21:mfa}, but, like PCA, is also a dimension reduction procedure \citep{ize08:ch7, ngu19:pcbi}, whereby a dataset, represented by a data matrix $\textbf{X}$ , $\textbf{X}=\textbf{X}_{s \times t}$ of space-time dimensions ($s, t$), can be reduced to or approximated by a product of factors such that $\textbf{X}\approx\textbf{XCS}=\textbf{X}_{s\times t}\textbf{C}_{t\times p}\textbf{S}_{p\times t} = \textbf{XC}_{s\times p}\textbf{S}_{p\times t}$, where the summation over the repeated indices $t$ and $p$ is implied. The factors, $\textbf{XC}_{s \times p}=\textbf{X}_{s\times t}\textbf{C}_{t\times p}$, are called archetypes. They are located on the convex hull of the dataset $\textbf{X}$ and are convex combinations of data points, the matrix $\textbf{C} = \textbf{C}_{t\times p}$ being a right-stochastic matrix with each row summing to 1, $\sum_{t} C_{tp}=\textbf{1}_{p},\, C_{tp}\ge 0,\, \forall p,t$. The number of archetypes, $p$, the cardinality of the archetypal representation, is chosen \textsl{a priori}. The other factors, $\textbf{S} = \textbf{S}_{p\times t}$, can be viewed as soft-clustering affiliation weights such that points in $\textbf{X}$ can be approximated by a convex combination of archetypes, $\textbf{X}\approx \textbf{XC}_{s\times p}\textbf{S}_{p\times t}$, with $\textbf{S}$ being also a right-stochastic matrix with $\sum_{p} S_{pt}=\textbf{1}_{t},\, S_{pt}\ge 0,\, \forall p,t$\footnote{Here, $\textbf{1}_{p}$ and $\textbf{1}_{t}$ stand for unit vectors of dimensions $p$ and $t$.}.

Both matrices, $\textbf{C}$ and $\textbf{S}$, are found by minimizing the square distance between the original dataset $\textbf{X}$ and its reduced representation $\textbf{XCS}$,
\begin{equation}
    \underset{\textbf{C},\textbf{S}}{\arg\min}\|\textbf{X} - \textbf{XCS}\|_{F}^{2},
    \label{eq:cost}
\end{equation}
where the distance function $\|\cdot\|_{F}$ stands here for the Froebenius norm, the convexity constraints on $\textbf{C}$ and $\textbf{S}$ being enforced in the minimization algorithm by adding penalty terms to Expression \ref{eq:cost} \citep{cut94:tec, mor12:neu}.

In their original paper, \cite{cut94:tec} proved through a \textsl{reductio ad absurdum}\footnote{\url{https://en.wikipedia.org/wiki/Reductio_ad_absurdum}.} argument that the archetypes, $\textbf{XC}$, can only be on the convex hull of the dataset under consideration. In broad terms, the main reasons for it lay in the convexity constraints such that the archetypes are a convex combination of data points and that the data points themselves are approximated by a convex combination of archetypes, and in the mathematical form of the cost function. This remarkable property can be used therefore to identify extreme configurations in geophysical observations \citep{han17:jcli, ris21:mwr, bla22:aies, cha22:ncom} and, conversely, to express observations in term of these extreme configurations. 

\begin{figure}[htbp]
    \centering
    \includegraphics[width=1.0\textwidth]{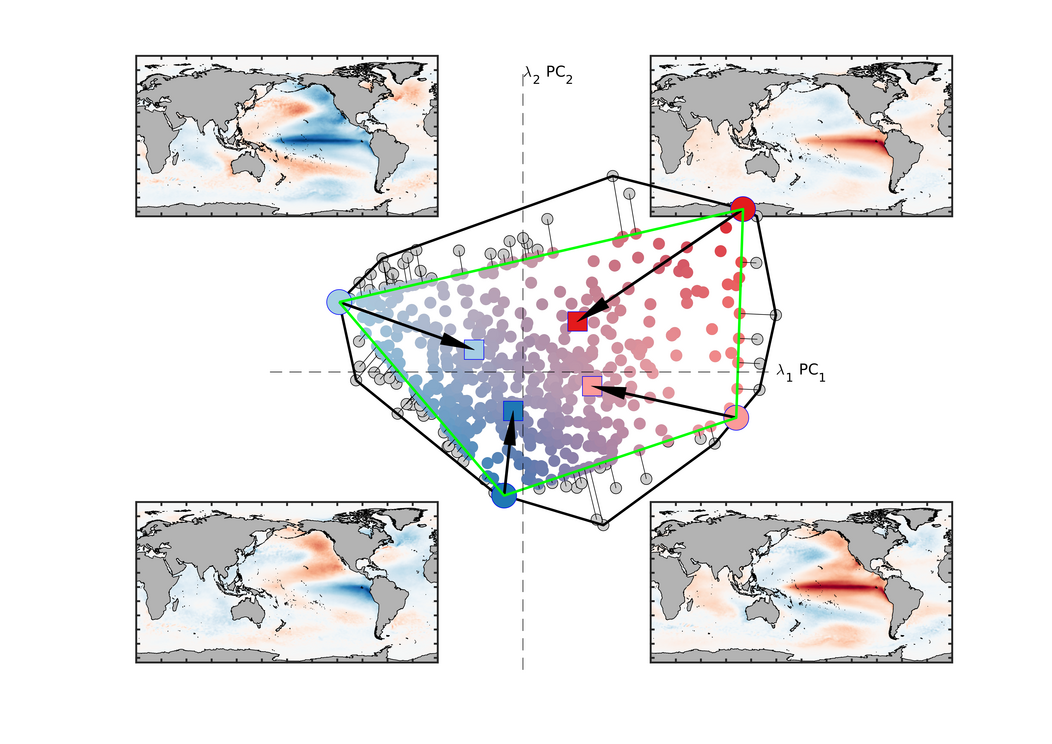}
    \caption{AA results for cardinality 4 based only on dSSTAs first 2 PCs multiplied by their respective eigenvalues, $\textbf{X}_{2 \times t}=[\lambda_{1}PC_{1}, \lambda_{2}PC_{2}]$.  The 2D convex hull is the black polygon. Points in the pointset are represented by coloured dots if they lay within the convex hull approximation - the green tetragon - which 4 vertices correspond to the 4 archetypes, $\textbf{XC}_{2 \times 4}$ (large coloured circles). Individual data points located outside the AA convex hull approximation are coloured grey, the data points within the AA approximation of the convex hull are coloured with a mixture of pure colours corresponding to the 4 archetypes. The respective archetypal maps corresponding to $\textbf{XC}_{2 \times 4}$ are reported in each quadrant. The coloured squares mark the composites of \textbf{X}, $\textbf{X}\Tilde{\textbf{S}}^{T}$ (Equation \ref{eq:fs}), constructed from the archetype mean probability of expression given by the \textbf{S}. The arrows link the archetypes $\textbf{XC}_{2 \times 4}$ to their respective composites $\textbf{X}\Tilde{\textbf{S}}^{T}$.}
    \label{fig:2pcs}
\end{figure}

To illustrate the nature of AA, Figure \ref{fig:2pcs} summarises its main characteristics for cardinality 4 and a highly truncated 2-dimensional dataset, where the data matrix $\textbf{X}_{2 \times t}=[\lambda_{1}PC_{1}, \lambda_{2}PC_{2}]$ is chosen to be the first 2 EOF/PC pairs of dSSTAs, illustrated in Figure \ref{fig:ssta_pca}b, rows 1 and 2. Its convex hull is represented by the black polygon encapsulating all the data points in the set. The archetypes $\textbf{XC}_{2 \times 4}$, the coloured circles located on the hull, correspond to extreme global configurations. The associated archetypal patterns are reported in each quadrant and recover the familiar warm/cold and eastern/central phases of ENSO. The green tetragon, whose vertices are the archetypes $\textbf{XC}_{2 \times 4}$, is an approximation of the dataset convex hull. It encapsulates all the points of the representation $\textbf{XCS}_{2 \times t}$. These points are found by solving Expression \ref{eq:cost} for \textbf{C} and \textbf{S} by minimising the distances, illustrated by the thin black segments in Figure \ref{fig:2pcs}, between the original data points in grey and their approximations in colour. The coloured squares stand for the composites of \textbf{X}, $\textbf{X}\Tilde{\textbf{S}}^{T}$, constructed from the archetype mean probability of expression given by the \textbf{S} as in Equation \ref{eq:fs}. In the Section S1 of the supplementary material, we show that this example can be extended to 3 dimensions (3D) with datasets such as the `Stanford Bunny'\footnote{\url{https://en.wikipedia.org/wiki/Stanford_bunny}}, a computer graphics 3D test model, and the first 3 EOF/PC pairs of dSSTAs, for selected cardinalities. In the rest of the paper, we use 492 PCs of dSSTAs and so a 492-dimensional dataset.

Hereafter, AA solutions will be compared to results from two commonly used pattern recognition or clustering methods: K-means (KM) and fuzzy C-means (FCM) \citep{ste57:baps,for65:bio,llo82:IEEE}. As in \cite{mor12:neu}, Table \ref{tab:ucm} summarises the main characteristics of KM and FCM factors with those of PCA and AA. The main difference between AA and both clustering methods, KM and its extension FCM, is that the KM and FCM cluster centres do not generally correspond to extreme configurations, as they are not necessarily located on the convex hull of the dataset.

\begin{table}[htbp]
    \centering
    \caption{Relation between the AA model and unsupervised methods such as SVD/PCA, fuzzy C-means, and K-means \citep{ste57:baps,for65:bio,llo82:IEEE} for the factorisation of $\textbf{X}$ adapted from \cite{mor12:neu}, where the symbol $\|\textbf{v}\|_{1} = \sum_{i}|v_{i}|$ denotes the Taxicab or Manhattan norm $l_{1}$ and $\|\cdot\|$ the Euclidean distance or $l_{2}$ norm. The $s \times t$ elements of the data matrix $\textbf{X}$ are denoted by $X_{ij}, \forall i=1,\ldots,s\,(\text{spatial dimension}), j=1,\ldots,t\,(\text{time dimension})$, whereas the index $p$ corresponds to the clustering cardinality, the exponent $m\in (1,+\infty)$ stands for the C-means `fuzzyfier' \citep{dun73:jc,bez84:cc}, $\textbf{x}_{t} = X_{\cdot t}$ and $\textbf{c}_{c}$ are the data records and the cluster centers in vector form, with $c=1,\ldots,p$. The symbols $\mathbb{R}$ and $\mathbb{B}$ respectively stand for the sets of real numbers and booleans, consisting of two elements, false-0 and true-1, used for categorical classification.}
    
	\begin{tabular}{llll}\hline \\
        SVD/PCA & AA/PCH & Fuzzy C-means\footnote{\url{https://en.wikipedia.org/wiki/Fuzzy_clustering}} & K-means\\ \\
	    \hline\hline
	    \\
	    $\textbf{C}\in \mathbb{R}$ & $\|\textbf{c}_{p}\|_{1}=1, \textbf{C}\ge 0$ & $C^{m}_{tc}=\frac{S_{ct}^{m}}{\sum_{t} S_{ct}^{m}}$ & $\|\textbf{c}_{p}\|_{1}=1, \textbf{C}\ge 0$ \\ \\	    
	    $\textbf{S}\in \mathbb{R}$ & $\|\textbf{s}_{t}\|_{1}=1, \textbf{S}\ge 0$ & $S^{m}_{ct}=\left\{ \sum_{k}^{p}\left(\frac{\left\|\textbf{x}_{t}-\textbf{c}_{c}\right\|}{\left\|\textbf{x}_{t}-\textbf{c}_{k}\right\|}\right)^{\frac{2}{m-1}} \right\}^{-1}$ & $\|\textbf{s}_{t}\|_{1}=1, \textbf{S}\in \mathbb{B}$ \\ \\
	    \hline
	\end{tabular}
    \label{tab:ucm}
\end{table}

Time-mean composites of any geophysical field can be straightforwardly constructed from the AA factors, $\textbf{C}$ and $\textbf{S}$, as weighted or conditional means, where the matrix elements $C_{tp}$ and $S_{pt}$ can be viewed as 1) the probability of a data record at time $t$ to contribute to the definition of archetype $p$ for $\textbf{C}$ or 2) the probability of archetype $p$ to be expressed in a data record at time $t$ for $\textbf{S}$, respectively. Let's recall the AA approximation,
\begin{equation*}
    \textbf{X}\approx\textbf{XCS},
\end{equation*}
with convexity constraints $\sum_{p} S_{pt}=\textbf{1}_{t}$, $S_{pt}\ge 0$ and $\sum_{t} C_{tp}=\textbf{1}_{p}$, $C_{tp}\ge 0,\, \forall p,t$. 
Comparing composites of the original dataset $\textbf{X}$ as in \cite{bla22:aies} Figures 6 to 8, $\textbf{XC}$ and $\textbf{X}\bar{\textbf{S}}^{T}$, separately built on $\textbf{C}$ and $\textbf{S}$ factors, is instructive as these composites may inform on the robustness of the AA factorisation; the more similar they are, the stronger the discrimination or separation between archetypes. Typically, for any given archetype $p$, the non-zero elements of the sparser matrix $C_{tp}$ at record $t$ should be reflected in the increased value of the element $S_{pt}$ at record $t$ for the denser matrix $\textbf{S}$ when compared to the time-mean $\bar{S}_{p}=\frac{1}{N}\sum_{t} S_{pt}, \, \forall p$, where $N$ is the total number of records. However, the reverse assertion does not necessarily hold true as the sparsity of $\textbf{C}$, resulting from the optimization, may well favour a single data snapshot as archetype even if the corresponding condition is repeated at multiple occasions across the time records considered.

For any spatiotemporal field, $\textbf{F} = \textbf{F}_{s\times t}$\footnote{The spatial dimensions of $\textbf{F}$ and $\textbf{X}$ do not need to be identical. We have kept the same symbol $s$ for $\textbf{F} = \textbf{F}_{s\times t}$ to simplify the notation.}, AA (time-)mean composite for each archetype $p$ can be constructed in similar fashion as weighted means
\begin{equation}
    \textbf{F}^{C}_{s\times p} = \textbf{FC} = \sum_{t} F_{st} C_{tp}
    \label{eq:fc}
\end{equation}
for \textbf{C}-composites, or as
\begin{equation}
    \textbf{F}^{S}_{s\times p} = \textbf{F}\bar{\textbf{S}}^{T} = \frac{\sum_{t} F_{st} S^{T}_{pt}}{\sum_{t} S_{pt}}
    \label{eq:fs}
\end{equation}
for $\textbf{S}$-composites. To generate seasonal composites, the summation over time in Equations \ref{eq:fc} and \ref{eq:fs} is only performed on the appropriate month-of-year records, hereafter December-January-February (DJF), March-April-May (MAM), June-July-August (JJA) and September-October-November (SON).

The corresponding `in-sample' AA composite uncertainty or spread is expressed as follows\footnote{\url{https://en.wikipedia.org/wiki/Weighted_arithmetic_mean#Reliability_weights}.}. With the AA composite weighted mean $\textbf{F}^{S}_{s\times p}$ constructed as above, $\sum_{t} S_{pt} = W^{S}_{p}$ and the weights $w_{pt} = S_{pt}/W^{S}_{p}$, the covariance matrix ${\textbf{C}^{S}_{F,s\times p}}$ reads
\begin{equation}
\textbf{C}^{S}_{F,s\times p} = \frac{\sum _{t}w_{pt}}{\left(\sum _{t}w_{pt}\right)^{2}-\sum _{t}w_{pt}^{2}} \sum _{t}w_{pt}\left(F_{st}-F^{S}_{sp}\right)^{T}\left(F_{st}-F^{S}_{sp}\right).
\label{eq:cs}
\end{equation}
A similar, but simpler expression, can be derived for the covariance matrix ${\textbf{C}^{C}_{F,s\times p}}$, corresponding to \textbf{C}-composites, 
\begin{equation}
\begin{split}
\textbf{C}^{C}_{F,s\times p} & = \frac{\sum _{t}C_{tp}}{\left(\sum _{t}C_{tp}\right)^{2}-\sum _{t}C_{tp}^{2}} \sum _{t}C_{tp}\left(F_{st}-F^{C}_{sp}\right)^{T}\left(F_{st}-F^{C}_{sp}\right),\\
                             & = \frac{1}{1-\sum _{t}C_{tp}^{2}}\sum _{t}C_{tp}\left(F_{st}-F^{C}_{sp}\right)^{T}\left(F_{st}-F^{C}_{sp}\right),
\end{split}
\label{eq:cc}
\end{equation}
as the weights reduce to $C_{tp}$ as $W^{C}_{p} = \sum_{t} C_{tp}=1, \forall p$. Of note is that similar expressions for the KM and FCM cluster centers and associated composites can be constructed.

The introduced `in-sample' measure of significance based on weighted covariance, Equations \ref{eq:cc} and \ref{eq:cs}, indicates whether or not the archetypal patterns discovered are repeatable or reoccur over the period considered in a conditional time-mean sense by comparing pattern signal-to-noise (or mean-to-spread) ratios. We note that the `in-sample' signal-to-noise ratios are a very conservative measure of reliability, especially for composites based of the affiliation weights, the stochastic matrix $\textbf{S}$. For `out-of-sample' AA composite uncertainty or spread determination, a less stringent measure of reliability or significance, the problem rapidly becomes computationally intractable for (spatial or reduced) dimensions larger than 9\footnote{\url{http://www.qhull.org}} \citep{bar96:acmtms}, as one is required to randomly sample the convex hull of the original or reduced dataset. An even less stringent approach as in \cite{cha22:ncom}, would be to employ a brute-force Monte-Carlo approach, 1) to generate an ensemble of random synthetic stochastic matrices to replicate the features of both the $\textbf{C}$ and $\textbf{S}$, 2) to compute the composites on these, and 3) to declare each point of the composites `significant' if not falling within a certain percentile range constructed from the synthetic ensemble. As archetypes correspond by construction to extreme configurations, the latter method mentioned is too permissive. The composites generated from random stochastic matrices are not necessarily located on the convex hull of the dataset and cannot therefore be considered `legitimate' samples as archetypes are located on the convex hull by construction.

%%%%%%%%%%%%%%%%%%%%%%%%%%%%%%%%%%%%%%%
%   Section Archetypal flavours of ENSO
%%%%%%%%%%%%%%%%%%%%%%%%%%%%%%%%%%%%%%%
\section{Archetypal flavours and indices of ENSO}\label{sec:flavours}
AA is applied to linearly detrended monthly mean SSTAs (dSSTA) over the 1982-2022 period. To increase computational speed, we implement the reduced space AA (RSAA) methodology reviewed in \cite{bla22:aies}, whereby a PCA dimension reduction step is performed on the dataset prior to apply AA, to factor out of the cost function -- Expression \ref{eq:cost}, the spatial dependency given by the orthonormal empirical functions $\textbf{U}$ in Table \ref{tab:aa}. AA is done on all $12\times 41 = 492$ PCs multiplied by their corresponding eigenvalues, $\Lambda\textbf{V}^{T}$, where 100\% of variance of dSSTA is retained for the ensuing AA step. The factors $(\textbf{XC})_{RSAA} = \textbf{U}(\Lambda\textbf{V}^T)\textbf{C}$ and $\textbf{S}_{RSAA}$ are now the archetypal patterns and affiliation probabilities. It can be shown that $(\textbf{XC})_{RSAA}$ and $\textbf{S}_{RSAA}$ are almost identical to the AA solutions of the non-reduced problem \citep[Figure 3]{bla22:aies}. To simplify the notation, the subscript $RSAA$ is omitted from this point onwards. An identical dimension reduction strategy is applied to both KM and FCM clustering algorithms. For a given cardinality or cluster number $p$ and in all three cases, AA, KM and FCM, multiple trials are necessary to make sure the solutions are close to optimal. Typically, 1000 trials are performed with different random initialisations and the solutions correspond to the lowest cost, Expression \ref{eq:cost}, across all trials.
%%  M-file
%   JCLI_fig2.m
%
\begin{figure}[htbp]
    \centering
    \includegraphics[width=0.95\textwidth]{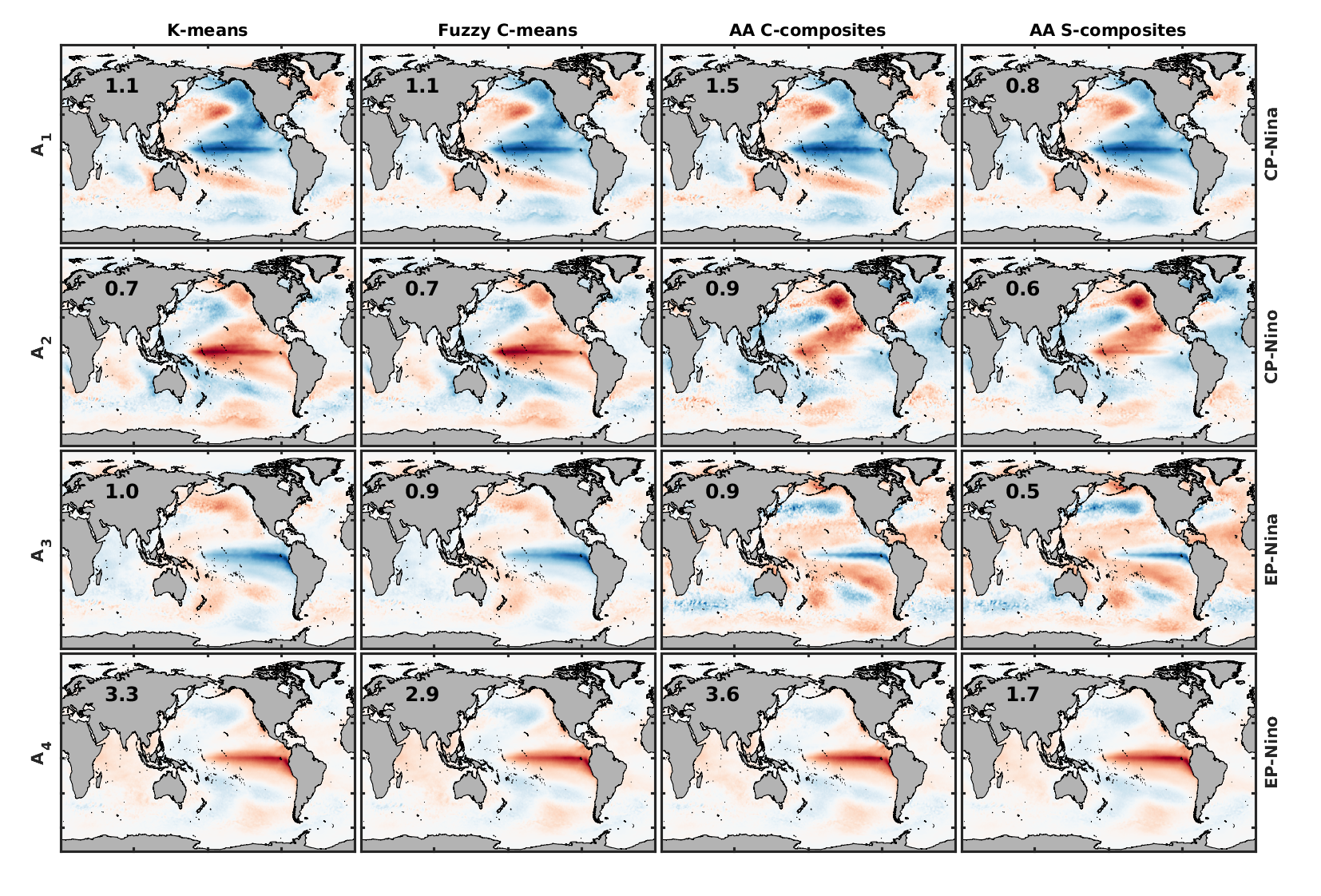}
    \caption{Reduced space (RS) detrended SSTA cluster centers for K-means, fuzzy C-means, AA $\textbf{C}$-composites $\textbf{X}\textbf{C}$ and $\textbf{S}$-composites $\textbf{X}\textbf{\~S}^{T}$ (columns). The cluster centers for each method (rows) are referenced to the AA results for a cardinality of 4 (last column) and assigned using the euclidean distance. Each pattern has been normalised by its maximum absolute value for increased contrast across clusters, reported on the Eurasian continent in \textdegree C. The `fuzzyfier' exponent $m$ was set to 1.05 for C-means. The labels $A_{p}$ in each row indicate the archetype ranks based on the time mean of the AA stochastic matrix $\textbf{S}_{p\times t}$, $\overline{S}_{p}=\frac{1}{N}\sum_{t} S_{pt}$ with $p=1,\ldots,4$, where $N$ is the total number of records.}
    \label{fig:kfa4}
\end{figure}

As in \cite{bla22:aies}, the dSSTA data matrix, $\textbf{X}_{s\times t}$, is weighted spatially by $\textbf{W}$ with elements $W_{ij} = \sqrt{cos(\theta_{i})}$, $\theta_{i}$ being the latitude of the grid cell $i=1,\ldots,s$ for all times $j=1,\ldots,t$ prior applying PCA, AA, KM and FCM to $\textbf{XW}_{s \times t} = \textbf{X}_{s \times t} \odot \textbf{W}_{s \times t}$\footnote{Element-wise or Hadamard matrix product, $(A \odot B)_{ij}=A_{ij}B_{ij}$.} rather than $\textbf{X}_{s\times t}$ in order to compensate for unequal grid spacing. Therefore the tropical data points are carrying more weight than mid- and high-latitude ones. As not to render our notation too cumbersome, we omit the weights, $W_{ij}$. However, when KM, FCM and AA composites are computed, unweighted snapshots based on the dSSTA cluster affiliation sequences are used.

The choice of the domain driving the analysis is motivated by minimizing the number of arbitrary choices with the explicit goal of capturing both the tropical character and remote teleconnection of the ENSO phenomenon at zero monthly lag. In Section 2 of the supplementary material, we have repeated the analysis for Pacific Ocean regions over a wide range of latitude bands ranging from $\pm 10^{\circ}$ to $\pm 90^{\circ}$ centered on the Equator with a fixed longitudinal sector from $120^{\circ}$E to $280^{\circ}$E and compared archetypes, $\textbf{XC}$, for cardinality 8 using pattern correlation between overlapping regions of the restricted and global domains in Tables S1 and S2. From the latitudinal band $\pm 50^{\circ}$ and higher, individual regional archetypes can be uniquely matched to the global ones, with high correlation values. Alternatively, when comparing archetypes from one latitudinal band to the next in increments of $\pm 5^{\circ}$, starting with the Pacific region $\pm 10^{\circ}$ latitudinal band as in Tables S3 and S4, individual regional archetypes can only be uniquely matched from the latitudinal band $\pm 55^{\circ}$ and higher, an indication that to discriminate between extreme dSSTAs conditions, the `latitudinal reach' has to be increased.

Both matching strategies show that, to discriminate between archetypes across bands, it is necessary to select Pacific regions which latitudinal extent larger than about $\pm 50^{\circ}$ centered on the Equator. As we strive for parsimony, but also would like to capture the global teleconnections (at lag zero) across oceanic basins, we have therefore opted for a global domain.

Figure \ref{fig:kfa4} compares KM, FCM and AA results for a cardinality of 4. At first glance, the 4 cluster centers and archetypes (rows), labelled $A_{i},\, i=1,\ldots,4$, share similarities across methods (columns). The patterns in first two rows resemble La Ni\~na and El Ni\~no Modoki -- or Central Pacific (CP) -- ENSO conditions with large anomalous SSTs located in the equatorial central Pacific, whereas the third and fourth rows display cold and warm anomalies in the Eastern Pacific, reminiscent of the canonical -- or Eastern Pacific (EP) -- ENSO patterns for La Ni\~na and El Ni\~no, respectively. The AA CP-Ni\~no pattern detected for cardinality 4, $A_{2}$ in Figure \ref{fig:kfa4}, shares commonalities with the so-called Pacific Meridional Mode (PMM) as defined by \cite{chi04:jcli}. A clearer discrimination between PMM- and CP-Ni\~no-like conditions will only be revealed for higher cardinalities. A closer inspection of these patterns for KM and FCM reveals both Central Pacific cold and warm patterns are almost anti-symmetric; a property shared neither by the AA $\textbf{C}$- nor $\textbf{S}$-composites, Equations \ref{eq:fc} and \ref{eq:fs} where $\textbf{F}$ has been replaced by $\textbf{X}$. 
%%  M-file
%   JCLI_fig3.m
%
\begin{figure}[htbp]
    \centering
    \includegraphics[width=1.0\textwidth]{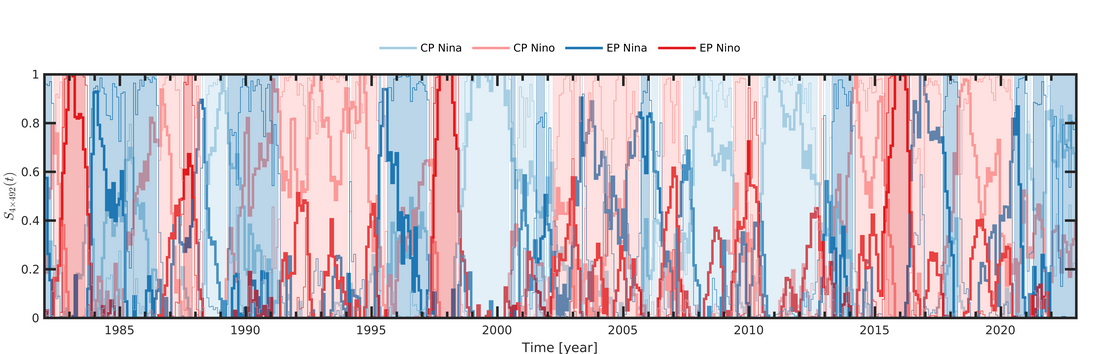}
    \caption{dSSTA cluster affiliation sequences for K-means, fuzzy C-means and AA. The coloured patches denote K-means `categorical' states corresponding to CP Ni\~na, CP Ni\~no, EP Ni\~na and EP Ni\~no intervals. Thick (thin) coloured lines stand for the AA (C-means) `soft' affiliation probabilities $S_{pt}\,(S^{m}_{pt}) \in [0, 1]$ with $p=1,\ldots,4$ corresponding to the spatial patterns depicted in rows of Fig. \ref{fig:kfa4}.}
    \label{fig:kfaa4}
\end{figure}
The asymmetry in amplitudes, spatial patterns, timing and teleconnection characteristics between the cold and warm phases of ENSO has been widely reported \citep[for early references on the subject] {hoe97:jcli,sar00:jcli,han03:cd} and is reflected by the affiliation probabilities reported in Figure \ref{fig:kfaa4}. However, the anti-symmetric nature of the KM and FCM clusters is an artefact of cardinality 4 as is evident for higher cluster cardinalities reported in Figure S7.

\begin{figure}[htbp]
    \centering
    \includegraphics[width=1.0\textwidth]{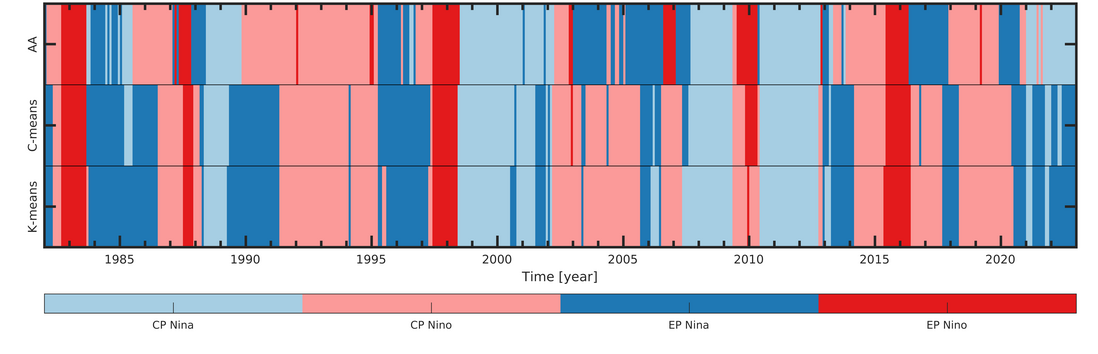}
    \caption{ENSO phase `hard' affiliation sequences for AA, fuzzy C-means, and K-means. The coloured patches denote `categorical' states corresponding to CP Ni\~na, CP Ni\~no, EP Ni\~na and EP Ni\~no intervals. For AA and fuzzy C-means clustering, the phases labeled by $i_{max}(t)$ and $i_{max}^{m}(t)$ correspond to highest affiliation probabilities, $S_{it}$ and $S^{m}_{it}$, are colour-coded.}
    \label{fig:kfai4}
\end{figure}

KM corresponds to a categorical clustering method where each data record is associated with a single pattern or cluster center with affiliation $\|\textbf{s}_{t}\|_{1}=1, \textbf{S}\in \mathbb{B}$ as in Table \ref{tab:ucm}. Figure \ref{fig:kfai4} compares KM to the soft-clustering methods, FCM and AA, whereby the ENSO phase categorical or `hard' affiliations are given by the states indexed by $i_{max}(t) = \underset{i={1,\ldots,p}}{\arg\max}(S_{it})$ and $i^{m}_{max}(t) = \underset{i={1,\ldots,p}}{\arg\max}(S^{m}_{it})$ corresponding to the maximum values of the AA and FCM affiliation time-series $S_{it}$ and $S^{m}_{it}$, respectively labeled by $i_{max}(t)$ and $i_{max}^{m}(t)$, and contributing with highest probability to the snapshot $X_{st}$ for each time record $t$\footnote{A tie, although always possible, is very unlikely given that the algorithms used implement double-precision arithmetic.}, indicating the `winning' archetype or FCM cluster - the archetype or FCM cluster corresponding to the highest probability of expression given by $i_{max}(t)$ or  $i^{m}_{max}(t)$.
%%  M-file
%   JCLI_fig4.m
%

\begin{figure}[htbp]
    \centering
    \includegraphics[width=1.0\textwidth]{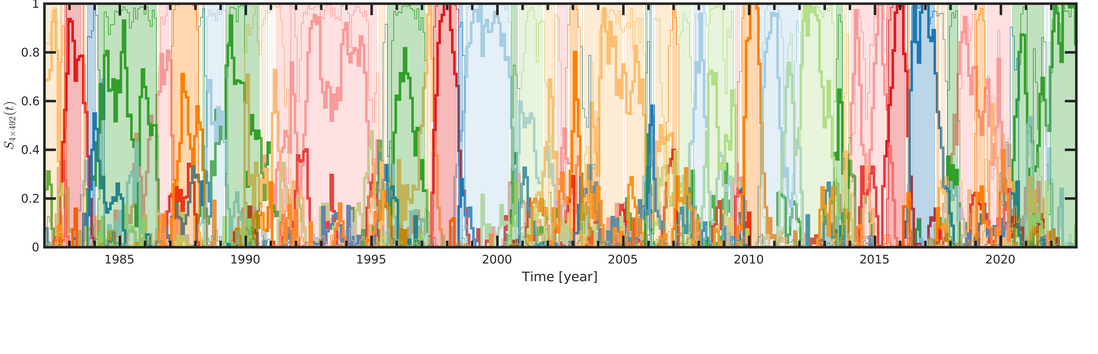}
    \caption{dSSTA cluster affiliation sequences for K-means, fuzzy C-means and AA for cardinality 8. The coloured patches denote K-means `categorical' states corresponding to CP Ni\~no, $A_{2}$, CP Ni\~na, $A_{4}$, EP Ni\~no, $A_{6}$, EP Ni\~na, and $A_{8}$ intervals. Thick (thin) coloured lines stand for the AA (C-means) `soft' affiliation probabilities $S_{pt} \in [0, 1]$ with $p=1,\ldots,8$ corresponding to the spatial patterns depicted in Fig. S7 rows.}
    \label{fig:kfaa8}
\end{figure}

\begin{figure}[htbp]
    \centering
    \includegraphics[width=1.0\textwidth]{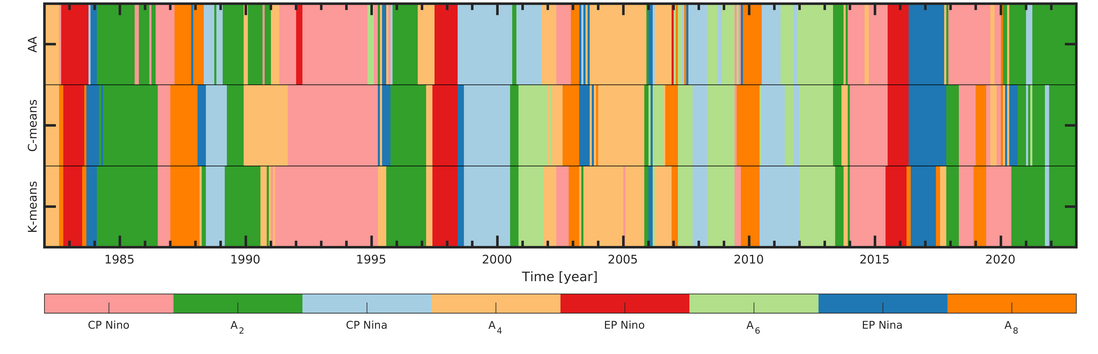}
    \caption{ENSO phase `hard' affiliation sequences for AA, fuzzy C-means, and K-means for cardinality 8. The coloured patches denote `categorical' states corresponding to CP Ni\~no, $A_{2}$, CP Ni\~na, $A_{4}$, EP Ni\~no, $A_{6}$, EP Ni\~na, and $A_{8}$ intervals. For AA and fuzzy C-means clustering, the phases corresponding to highest affiliation probabilities are colour-coded.}
    \label{fig:kfai8}
\end{figure}

Overall, the `hard' ENSO state affiliation mostly agrees across methods for cardinality 4, especially over EP Ni\~no (dark red) and CP Ni\-na (light blue) intervals in Figure \ref{fig:kfai4}. Of interest is the 2001-2007 interval, where both the KM and its soft-clustering extension FCM disagree with the AA categorical representation. However, increasing the cardinality improves the concordance between KM, FCM and AA as shown for the patterns displayed in Figure S7, where not only the issue of KM and FCM phase anti-symmetry raised previously is resolved, but also a better categorical cluster agreement across methods is achieved, as shown in Figures \ref{fig:kfaa8} and \ref{fig:kfai8}.

Due to the archetypal `nestedness'\footnote{Archetypal `nestedness', although not expected \textsl{apriori} in the original formulation of \cite{cut94:tec}, describes the remarkable property that archetypes can be matched across cardinality or number of clusters.} originally reported by \cite{bla22:aies} for dSSTAs, the 4 new distinct archetypes in Figure S7 for cardinality 8, not only adds more diversity to the ENSO states already identified in Figure \ref{fig:kfaa4}, but also reveals the complexity of the ENSO phenomenon and its sequential evolution over the 1982-2022 period. The 2001-2007 interval, where substantial discordance between KM, FCM and AA categorical affiliation was observed, shows now a better agreement across methods.

%%  M-files
%   JCLI_fig6.m and JCLI_fig7.m
%

Figures S15a and S15b, updated from \cite{bla22:aies}, illustrate the impact of cardinality on the affiliation probabilities expressed by the matrix $\textbf{S}$ and the patterns `nestedness', demonstrating that no spurious blending occurs across dSSTA states or clusters when extreme conditions are `well-expressed', as can be seen when a single archetype corresponding to the 3 major EP Ni\~no intervals of 1982-1983, 1997-1998 and 2015-2016, matches across 2-8 cardinalities as the dark red coloured intervals indicate. For a cardinality of 2 (the supplementary material, top row in Figure S15a, the other records when not corresponding to pure CP La Ni\~na intervals in light blue, being convex combinations of archetypes, have to be expressed as `blended' patterns by construction. 

Taking advantage of the convexity constraint on $\textbf{S}$, the extent of how `well-expressed' or not an archetype is, can be measured by a discrimination score, introduced for AA by \cite{ris21:mwr}, 
\begin{equation}
    \Delta^{p}_{i_{max}}(t)= 1 - \frac{1}{p-1} \left\{\frac{1}{\underset{i=1,\ldots,p}{\max} (S_{it})} - 1\right\} = 1 - \frac{1}{p-1}\left\{\frac{1}{S_{i_{max}t}} - 1\right\},
    \label{eqn:delta}
\end{equation}
where $\Delta^{p}_{i_{max}}(t) \in [0, 1]$ with $S_{i_{max}t} = \underset{i=1,\ldots,p}{\max} (S_{it})$ is obtained from maximum values of $\textbf{S}_{p\times t}$ corresponding to the archetype labelled by $i_{max} = i_{max}(t)$ contributing with highest probability to the snapshot $\textbf{X}_{s\times t}$ for each time record $t$\footnote{A similar expression can be devised for FCM affiliation probabilities $S^{m}_{pt}$ in Table \ref{tab:ucm}.}. Here, the symbol $i_{max}$ indicates that the discrimination score not only corresponds to a value ranging from 0 to 1, but also to the label of the `winning' archetype. A low discrimination score, $\Delta^{p}_{i_{max}}(t)$ close to zero, corresponds to a mixed or blended state at time $t$ with probabilities being approximately the same and equal to $1/p$ for all archetypes for a given cardinality $p$, whereas a discrimination score close to 1 indicates a pure archetype or `well-expressed' state at time $t$.

\begin{figure}[htbp]
    \includegraphics[width=\textwidth]{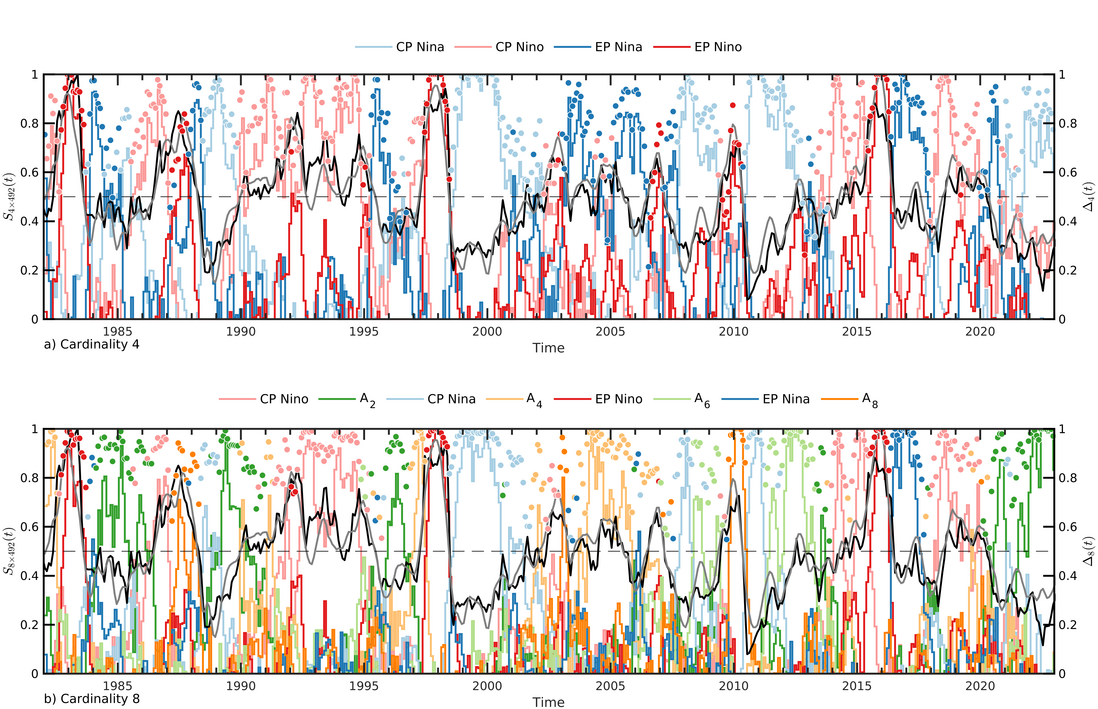}
    \caption{Comparison of AA affiliation sequences for cardinalities of a) $p =4$ and b) $p=8$. The discrimination scores (\cite{ris21:mwr}) $\Delta^{p}_{i_{max}}(t) \in$ [0, 1], $\, \forall i=1,\ldots,p$ for all monthly records $t$ are represented by archetypal colour-coded dots of the winning archetype. A higher discrimination score indicates less blended state. Both MEI (black) and ONI (grey) time series were scaled to fit within the $S_{4/8 \times 492}$ and $\Delta_{4/8}$ bounds.}
    \label{fig:stairs}
\end{figure}

\begin{figure}[htbp]
    \centering
    \includegraphics[width=1.0\textwidth]{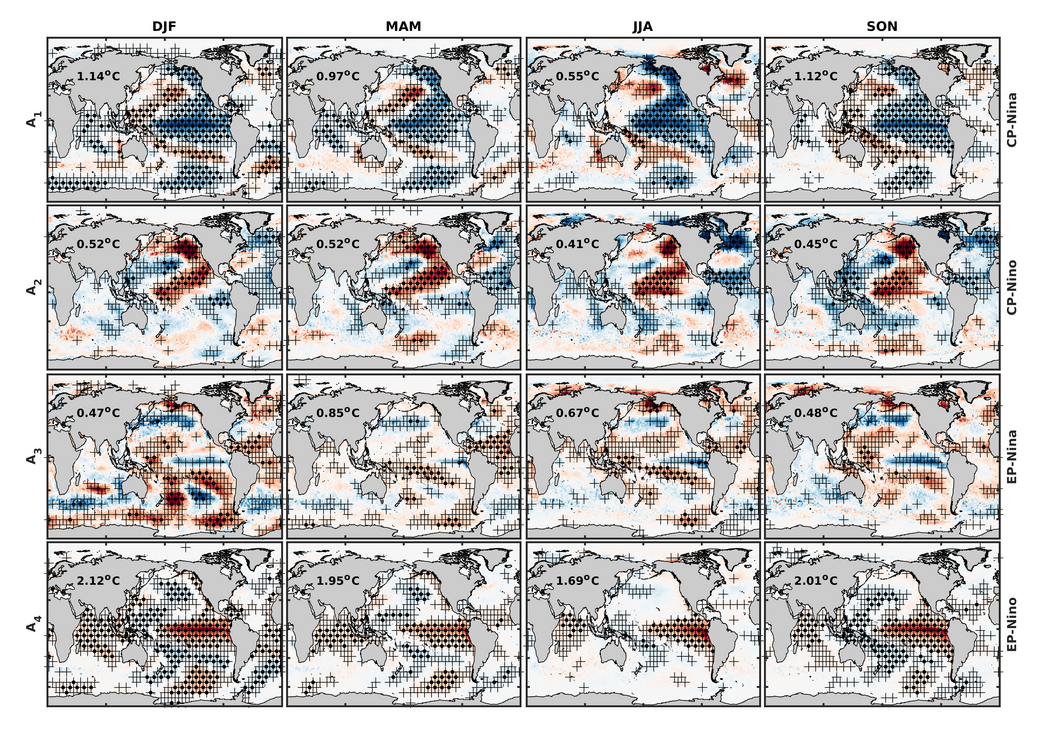}
    \caption{AA seasonal lag-0 composites (columns) and signal-to-noise ratios ($\textbf{+} > 0.15, \bullet > 0.3$, $(F^{2}_{S}/C^{S}_{F})^{1/2}$) based on time averages of the $\textbf{S}$-affiliation matrix over the 1982-2022 period for a cardinality of 4. As each individual subplot is scaled to increase contrast across archetypal patterns (rows), the boldfaced numbers on the Eurasian continent label the maximum absolute SSTA for each archetypal pattern $\textbf{A}_{i}, i = 1,...,4$ and seasons (DJF, MAM, JJA and SON).}
    \label{fig:aasc4}
\end{figure}

Figures \ref{fig:stairs}a and \ref{fig:stairs}b compare AA affiliation sequences, the AA ENSO indices, for cardinalities $p=4$ and $p=8$. The discrimination scores, $\Delta^{p}_{i_{max}}(t) \in [0, 1]$ for all monthly records $t$, are represented by archetypal colour-coded dots of the archetype $i_{max}(t)$. Increasing the cardinality from 4 to 8 leads to overall higher discrimination scores where they were initially low for cardinality 4 over the periods 1983-1997, 1998-2015 and 2016-2022, and introduces 4 new and distinct patterns corresponding archetypes $A_{2}$, $A_{4}$, $A_{6}$ and $A_{8}$ in Figures S7, for overall time-mean, and S8, for seasonal composites, respectively. As such, increasing the cardinality tends to improve the discrimination between states and reduces the number of `blended' states. When checked against Figure \ref{fig:aasc4}, the CP and EP ENSO patterns previously identified for cardinality 4 remain. However, the onset and decay of the 3 major EP Ni\~no intervals of 1982-1983, 1997-1998 and 2015-2016, are now different with $A_{4}$ $\rightarrow$ EP Ni\~no $\rightarrow$ EP Ni\~na, $A_{4}$ $\rightarrow$ EP Ni\~no $\rightarrow$ CP Ni\~na, and CP Ni\~no $\rightarrow$ EP Ni\~no $\rightarrow$ EP Ni\~na, respectively. Apart from the 1998-2000 period, the long lasting CP Ni\~na intervals detected for cardinality 4 over 1988-1989, 2007-2008, 2010-2012, and 2021-2022, are now further split by the introduction of an EP Ni\~na-like and EP Ni\~no-like patterns, $A_{2}$ and $A_{6}$. The interval commonly referred in the literature \citep{cap15:bams, wie21:fc} as the (Ni\~no) Modoki period over 2001-2007 is now mainly captured by CP Ni\~no-like archetypes $A_{4}$ and $A_{8}$.

%%  M-files
%   JCLI_fig8.m
%

The dSSTA composites based on time averages of the $\textbf{S}$-affiliation matrix over the 1982-2022 period for a cardinality of 4, Figure \ref{fig:aasc4}, show that archetypal patterns across seasons resemble the annual $\textbf{S}$- and $\textbf{C}$-composites displayed in Figure \ref{fig:kfa4}, but vary in amplitude from the tropics to higher latitudes. Of note are the strong seasonal differences expressed by archetypes corresponding CP-Ni\~no ($A_{2}$) and EP-Ni\~na ($A_{3}$), possibly indicative of more transitory -- but still extreme -- global SST conditions. For CP-Ni\~no, the positive SST anomalies approximately centered around the date-line in JJA, migrate westwards in SON, DJF and MAM, and are accompanied by a strengthening, relative to the Tropics that is, of a PMM-like pattern in the North Pacific \citep{chi04:jcli, ama19:cccr, ama20:ncom, ric22:npj}. This is reminiscent of the North Pacific SST conditions of 2013-2015 preceding the 2015-2016 EP Ni\~no event, as Figures \ref{fig:stairs}a and \ref{fig:stairs}b attest, and also reported by \cite{tse17:erl}, for example. The 2013-2015 CP-Ni\~no (PMM-like) conditions persist across AA cardinalities ranging from 3 to 8 as can be seen on Figures S15a (rows 2 to 8) and S15b (row 1). The EP-Ni\~na ($A_{3}$) phases are not as persistent for cardinality 4. An extra EP-Ni\~na like pattern, $A_{2}$, is now expressed for cardinality greater than 4, with the original EP-Ni\~na previously detected for a cardinality 4, now solely well expressed over 2016-2017 interval (dark blue line) in Figure \ref{fig:stairs}b.
%%  M-files
%   JCLI_fig10.m and JCLI_fig11.m
%

In summary, the comparison between AA results across cardinalities shows that the ENSO phenomenon takes multiple guises. AA, applied here to global dSSTAs, is able to distinguish between both recurrent patterns such as those observed over the well-known, but short, EP-Ni\~no 1982-1983, 1997-1998 and 2015-2016 intervals, and a series of intermittent ones between these.

Recent work by \cite{sch23:arX} derived from fuzzy unsupervised clustering also classifies the AA-detected EP-Ni\~nos as Extreme El Ni\~nos and our results, based on the detection of extreme configurations in global SSTA, corroborate these findings. Over the period 1982-2022 under study, the EP-Ni\~no phases are typically followed by Ni\~na-like conditions of various types and interspersed by longer periods of somewhat muted CP-Ni\~no conditions. The archetypal sequences over these `inter'--EP-Ni\~no intervals are all different as can be seen on Figures \ref{fig:kfai8} and \ref{fig:stairs}b. This suggests that ENSO likely corresponds to a `continuum' of types as advocated by \cite{joh13:jcli} using a self-organizing-map (SOM, \cite{koh90:ieee}) or by \cite{die21:cec} when analysing changes in the distributions of ENSO phases' location and intensity. The question of statistical distinguishability of these patterns raised in \cite{joh13:jcli} is a difficult one to assess given the short interval of high-quality satellite data with global coverage available. Three EP-Ni\~no events over 41 years can barely be considered as a large enough sample size for statistical inferences. However, we will show in Section \ref{sec:tele} that archetypes, expressing extreme global dSSTA conditions, correspond to distinct teleconnection composite patterns in atmospheric, surface and sub-surface fields. From this point onward, archetypal patterns and geophysical field composites pertaining to a cardinality of 8, will be given in the supplementary material.

An attentive reader would notice the unusual feature of AA that the event categories do not include  a `neutral’ condition. The `neutral' concept stems mainly from the adoption of tercile-type (negative/neutral/positive) classification based on univariate indices used by operational centers to qualify the current state of the climate. These have the advantage of simplicity in the messaging, but the over-reliance on these can be fraught and controversial as recently illustrated by the difficulty of identifying the 2023 El Ni\~no event.

The `neutral' ENSO condition can easily be implemented with AA by inspecting the discrimination score $\Delta^{p}_{i_{max}}(t)$, Equation \ref{eqn:delta}, implemented for AA in \cite{ris21:mwr} and \cite{cha22:ncom}, where archetypes dominance and persistence are illustrated. This score is a measure how representative archetypes are in any given data record $t$, ranging from 0 -- for low discrimination or a `neutral' ENSO state, the record corresponding then to a blend of archetypes with the equal probability of expression ($1/p$ for p archetypes) -- to 1 -- for high discrimination, the record being expressed by a single archetype $i_{max}$ or a `pure' ENSO state, with (maximum) probability of expression equal to 1. As with any neutral condition identification, thresholds selection based on expert consensus will have to be tested and implemented.

Together with a persistence criterion, neutral ENSO states could be defined as period of low archetype discrimination below a certain threshold over a certain length of time. However, we note that the power discrimination across archetypes from cardinalities ranging from 2 to 20 increases monotonically with cardinality as illustrated in Figure S9 of the supplementary material, making what one considers `neutral' somewhat moot.

%%%%%%%%%%%%%%%%%%%%%%%%%%%%%%%%%%%%%%%%%%%%%%%%%%%%%%%%%%%%%%
%   Section Traditional approaches or other clustering methods
%%%%%%%%%%%%%%%%%%%%%%%%%%%%%%%%%%%%%%%%%%%%%%%%%%%%%%%%%%%%%%
\section{Traditional approaches and other clustering methods}\label{sec:tradition}
Several studies \citep[for recent examples]{cap15:bams, cap20:ch4, wie21:fc, sch23:arX} have compared different approaches in the characterisation and identification of ENSO phases based on SSTAs. These approaches can be broadly categorised as methods based on 1) principal component or multiple correspondence analysis, as in \cite{chi04:jcli}, \cite{tak11:grl}, and \cite{feng20:cd}, 2) composites based on region of influence or box-type indices, as in \cite{ash07:jgr}, \cite{kug09:jcli},  \cite{kar13:grl}, \cite{cai09a:grl}, and \cite{sul16:sr}, and 3) pattern recognition, as in \cite{joh13:jcli}, \cite{bla22:aies}, \cite{wie21:fc} and \cite{sch23:arX}.

\begin{table}[htbp]
    \centering
    \caption{Adapted from \cite{sch23:arX}, classification of El Ni\~no events over the 1982-2022 DJF interval. The approaches of \cite{kug09:jcli}, \cite{tak11:grl}, \cite{sch23:arX}, and AA are compared. Similar to \cite{sch23:arX}, AA categorize El Ni\~no events based on the archetype corresponding to the highest affiliation probabilities for cardinalities 4 and 8, with archetypes $A_{1,4,8}$ standing for CP Ni\~no and $A_{5}$ for EP Ni\~no, as in Figures S7 and S8 for cardinality 8. The EP Ni\~no archetype, $A_{5}$ for cardinality 8, corresponds to the \textsl{Extreme El Ni\~nos} type, EP*, identified by \cite{sch23:arX}.
    The last column indicates when AA8 aligns with the consensus characterisation of El Ni\~no events across the 3 other methods listed.}
    \begin{tabular}{c|lllllll}
        Ni\~no    &  Kug et al.    & Takahashi et al. & Schl\"or et al.      & AA4   & AA8   & Consensus \\
        Events    &  2009          & 2011             & 2023                 &       &       &           \\
        \hline
        1982–1983 &  EP             &  EP               &  EP*               & EP    & EP    & \textbf{Yes}      \\
        1986–1987 &  EP             &  EP               &  EP                & CP/EP & CP    & No       \\
        1987–1988 &  CP             &  CP               &  EP/CP             & EP    & CP    & \textbf{Yes}      \\
        1991–1992 &  EP             &  CP               &  EP                & CP/EP & CP/EP & \textbf{Yes}      \\
        1994–1995 &  CP             &  CP               &  EP                & CP    & CP    & \textbf{Yes}      \\
        1997–1998 &  EP             &  EP               &  EP*               & EP    & EP    & \textbf{Yes}      \\
        2002–2003 &  CP             &  CP               &  CP/EP             & CP/EP & CP    & \textbf{Yes}      \\
        2004–2005 &  CP             &  CP               &  CP                & CP    & CP    & \textbf{Yes}      \\
        2006–2007 &  CP             &  CP               &  EP                & EP    & CP    & \textbf{Yes}      \\
        2009–2010 &  CP             &  CP               &  CP/EP             & CP/EP & CP    & \textbf{Yes}      \\
        2014–2015 &  CP             &  -                &  CP                & CP    & CP    & \textbf{Yes}      \\
        2015–2016 &  EP             &  CP               &  EP*               & EP    & EP    & \textbf{Yes}      \\
        2018–2019 &  CP             &  CP               &  EP/CP             & CP/EP & CP    & \textbf{Yes}      \\
        2019-2020 &  CP             &  CP               &  CP                & CP    & CP    & \textbf{Yes}      \\
    \hline
    \hline
    \end{tabular}
    \label{tab:nino}
\end{table}

Table \ref{tab:nino} compares the classification of El Ni\~no conditions over the satellite era 1982 to 2020 for the region of influence, Ni\~no 3 -- Ni\~no 4, of \cite{kug09:jcli}, the original EOF/PC based approach of \cite{tak11:grl}, the unsupervised pattern recognition methods of \cite{sch23:arX} and AA for cardinalities 4 and 8. Both unsupervised pattern recognition methods have been `rendered' categorical for comparison by only listing the pattern or archetype corresponding to the highest and second highest affiliation probabilities for \cite{sch23:arX} and the `winning' archetype based on the discrimination scores $\Delta^{4}_{i_{max}}$ and $\Delta^{8}_{i_{max}}$, i.e. the archetype contributing to any given record with the highest probability of expression. The primary aim of using global SSTAs is not only to categorise the ENSO phenomenon in the Tropics, but also via its impacts through teleconnections. Although, the choice of going global with AA may incur the risk of `blending' different processes and be potentially prone to confounding effects, it leads to similar types when compared to other categorization methods and associated indices, which are all based on SSTAs in the tropical and sub-tropical Pacific. These indices are derived from separated, but at time overlapping, domains and linearly combined to form EP- and CP-like ENSO indices. It could be argued that such combinations are not necessarily justifiable from a dynamical standpoint as they mix values, coincident in time, from spatially separated regions. Climate and weather research practitioners also rely on prior knowledge gathered from historical records for the choice and extent of the domains used to construct these metrics; this corresponds to a form of supervised learning. AA does not suffer from these drawbacks as the resulting patterns are close to the observed (global) dSSTA snapshots without any non-local manipulation needed to discriminate between CP and EP conditions or \textsl{a priori} choices of domains. It is therefore expected that noticeable differences may emerge when comparing AA categorical ENSO phase affiliation to other methods. The non-local manipulations will also impact teleconnection patterns and associated dynamical inferences derived typically from correlation maps, normalised expectations in statistical terms, where the contributing records of both the ENSO indices and geophysical fields under investigation are equally weighted. This is not the case with AA composites, where teleconnection patterns of any fields built on AA indices ($\textbf{S}$-matrix), correspond to weighted averages, or conditional expectations, of field records multiplied by the probabilities of expression of the associated dSSTA archetypes as per Equation \ref{eq:fs}.

Interestingly, the majority of the comparative studies mentioned earlier only focus on differentiating between El Ni\~no conditions and rarely attempt to classify La Ni\~na flavours, with \cite{cai09a:grl} and \cite{bla22:aies} being notable exceptions. Results discussed in \cite{cai09a:grl} highlight the importance of CP-Ni\~na or Modoki conditions affecting Australian rainfall. \cite{bla22:aies} work reveals the ENSO teleconnection global imprint on surface and atmospheric fields, unequivocally showing in particular that CP-Ni\~na and EP-Ni\~no conditions strongly influence the circulation at mid- and high-latitude as identified by large amplitude composites when compared to opposite CP-Ni\~no and EP-Ni\~na phases. Using a reduced geographical domain, \cite{cha22:ncom} recently illustrate the value of AA methodology in potentially disentangling the impacts of local and remote large-scale drivers when applied to marine heatwaves (MHWs) and cold spells (MCSs) in the Australasian region by comparing AA to the conventional geographical point-based methods described in \cite{oli21:arms}. They show that AA unambiguously identifies concurrent CP- and EP-Ni\~na conditions associated with MHWs in South-Eastern Indian Ocean of the Western Australian coast and in South Pacific close to New Zealand, respectively. As with Ni\~na flavours, the different Ni\~no types have varied teleconnection impacts. These will be further examined in Section \ref{sec:tele}.

%%%%%%%%%%%%%%%%%%%%%%%%%%%
%   Section On-line updates
%%%%%%%%%%%%%%%%%%%%%%%%%%%
\section{AA on-line updates}\label{sec:online}
AA can lend itself to the characterisation of forecasted ENSO conditions and corresponding teleconnections in S2S or multi-year prediction systems by first training AA on observed or reanalysed historical global SST patterns for a given cardinality $p$ as follows. For $M$ historical time records, AA solutions for the dataset $\textbf{X}$, here dSSTAs, are first computed,  
\begin{equation*}
        X_{st} \approx \sum_{r=1}^{M}\sum_{i=1}^{p} X_{sr} C_{ri} S_{it} = \sum_{j=1}^{p}XC_{sj} S_{jt}, \forall t \in 1,\ldots,M
\end{equation*}
where $\textbf{XC} = \sum_{r=1}^{M} X_{sr} C_{rj}$ are the previously found archetypal patterns. The on-line affiliation probability updates, $\hat{S}_{it'},\, \forall t' \in M+1,\ldots,M+n$ for $n$ new forecasted records of $Y_{st'}$, are then approximated by the solutions of
\begin{equation}
    \underset{\hat{\textbf{S}}}{\arg\min}\|Y_{st'} - \sum_{i=1}^{p} XC_{si}\hat{S}_{it'}\|_{F},
    \label{eq:update}
\end{equation}
where $p$ stands for the archetype cardinality considered \textsl{a priori}. The constraints, $\sum_{j=1}^{p} \hat{S}_{jt'}=\textbf{1}_{t'}$ and $\hat{\textbf{S}}\ge 0$, still apply. By modifying \cite{mor12:neu} algorithm, Expression \ref{eq:cost} reduces to the simpler least squares problem with convexity constraints of finding the minimum for Expression \ref{eq:update} as a function of $\hat{\textbf{S}}$ for given archetypes computed over the training, here $\textbf{XC}$, and the forecasted records $Y_{st'}$.

         \tiny
         \begin{table}[t]
             \centering
             \begin{tabular}{|l|*{12}{c}|}
             \hline
             $\hat{\textbf{S}}_{4\times 12}$ & 1 & 2 & 3 & 4 & 5 & 6 & 7 & 8 & 9 & 10 & 11 & 12 \\ \hline
             $\hat{S}_{1,\cdot}$ & 0.13 & 0.26 & 0.31 & 0.36 & 0.45 & 0.48 & 0.40 & 0.30 & 0.21 & 0.06 & 0.00 & 0.07 \\
             $\hat{S}_{2,\cdot}$ & 0.62 & 0.56 & 0.56 & 0.41 & 0.42 & 0.36 & 0.38 & 0.19 & 0.39 & 0.69 & 0.68 & 0.68 \\
             $\hat{S}_{3,\cdot}$ & 0.25 & 0.19 & 0.08 & 0.22 & 0.13 & 0.09 & 0.14 & 0.47 & 0.35 & 0.21 & 0.32 & 0.25 \\
             $\hat{S}_{4,\cdot}$ & 0.00 & 0.00 & 0.05 & 0.00 & 0.00 & 0.07 & 0.08 & 0.03 & 0.06 & 0.04 & 0.00 & 0.00 \\
             \hline \hline
             $\textbf{S}_{4\times 12}$ & 1 & 2 & 3 & 4 & 5 & 6 & 7 & 8 & 9 & 10 & 11 & 12 \\ \hline
             $S_{1,\cdot}$ & 0.13 & 0.26 & 0.31 & 0.36 & 0.45 & 0.48 & 0.40 & 0.29 & 0.19 & 0.04 & 0.00 & 0.08 \\
             $S_{2,\cdot}$ & 0.65 & 0.57 & 0.57 & 0.43 & 0.44 & 0.39 & 0.40 & 0.22 & 0.44 & 0.78 & 0.75 & 0.77 \\
             $S_{3,\cdot}$ & 0.22 & 0.17 & 0.07 & 0.21 & 0.11 & 0.06 & 0.12 & 0.45 & 0.31 & 0.13 & 0.25 & 0.16 \\
             $S_{4,\cdot}$ & 0.00 & 0.00 & 0.05 & 0.00 & 0.00 & 0.07 & 0.08 & 0.03 & 0.06 & 0.04 & 0.00 & 0.00 \\
             \hline \hline
             $|\Delta \textbf{S}_{4\times 12}|$ & 1 & 2 & 3 & 4 & 5 & 6 & 7 & 8 & 9 & 10 & 11 & 12 \\ \hline
             $|S_{1,\cdot} - \hat{S}_{1,\cdot}|$ & 0.00 & 0.00 & 0.00 & 0.00 & 0.00 & 0.00 & 0.01 & 0.01 & 0.01 & 0.02 & 0.00 & 0.00 \\
             $|S_{2,\cdot} - \hat{S}_{2,\cdot} |$ & 0.03 & 0.02 & 0.01 & 0.02 & 0.02 & 0.03 & 0.03 & 0.03 & 0.05 & 0.09 & 0.07 & 0.09 \\
             $|S_{3,\cdot} - \hat{S}_{3,\cdot}|$ & 0.03 & 0.02 & 0.01 & 0.01 & 0.02 & 0.03 & 0.03 & 0.02 & 0.04 & 0.08 & 0.07 & 0.09 \\
             $|S_{4,\cdot} - \hat{S}_{4,\cdot}|$ & 0.00 & 0.02 & 0.00 & 0.00 & 0.00 & 0.00 & 0.00 & 0.00 & 0.00 & 0.01 & 0.00 & 0.00 \\
             \hline
             \end{tabular}
             \caption{$|\Delta \textbf{S}_{4\times 12}|$ absolute differences for the 2021 monthly records between the `truth' based on 1982-2021 AA, $\textbf{S}_{4\times 12}$, and the `approximation' update based on 1982-2020 AA, $\hat{\textbf{S}}_{4\times 12}$.}
             \label{tab:online}
         \end{table}
\normalsize
Table \ref{tab:online} compares absolute differences, $|\Delta \textbf{S}_{4\times 12}|$, for the 2021 monthly records between the `truth' based on 1982-2021 AA, $\textbf{S}_{4\times 12}$, and the `approximated' update based on 1982-2020 AA, $\hat{\textbf{S}}_{4\times 12}$. The 12 month updates for 2021, based on 1982-2020 archetypes, were computed with a single initialization trial and application of the algorithm. The 2021 monthly records set aside for testing, $Y_{st'}$, are not used for training. The largest absolute mismatches of the affiliation matrix $\textbf{S}$ elements over the forecast period are about 0.09 and are mainly seen on affiliation time-series for CP- and EP-Ni\~na conditions\footnote{As in \cite{bla22:aies} for AA based on dSSTAs over the 1982-2020 period, archetypes $A_{1}$ and $A_{2}$ correspond to CP-Ni\~no and CP-Ni\~na, respectively.}, which are difficult to classify for cardinality 4 as illustrated in Section \ref{sec:flavours}. When applied to S2S forecasts, the AA on-line update is likely to degrade as model representation and observations may diverge when the forecast lead-time increases. The AA on-line update procedure presented here is a proof-of-concept only. It requires thorough benchmarking for S2S applications. Such an exercise would not only include characterising S2S forecasts with AA, but also would test the ability of any S2S OAGCM to represent ENSO, and is beyond the scope of this work. As an aside, we note that AA could be added to the suite of diagnostics commonly used to check the OAGCM representation fidelity of various climate modes across geophysical realms, an approach illustrated recently by \cite{cha22:ncom} for 2500-year-long control run of the GFDL Climate Model 2.1 (CM2.1) Coupled Model Intercomparison Project (CMIP) phase 3 -- class OAGCM, used in older versions of the Australian Community Climate and Earth System Simulator (ACCESS).

%%%%%%%%%%%%%%%%%%%%%%%%%%%
%   Section Teleconnections
%%%%%%%%%%%%%%%%%%%%%%%%%%%
\section{Teleconnections}\label{sec:tele}
An indirect way to test the reliability and relevance of the AA classification of ENSO types, is to ascertain that the extreme configurations detected in dSSTAs correspond to robust and explainable teleconnection patterns of various atmospheric and oceanic fields. In this section, we explore the archetypes spatial and temporal imprints, and their linkages to extra-tropical atmospheric circulation and rainfall. Composites of depth-integrated quantities such as SLA and sub-surface ocean temperature anomaly profiles in the Tropics will provide additional evidences of the realistic AA representation of ENSO teleconnections. Only results pertaining to the AA cardinality of 4, will be shown, with signal-to-noise ratios for individual fields and results for cardinality 8 given in the supplementary material.

%%  M-files
%
%   JCLI_fig14.m and JCLI_fig15.m
%
\begin{figure}[htbp]
    \centering
    \includegraphics[width=1.0\textwidth]{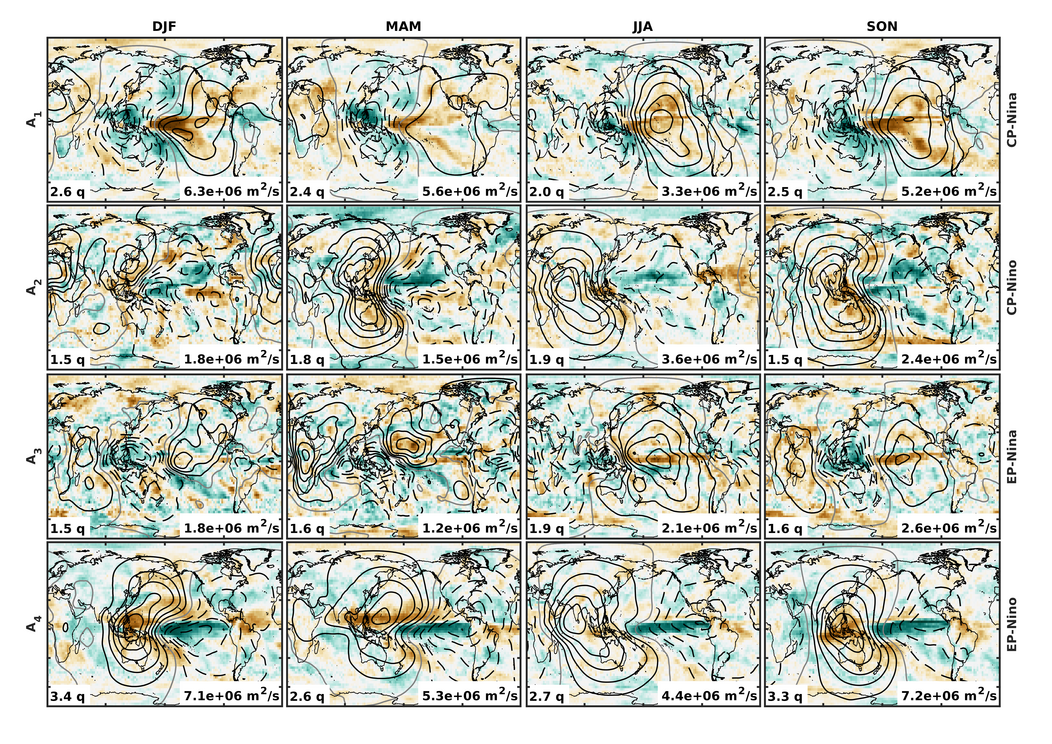}
    \caption{AA seasonal lag-0 composites (columns) of GPCP precipitation centered undeciles (shaded) overlaid with anomalies of velocity potential differences between isobaric pressure levels of 150 and 850 hPa (contours) based on time averages of the $\textbf{S}$-affiliation matrix over the 1982-2022 period for a cardinality of 4. Each individual subplot is scaled to increase contrast across archetypal patterns (rows) and seasons (column). The scaling factors are given at the bottom of each panel on the right for rainfall undeciles and on the left for velocity potential anomalies. Green (brown) shadings correspond to enhanced (reduced) precipitation above (below) the median rainfall in white. Dashed (full) contours correspond to enhanced (reduced) convection as diagnosed by $\Delta\chi_{150-850}$. Signal-to-noise ratios $(F^{2}_{S}/C^{S}_{F})^{1/2}$ for both fields are reported in Figures S13 and S14 of the supplementary material.}
    \label{fig:aavpot4}
\end{figure}
To begin with, ENSO conditions lead to enhanced or reduced atmospheric convective activity over the tropical and associated convergence zones, not only modulating rainfall over these regions, but also interacting with other modes of variability such as the Madden-Julian Oscillation (MJO) \citep{mad72:jas}, for example. Figure \ref{fig:aavpot4} displays the lag-0 seasonal imprints of ENSO archetypes on monthly GPCP precipitation centered undeciles\footnote{Centered undeciles or elftiles label the rank order of values of a distribution partitioned in 11 frequency classes of equal width measured with respect to `median' class 6, hence ranging from $-5$ to $+5$.} (shaded) overlaid with anomalies of JRA-55 velocity potential difference (contours) between isobaric pressure levels of 150 and 850 hectoPascal (hPa), $\Delta\chi_{150-850}$, based on time averages of the $\textbf{S}$-affiliation matrix over the 1982-2022 period for a cardinality of 4. As reported by \cite{ada14:jas} Table 2, a MJO index based on the first two PCs of $\Delta\chi_{150-850}$ explains 57\% of its total variance is not only an excellent proxy for probing the propagation of tropical Kelvin and Rossby-waves and their impact on extra-tropical circulation, but also marks regions of enhanced and reduced convective activity. Based on these findings, we are using the velocity potential difference $\Delta\chi_{150-850}$ anomalies to diagnose the imprint of ENSO phases on vertical atmospheric convection. Overall, both the precipitation and the velocity potential composites agree well with the expected behaviour associated with SST anomalies driven by ENSO. EP- and CP-Ni\~no phases. Associated SST signatures are characterised by enhanced precipitation (in blue) in the Eastern and Central Pacific, respectively, collocated with enhanced upward convection marked by dashed contours on Figure \ref{fig:aavpot4}. On the western side of the Pacific Basin, drier conditions (in red) prevail and the convective activity reduces as shown by full contours in the $\Delta\chi_{150-850}$ composites. The situation is reversed for EP- and CP-Ni\~na phases. Marked seasonal differences are also observed in both fields, but especially in the $\Delta\chi_{150-850}$ composites, where the enhanced and reduced convective center of action locations change with latitudes and longitudes. Clear extra-tropical signatures along the Pacific Rim impacting the North- and South-American, the Australian and Far-East Asian continents, extending at times to the Indian sub-continent, the African and the European side of the Eurasian continents, are apparent for CP-Ni\~no and EP-Ni\~na composites. The corresponding composites from cardinality 8 are reproduced in the supplementary material, Figure S9.

%%  M-file
%   JCLI_fig16.m
%
\begin{figure}[htbp]
    \centering
    \includegraphics[width=1.0\textwidth]{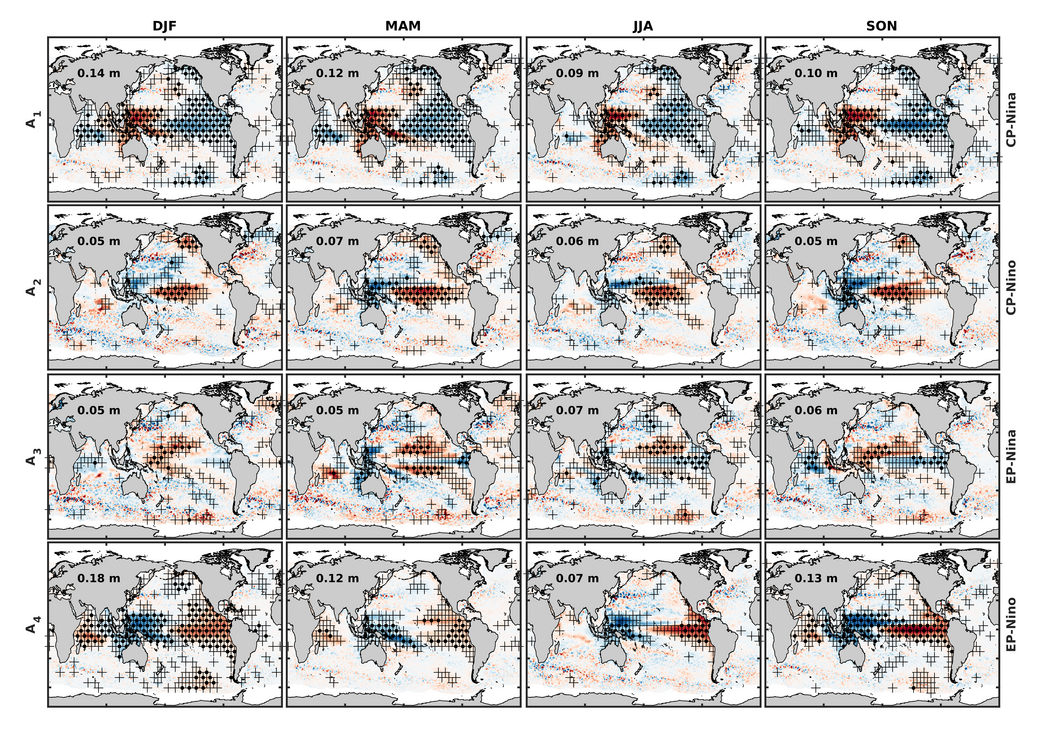}
    \caption{AA seasonal lag-0 composites (columns) of detrended sea-level anomalies and signal-to-noise ratios ($\textbf{+} > 0.15, \bullet > 0.3$, $(F^{2}_{S}/C^{S}_{F})^{1/2}$) based on time averages of the $\textbf{S}$-affiliation matrix over the 1993-2021 period for a cardinality of 4. Each individual subplot is scaled to increase contrast across archetypal patterns (rows). The maximum absolute value of the detrended sea-level anomaly composites in a $\pm 20^{\circ}$ tropical band is reported in meter on the Eurasian continent for each panel.}
    \label{fig:aadsla4}
\end{figure}

\begin{figure}[htbp]
    \centering
    \includegraphics[width=1.0\textwidth]{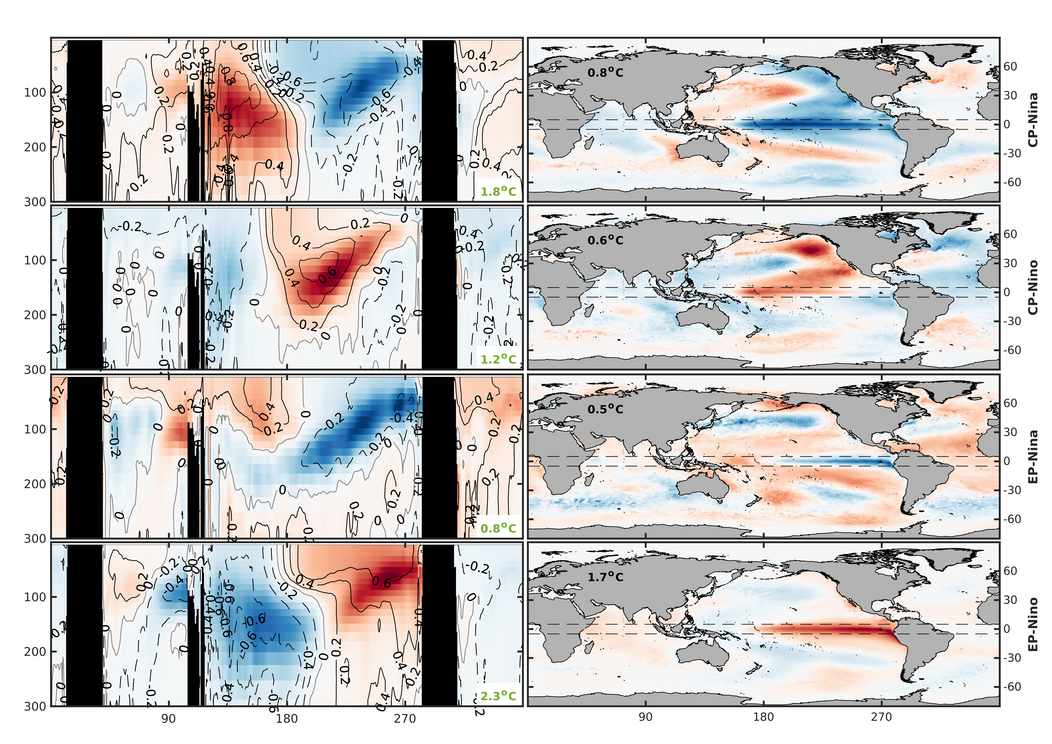}
    \caption{Detrended anomalies of temperature AA lag-0 composites longitudinal profiles (left) from SODA v3.15.2 reanalysis averaged over the $-5^{\circ}$ to $5^{\circ}$ latitude band (dashed lines) and corresponding detrended SSTA archetypes (right) for a cardinality of 4. The shadings on the left correspond to time-mean temperature composites and the contours to the (time) mean to standard deviation ratio $(F^{2}_{S}/C^{S}_{F})^{1/2}$ in 0.2 increment and the green values in the lower right corners are the maximum absolute temperature anomalies in $^{\circ}$Celsius. Each individual subplot is scaled by its maximum amplitude to increase contrast across archetypal patterns (rows).}
    \label{fig:aasoda4lat}
\end{figure}

\begin{figure}[htbp]
    \centering
    \includegraphics[width=1.0\textwidth]{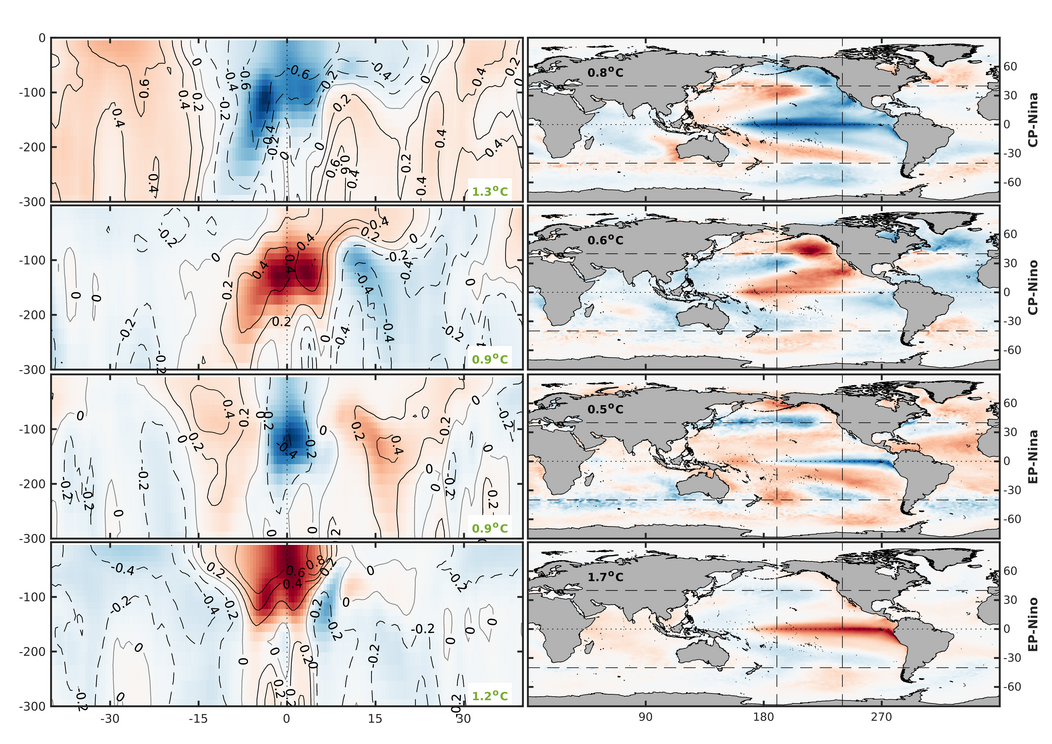}
    \caption{Detrended anomalies of temperature AA lag-0 composites latitudinal profiles (left) from SODA v3.15.2 reanalysis averaged over the 190$\degree$ to 240$\degree$ (Ni\~no 3.4 longitudes) longitude band (dashed lines) and corresponding detrended SSTA archetypes (right) for a cardinality of 4. The shadings on the left correspond to time-mean temperature composites and the contours to the (time) mean to standard deviation ratio $(F^{2}_{S}/C^{S}_{F})^{1/2}$ in 0.2 increment and the green values in the lower right corners are the maximum absolute temperature anomalies in $^{\circ}$Celsius. Each individual subplot is scaled by its maximum amplitude to increase contrast across archetypal patterns (rows).}
    \label{fig:aasoda4lon}
\end{figure}

In the tropics, detrended sea-level anomalies (dSLAs) provide a first-order approximation for (detrended) steric height. As shown by  \cite{gas17:po} and \cite{izu19:cd}, the dSLA signal is mainly driven by sub-surface anomalies in the thermohaline expansion of seawater integrated from the surface to below the thermocline down to about a depth of 300 meters. The seasonal dSLA $\textbf{S}$-composites displayed in Figure \ref{fig:aadsla4} discriminate between CP and EP ENSO sub-surface expressions, not only along the Equator, but also in the inter-tropical convergence and South-Pacific convergence zones (ITCZ and SPCZ), and even further afield, in the tropical Indian Ocean basin. The largest anomalies are observed for the CP-Ni\~na and EP-Ni\~no conditions peaking in DJF, with values twice as high as their CP-Ni\~no and EP-Ni\~na counterparts, with their maximum values now expressed in MAM and JJA, respectively. The behaviour is confirmed by sub-surface SODA profiles in potential temperature, $\theta$, averaged 1) in latitudes over the $\pm 5\degree$ tropical band, Figure \ref{fig:aasoda4lat}, and 2) in longitudes over the Ni\~no 3.4, 190$\degree$- to 240$\degree$-East, bounds, Figure \ref{fig:aasoda4lon}. The signal-to-noise ratios between the composite means and standard deviations, $(\theta^{2}_{S}/C^{S}_{\theta})^{1/2}$, are 2 or 3 time larger than the GPCP precipitation composite ratios displayed on Figures S13 and S18, for example, and provide an indication of the strength of the influence of ENSO cycles (as determined by the AA metric), especially between 50- to 250-meter depth. For EP-Ni\~no, the strong anomalies in ocean temperature reach all the way to the surface and have the largest signal-to-noise ($>$0.6) across all archetypes in the Ni\~no 3.4 longitude band displayed in Figure \ref{fig:aasoda4lat}.

%%  M-file
%   JCLI_fig18.m
%

%%  M-file
%   JCLI_fig20.m
%
The larger signal-to-noise results, obtained for both sea-level height and potential temperature composites, indicate that AA based on ocean fields is likely to be less prone to atmospheric weather noise and, therefore, possibly more suited to the detection and classification of ENSO conditions.

\begin{figure}[htbp]
    \centering
    \includegraphics[width=1.0\textwidth]{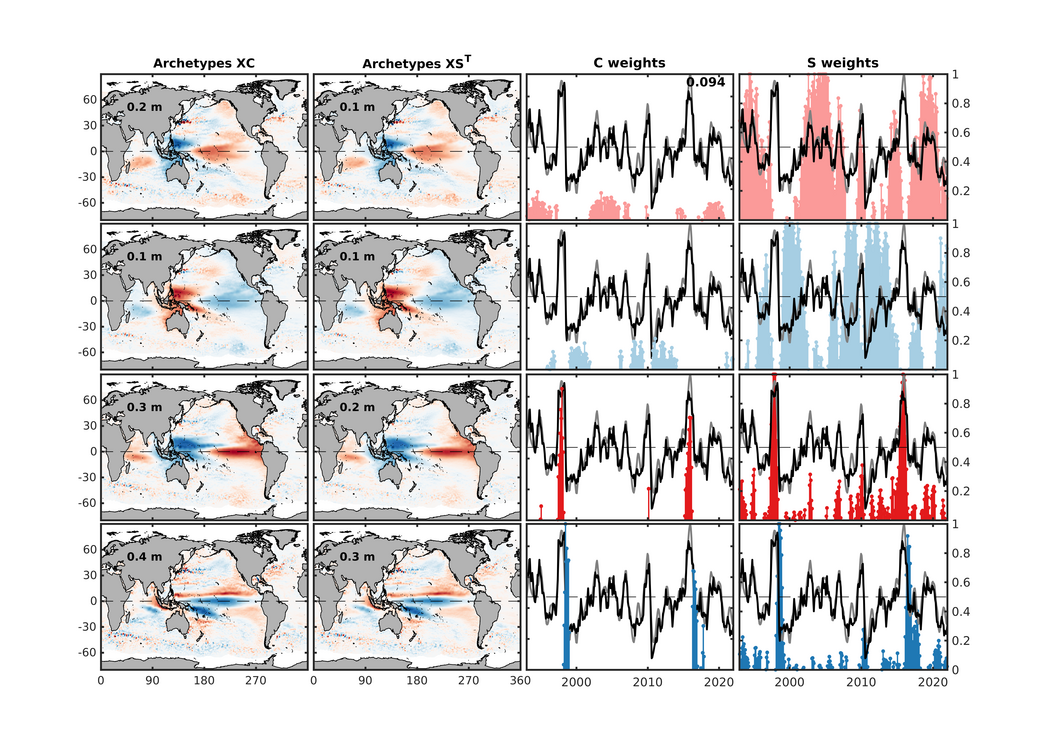}
    \caption{AA spatial pattern and time series results using detrended monthly SLA anomalies over 1982-2021 for cardinality 4. The two left columns show archetypes and dSLA composites, $\textbf{XC}$ and $\textbf{X}\overline{\textbf{S}}^T$, with $\overline{S}_{p\times t}=S_{p\times t}/\sum^{t}_{i=1} S_{pi}$, respectively. The two right columns show $\textbf{C}$ and $\textbf{S}$ matrix time series, respectively, with MEI (black) and ONI (grey) time series overlaid on both. The values in the upper left corner of each panel, in the first two columns, correspond to the maximum dSLA anomalies range scale expressed in meters, whereas the value in bold in third column first row is the maximum $\textbf{C}$-weights across all archetypes. Both MEI (black) and ONI (grey) time series were scaled to fit within the $\textbf{C}$ and $\textbf{S}$ weights bounds.}
    \label{fig:aa4xcxs}
\end{figure}

To test this hypothesis, we performed AA on detrended SLAs directly. Figure \ref{fig:aa4xcxs} displays AA spatial pattern and time series results using detrended monthly SLA anomalies over 1982-2021 for cardinality 4. The first two columns on the left show archetypes and dSLA composites, $\textbf{XC}$ and $\textbf{X}\overline{\textbf{S}}^T$, with $\overline{S}_{pt}=S_{pt}/\sum^{t}_{i=1} S_{pi}$. The archetypal patterns are ranked, from top-to-bottom, according to the time mean of $\textbf{S}$, $\frac{1}{N}\sum^{t}_{i=1} S_{pi}, \forall p$, the mean probability of expression of archetype $p$ over all $N$ records. The last two columns on the right show $\textbf{C}$ and $\textbf{S}$ matrix time series, respectively, with MEI and ONI time series overlaid on both. The value in bold, third column top row, corresponds to the maximum $\textbf{C}$-weights across all archetypes. A cursory comparison with Figure \ref{fig:aadsla4} indicates that the seasonal dSLA composites at 0-lag built on dSSTA AA affiliation weights resemble the dSLA archetypes. The time-mean affiliation weights, $\frac{1}{N}\sum^{t}_{i=1} S^{SLA}_{pi}$ for dSLA and $\frac{1}{N}\sum^{t}_{i=1} S^{SST}_{pi}$ for dSSTA, are nevertheless different, so archetype patterns do not follow the same top-to-bottom row order in Figures \ref{fig:aadsla4} and \ref{fig:aa4xcxs}.
%%  M-file
%   JCLI_fig22/23.m for dSLAs
%

\begin{figure}[htbp]
    \includegraphics[width=\textwidth]{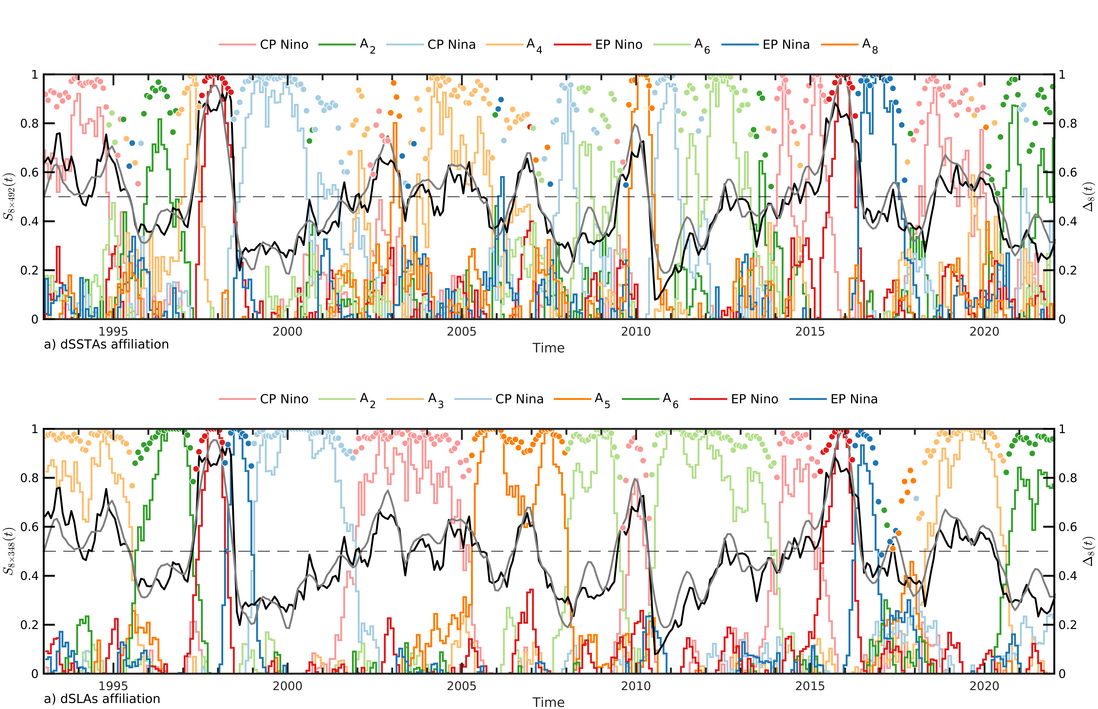}
    \caption{Comparison of AA affiliation sequences for a) dSSTAs and b) dSLAs. The discrimination scores (\cite{ris21:mwr}) $\Delta^{p}_{i_{max}}(t) \in$ [0, 1], $\, \forall i=1,\ldots,8$ for all monthly records $t$ are represented by archetypal colour-coded dots of the winning archetype. Here, the color codes for dSLA archetypes are tentatively matched to those of dSSTA archetypes. The corresponding archetypal patterns are given in Figures S7, columns 3 and 4 for dSSTAs, and S13, columns 1 and 2 for dSLAs. Both MEI (black) and ONI (grey) time series were scaled to fit within the $S_{8 \times 492}$ and $\Delta_{8}$ bounds.} 
    \label{fig:dSSTA_dSLA_8}
\end{figure}

Figures \ref{fig:dSSTA_dSLA_8}a and \ref{fig:dSSTA_dSLA_8}b try to match the affiliation sequences for both dSSTAs and dSLAs AA based on the `nestedness' information displayed in Figures S15 and S16. However, a perfect match between dSSTA and dSLA archetypes is not to be expected. The main differences are found in the archetype persistence across time records expressed by the discrimination scores, $\Delta^{8}_{i_{max}}$. In any given phases, the scores are more elevated and persist over longer periods for dSLA archetypes when compared to those of dSSTA. The dSLA score persistence indicates 1) that sub-surface quantities are less exposed to the vagaries of the weather and associated drivers over both the Tropics and Sub-Tropics, and 2) that the ENSO phenomenon operates over longer timescales than evident from surface-based representations. There are clear periods where the ENSO state is poorly expressed in dSSTA, but well expressed in dSLA in Figures \ref{fig:dSSTA_dSLA_8}a and \ref{fig:dSSTA_dSLA_8}b.  For example, the period 2001--2004 where there is a mix of archetypes expressed in dSSTA but a well expressed preference for CP Nino in dSLA.  More generally, when the MEI and ONI indices have small amplitude variation, dSSTA ENSO types are poorly expressed, but there is still a well expressed dSLA type.

Although all flavours of El Ni\~no may have similar early subsurface origins at long lead times as claimed by \cite{ram13:ncc}, the extreme configurations revealed by AA indices in both $\textbf{S}$ time-series on Figure \ref{fig:dSSTA_dSLA_8} and in sub-surface composites at 0-lag on Figures S11, S12 and S13, not only point to significantly different expressions across ENSO types, but also to fast -- over 6 to 12 months -- transitions for both the initiation and termination of events across all flavours. These extreme configurations are interspersed by longer lasting CP Ni\~no and Ni\~na intervals, especially evident in the sub-surface.

Many different extra-tropical modes have been hypothesized as potential ENSO precursors, both in the northern and southern hemispheres such as in \cite{yu10:jcli}, \cite{wan12:grl} and \cite{zha14:jcli}. A statistical relationship and the underlying mechanisms are very difficult to separate from each other and from ENSO itself, as all modes are correlated and not independent one from another, as nicely illustrated by \cite{peg20:cd} for example. \cite{peg20:cd} and \cite{cap21:scirep} work supports the global approach adopted in this study. By extending our analysis domain to global dSSTAs, we attempt to follow their lead, not only including the extra-tropics, but also the Indian and Atlantic Ocean basins, where extreme dSSTA configurations do have an imprint. Amongst other influencing factors, the so-called `capacitor effect' of the Indian and Atlantic Oceans discussed in \cite{oku10:jcli}, \cite{an18:grl} and \cite{an21:book}, where the ENSO driven anomalies imparted to these basins via the atmospheric bridge are imprinted and eventually feeding back to the Pacific, may play a role.

%%  M-files
%   JCLI_fig8.m
%

The results presented in Figures \ref{fig:dSSTA_dSLA_8}a and \ref{fig:dSSTA_dSLA_8}b not only reaffirm and extend the earlier observations of \cite{wyr75:jpo} that no ENSO events are quite the same across geophysical realms, but also show for a given cardinality that dSLA archetypes provide a clearer representation of ENSO types and that they are less prone to mischaracterisation than dSSTA archetypes. For both dSSTA and dSLA archetypes, increasing the AA cardinality adds variety to the phenomenon and sharpens the discrimination across ENSO types.

\section{ENSO phase transitions}\label{sec:transition}
Although hampered by the short sample size of the well-observed SST records over the satellite area, the phase evolution of ENSO can be studied by counting transitions from one well-expressed archetype to the next. Tables \ref{tab:tp4} and \ref{tab:tp8} tally 491 monthly transitions, including transitions to oneself, $\pi_{ij} = \frac{1}{N-1}\sum_{t=1}^{N-1} \{\underset{i=1,\ldots,4/8}{\arg\max}\,S_{it} \longrightarrow \underset{j=1,\ldots,4/8}{\arg\max}\,S_{jt+1}\}$, across the categorical AA representation of dSSTAs for cardinalities 4 and 8. $N=492$ stands for the total number of months in the 41-year long record and $\pi_{ij}$ the time-mean transition probability from archetypes $i$ to $j$. The diagonal entries in both tables correspond to the number of self-transitions. The individual phase mean persistence can be easily computed as the diagonal entries divided by the number of separate intervals of a given phase. For a cardinality of 8 for example, the EP-Ni\~no intervals, labeled $A_{5}$ in Table \ref{tab:tp8},  persist in average for about $33/4\approx 7$ to 8 months, where 4 corresponds to the number of separate EP-Ni\~no intervals as shown in Figure \ref{fig:stairs}a.

\begin{table}[htbp]
    \centering
    \begin{tabular}{c|cccc|l}
    & $\textbf{A}_{1}$ & $\textbf{A}_{2}$ & $\textbf{A}_{3}$ & $\textbf{A}_{4}$ & Phase\\
    \hline
    $\textbf{A}_{1}$ & 129 &    8 &    7 &    1 & CP Ni\~na\\
    $\textbf{A}_{2}$ &   3 &  143 &    8 &    8 & CP Ni\~no\\
    $\textbf{A}_{3}$ &  12 &    8 &   95 &    3 & EP Ni\~na\\
    $\textbf{A}_{4}$ &   2 &    3 &    7 &   54 & EP Ni\~no\\
    \hline
    \end{tabular}
    \caption{Number of the transitions from winning $i$ (row) to winning $j$ (column) archetypes, $ (N-1)\pi_{ij} = \sum_{t=1}^{N-1} \{ \underset{i=1,\ldots,4}{\arg\max} \,S_{it} \longrightarrow \underset{j=1,\ldots,4}{\arg\max}\,S_{jt+1} \} $, where $N=492$ stands for the total number of months in the 41-year long record and $\pi_{ij}$ the transition probability from archetype $i$ to $j$. The table diagonal elements correspond here to the numbers of self transitions omitted in Figure \ref{fig:aatp4}. The winning archetypes are the archetypes contributing to any given record with the highest probability of expression, $\underset{i=1,\ldots,4}{\arg\max} \,S_{it},\, \forall t$.}
    \label{tab:tp4}
\end{table}
\begin{table}[htbp]
    \centering
    \begin{tabular}{c|cccccccc|l}
    & $\textbf{A}_{1}$ & $\textbf{A}_{2}$ & $\textbf{A}_{3}$ & $\textbf{A}_{4}$ & $\textbf{A}_{5}$ & $\textbf{A}_{6}$ & $\textbf{A}_{7}$ & $\textbf{A}_{8}$ & Phase\\
    \hline
    $\textbf{A}_{1}$ & 89 &    4 &    1 &    2 &    3 &    2 &    1 &    3 & CP Ni\~no\\
    $\textbf{A}_{2}$ &  6 &   83 &    3 &    6 &    0 &    0 &    1 &    0 & --\\
    $\textbf{A}_{3}$ &  0 &    5 &   62 &    2 &    0 &    4 &    2 &    1 & CP Ni\~na\\
    $\textbf{A}_{4}$ &  6 &    5 &    0 &   58 &    2 &    0 &    2 &    0 & --\\
    $\textbf{A}_{5}$ &  1 &    0 &    2 &    0 &   33 &    1 &    1 &    0 & EP Ni\~no\\
    $\textbf{A}_{6}$ &  3 &    1 &    3 &    0 &    0 &   33 &    0 &    1 & --\\
    $\textbf{A}_{7}$ &  0 &    1 &    3 &    4 &    0 &    0 &   20 &    2 & EP Ni\~na\\
    $\textbf{A}_{8}$ &  0 &    1 &    2 &    0 &    0 &    1 &    3 &   22 & --\\
    \hline
    \end{tabular}
    \caption{Number of the transitions from winning $i$ (row) to winning $j$ (column) archetypes, $ (N-1)\pi_{ij} = \sum_{t=1}^{N-1} \{ \underset{i=1,\ldots,8}{\arg\max} \,S_{it} \longrightarrow \underset{j=1,\ldots,8}{\arg\max} \,S_{jt+1} \} $, where $N$ stands for the number of months in the 41-year long record. The table diagonal elements correspond here to the numbers of self transitions omitted in Figure S14.}
    \label{tab:tp8}
\end{table}
\begin{figure}[htbp]
    \centering
    \includegraphics[width=1.0\textwidth]{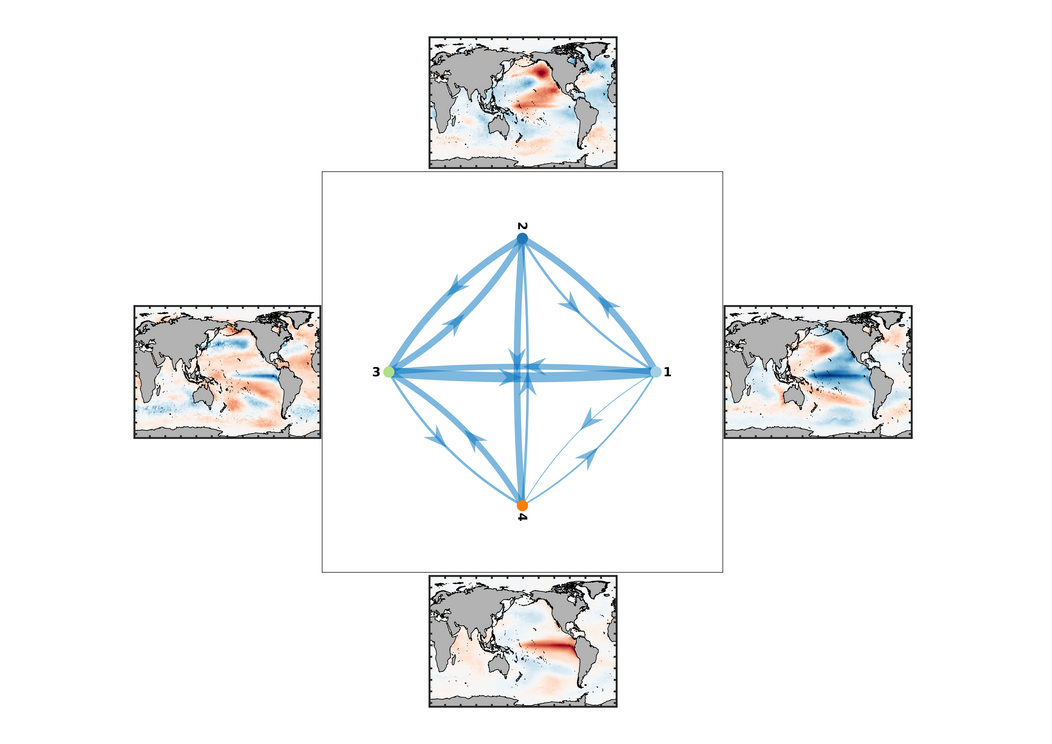}
    \caption{The directed graph representation of time-mean states transition probabilities from winning to winning archetypes, $\pi_{ij} = \frac{1}{N-1}\sum_{t=1}^{N-1} \{\underset{i=1,\ldots,4}{\arg\max}\,S_{it} \longrightarrow \underset{j=1,\ldots,4}{\arg\max}\,S_{jt+1}\}$, where $N$ is the total number of monthly records. The arrows thicknesses are proportional to the transition probabilities. Self transitions, $\pi_{ii}$ with $i=1,\ldots,4$, have been omitted. The colour-coded graph nodes correspond to the ENSO phase colours in Fig. \ref{fig:kfaa4} and \ref{fig:kfai4}.}
    \label{fig:aatp4}
\end{figure}
In Figures \ref{fig:aatp4} and S14, the time-mean probability of transitions between ENSO phases is represented by directed graphs illustrating the transitions from `winning' to `winning' archetypes, $\pi_{ij} = \frac{1}{N-1}\sum_{t=1}^{N-1} \{\underset{i=1,\ldots,4}{\arg\max}\,S_{it} \longrightarrow \underset{j=1,\ldots,4}{\arg\max}\,S_{jt+1}\}$, where $N$ is the total number of monthly records. The arrows thicknesses are proportional to the transition probabilities. Self transitions, $\pi_{ii}$ with $i=1,\ldots,4$, have been omitted given the persistence of ENSO conditions. The colour-coded graph nodes correspond to the ENSO phase colours in Figures \ref{fig:kfaa4} and \ref{fig:kfai4}. Such a representation suggests that AA could be used as a dimension reduction strategy, whereby the evolution of the ENSO phenomenon (and its teleconnection impacts) is reduced to and represented by the $\textbf{S}$-matrix time-series alone, our AA ENSO indices. Based on these indices, machine learning methods such long short-term memory (LSTM) deep learning methods \citep{ho97:nc, cho21:book} or Hidden Markov Models (HMM) \citep{bau66:ams, rab89:ieee} could be deployed to forecast ENSO, keeping in mind that the power\footnote{Here, `power' is understood as the predictive accuracy of the statistical algorithms employed as in \cite{bre01:ss}.} of any statistical forecasting method is contingent on the number of high-quality historical records available for training. In principle, the approach could be tested on long OAGCM simulations for models with an adequate ENSO representation.

\section{Summary and conclusion}\label{conclusion}
Since the discovery of the Southern Oscillation by Sir Gilbert Walker, our understanding of the ENSO has advanced in leaps and bounds. As illustrated in the introduction, every unusual event departing from the accepted historical paradigm at the time, triggered a surge of activity to further research, and lead to an extension of ocean-atmosphere observing systems in order to better capture the intricacies of the phenomenon given its global environmental and societal impacts. These successive changes in the understanding of ENSO and, not only associated with, but also driving, improvements in spatiotemporal observational coverage in both the atmosphere and oceans, all contributed to the current state of ENSO research. To date, ENSO events are commonly categorized by anomalous patterns of SST and their centers of action (regions of absolute anomaly maxima) in the tropical Pacific Ocean; these typically correspond to the central and eastern Pacific El Ni\~no and La Ni\~na conditions.

Built on recently published work by \cite{han17:jcli}, \cite{bla22:aies} and \cite{cha22:ncom}, we show that the approach based on archetypal analysis in Sections \ref{aa} and \ref{sec:flavours} and is able to recover the CP- and EP-like phases of ENSO using detrended global SST anomalies. The favorable comparison of AA to the categorical and soft clustering methods, K-means and fuzzy-C-means, underlines the robustness of the method. 

The choice of going global for the analysis domain is intentional as both tropical and extra-tropical information allows to capture the essence of the phenomenon and, importantly, its remote impact. We posit in Section \ref{sec:online} that AA, when applied to S2S forecasts, may complement and possibly sharpen the advice given by operational centers by further characterising ENSO teleconnections. Given the `nestedness' properties as a function of AA cardinality, extra flavours can be hierarchically introduced to study onset and decay of different phases, without changing the archetypal patterns uncovered for lower cardinalities as shown for dSSTAs and dSLAs in the supplementary material S15 and S16; the higher the cardinality, the more discriminatory AA becomes, as illustrated in Figure S17 for dSSTAs.

The archetype affiliation probabilities, the $\textbf{S}$-matrix, associated with each SST records, allow forecasting centers to set explicit criteria for onset, persistence, and decay of events. They correspond to the archetypal flavours and ENSO indices. We consider the probabilistic nature of the AA factorisation as a strong point. It allows the identification of teleconnections based on conditional expectations, in both mean and variance, by deferentially weighing the records of any geophysical fields of interest as a function of how well-expressed dSSTA global configurations are.

The non-stationarity in event sequences can be captured by changes in event probabilities of expression throughout the period under study as illustrated by Figures \ref{fig:stairs}b, \ref{fig:dSSTA_dSLA_8}a and \ref{fig:dSSTA_dSLA_8}b, showing that no two identical sequences of ENSO events are observed over the 1982 to 2022 period, even after removing a linear trend at every point approximately representing anthropogenic forcing. The classification of events via AA is reasonably robust to changes in the selection of time periods and geographical domains. The identification of new archetypes, where and when the dynamics or the underlying observing systems have changed, allows for data-driven regime shift detection and identification, either real or spurious, as AA captures to extreme configurations. AA performed on ocean sub-surface fields, such as dSLAs, provides a clearer expression of ENSO types, especially when the categorisation based on surface fields lacks discrimination. 

Finally, when considered as a dimension reduction strategy in Section \ref{sec:transition}, AA results, the affiliation probability matrix $\textbf{S}$ in particular, can be used as input to statistical prediction models applied to time-series forecasting to speed up deep learning algorithms. 

%%%%%%%%%%%%%%%%%%%%%%%%%%%%%%%%%%%%%%%%%%%%%%%%%%%%%%%%%%%%%%%%%%%%%
% ACKNOWLEDGMENTS
%%%%%%%%%%%%%%%%%%%%%%%%%%%%%%%%%%%%%%%%%%%%%%%%%%%%%%%%%%%%%%%%%%%%%
\acknowledgments
This work is supported the Australian Climate Service, a partnership of the Australian Bureau of Meteorology, Geoscience Australia, Commonwealth Scientific and Industrial Research Organisation (CSIRO) and Australian Bureau of Statistics (\url{https://www.acs.gov.au}). The collaboration between Abdelwaheb Hannachi and Didier Monselesan was fostered by the International Meteorological Institute at the University of Stockholm through its visiting scientist program and Didier Monselesan would like to thank Henriette Vals\"o at Stockholm University -- Department of Meteorology and Leesa MacKay at the CSIRO -- Environment, for facilitating international travel to Sweden. Sophie Cravatte, Bastien Pagli and Takashi Izumo contributions to the paper were supported in part by the Franco-Australian Hubert Curien Program (FASIC). FASIC is implemented jointly by the
Ministry of Europe and Foreign Affairs (MEAE) and the Ministry of Higher Education and Research (MESR) in France, and by research partners in Australia. Christopher Chapman was supported by the CSIRO Environment Climate Atmosphere Oceans Interaction Program. All coauthors have contributed to the work since its inception. No artificial intelligence (AI) chat-bots were deployed to either conceptualise, code or generate content for this work. 

%%%%%%%%%%%%%%%%%%%%%%%%%%%%%%%%%%%%%%%%%%%%%%%%%%%%%%%%%%%%%%%%%%%%%
% DATA AVAILABILITY STATEMENT
%%%%%%%%%%%%%%%%%%%%%%%%%%%%%%%%%%%%%%%%%%%%%%%%%%%%%%%%%%%%%%%%%%%%%
\datastatement
Sea surface temperature (SST) data are from the Optimum Interpolation Sea Surface Temperature (OISST) v2.1 high resolution dataset provided by the NOAA/OAR/ESRL Physical Sciences Laboratory. This data is available at \url{https://psl.noaa.gov}. The multi-variate ENSO index (MEI) was retrieved from \url{https://psl.noaa.gov/enso/mei}. The Oceanic Ni\~no Index (ONI) from the Climate Prediction Center (CPC)
is provided by NOAA/PSL downloaded from \url{http://www.cpc.ncep.noaa.gov/data/indices/oni.ascii.txt}. Global atmospheric reanalysis data used to relate extreme events to large-scale climate modes comes from the Japanese 55-year Reanalysis (JRA-55) project carried out by the Japan Meteorological Agency (JMA). JRA-55 data is available at \url{https://jra.kishou.go.jp}. Global precipitation composites use Global Precipitation Climatology Project (GPCP) Climate Data Record (CDR), Version 2.3 (Monthly), from the National Centers for Environmental Information, doi:10.7289/V56971M6, whereas Australian precipitation composites are generated from the Australian Gridded Climate Data (AGCD), the Australian Bureau of Meteorology's official dataset for climate analyses covering the variables of rainfall, temperature (maximum and minimum) as well as vapour pressure at daily and monthly timescales accessible at \url{https://geonetwork.nci.org.au/geonetwork/srv/eng/catalog.search#/metadata/f7421_3667_8543_6351}. The Ssalto/Duacs altimeter products were produced and distributed by the Copernicus Marine and Environment Monitoring Service (CMEMS), accessible at \url{http://www.marine.copernicus.eu}. The Simple Ocean Data Assimilation (SODA) dataset was retrieved from \url{https://climatedataguide.ucar.edu/climate-data/soda-simple-ocean-data-assimilation} on 2023-05-23.
\clearpage
\bibliographystyle{ametsocV6}
\bibliography{aa_flavours_00}
\end{document}

\nolinenumbers
\maketitle

\section{The archetypal analysis method}
For any set of data points $\textbf{X}=\textbf{X}_{s\times t}$\footnote{Where the indices $s$ and $t$ stand for the spatial and temporal dimensions} or pointset\footnote{\url{https://en.wiktionary.org/wiki/pointset}}, \citeauthor{cut94:tec}~(\citeyear{cut94:tec}) show 1) that AA detects `extreme configurations', $\textbf{XC}_{s\times p} = \sum^{t}_{i=1}X_{si}C_{ip}$, the so-called archetypes, which are themselves expressed as a convex combination\footnote{\url{https://en.wikipedia.org/wiki/Convex_combination}} or the product of the data points $\textbf{X}$ with a right-stochastic matrix\footnote{\url{https://en.wikipedia.org/wiki/Stochastic_matrix}} $\textbf{C}=\textbf{C}_{t\times p}$, where $p$ stands for the number of archetypes set \textsl{a priori}, and 2) that the original dataset can be approximated as a convex combination of these archetypes or the product of $\textbf{XC}_{s\times p}$ with an other right-stochastic matrix $\textbf{S}=\textbf{S}_{p\times t}$, $\textbf{XCS}_{s\times t} = \sum^{p}_{j=1} XC_{sj}S_{jt}$. Both matrices, $\textbf{C}$ and $\textbf{S}$, are by minimizing the distance between this representation and the original dataset. For data matrices, the distance function used here corresponds to the Froebenius norm\footnote{\url{https://en.wikipedia.org/wiki/Matrix_norm\#Frobenius_norm}}, 
\begin{equation*}
    \|\textbf{X}_{s\times t}\|_{F} = \sqrt{\sum_{i=1}^{s}\sum_{j=1}^{t} |X_{ij}|^2},
\end{equation*}
as in Equation 1 in the submitted manuscript. To paraphrase Wikipedia on convex combination, each data point (or archetype) is equivalent to a standard weighted average of the archetypes (or data points), whose weights are expressed as a percent of the total weight, here equal to one, instead of as a fraction of the count of the weights as in a standard weighted average. Left (or column) - stochastic or right (or row) - stochastic matrix are accepted terms in statistics. They refer to positive definite matrices, $\textbf{M}_{i \times j}$, whose entries are all greater or equal to zero and where rows (right) entries $M_{ij}$ sum to 1, $\sum_{j} M_{ij} = 1,\, \forall i$, and where column (left) entries sum to 1, $\sum_{i} M_{ij} = 1,\, \forall j$. Note that sometimes the definitions are reversed, i.e. right for left and row for column. Here we follow the convention $\textbf{M}_{i \times j} = \textbf{M}_{row \times col}$.

\begin{figure}[htbp]
  \centering
  \noindent
  \begin{subfigure}[b]{0.49\textwidth}
    \centering  
    \includegraphics[width=\textwidth]{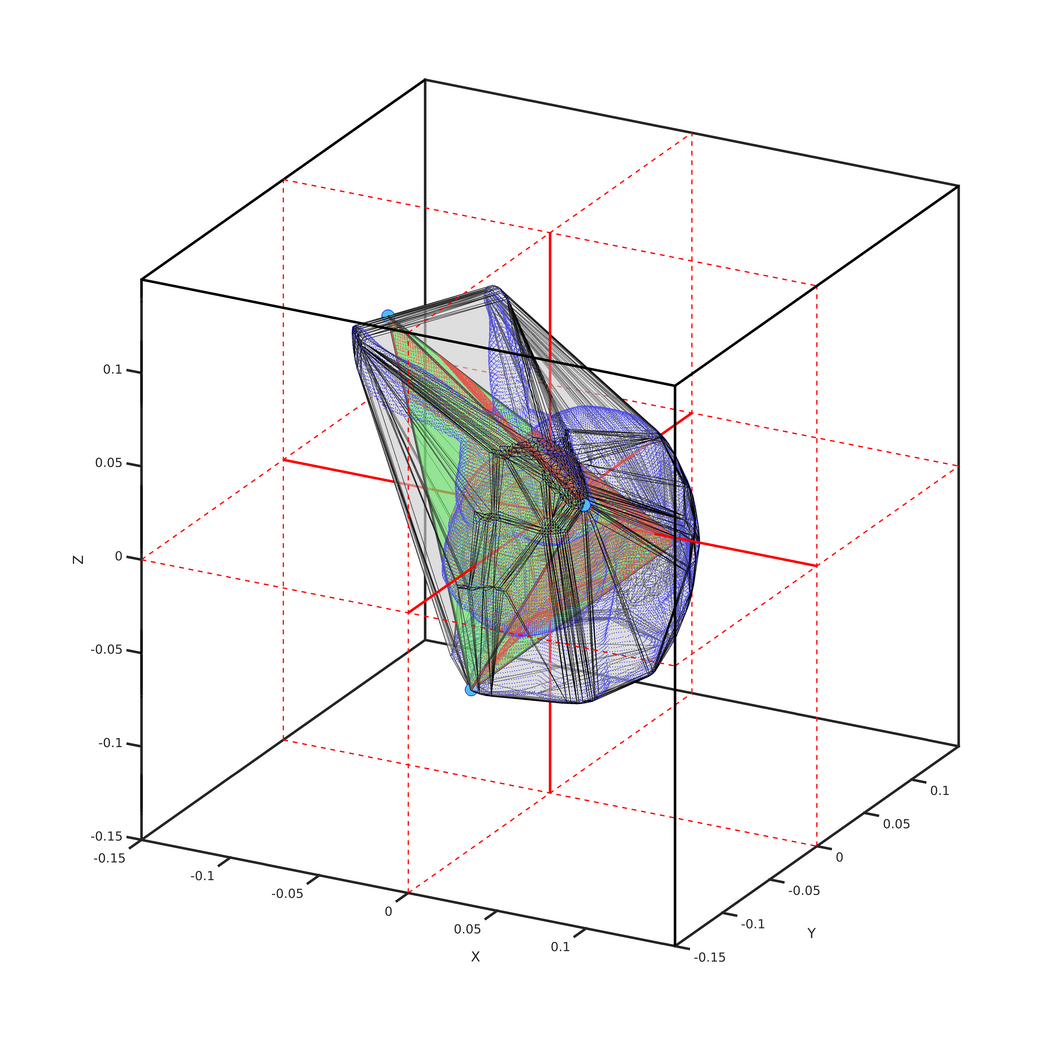}
    \caption{Cardinality 4}
  \end{subfigure}
  \begin{subfigure}[b]{0.49\textwidth}
    \centering  
    \includegraphics[width=\textwidth]{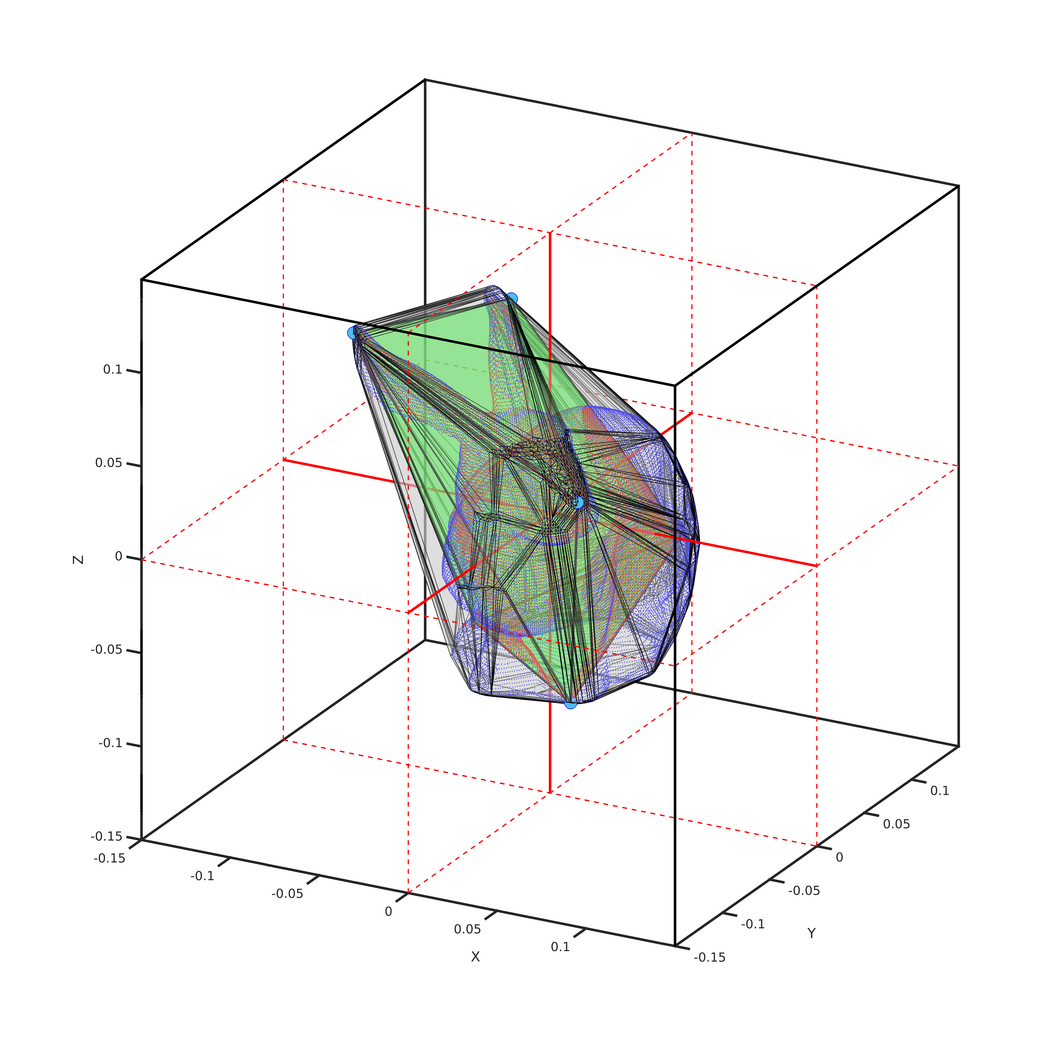}
    \caption{Cardinality 6}
  \end{subfigure}
  \centering
  \noindent
  \begin{subfigure}[b]{0.49\textwidth}
    \centering  
    \includegraphics[width=\textwidth]{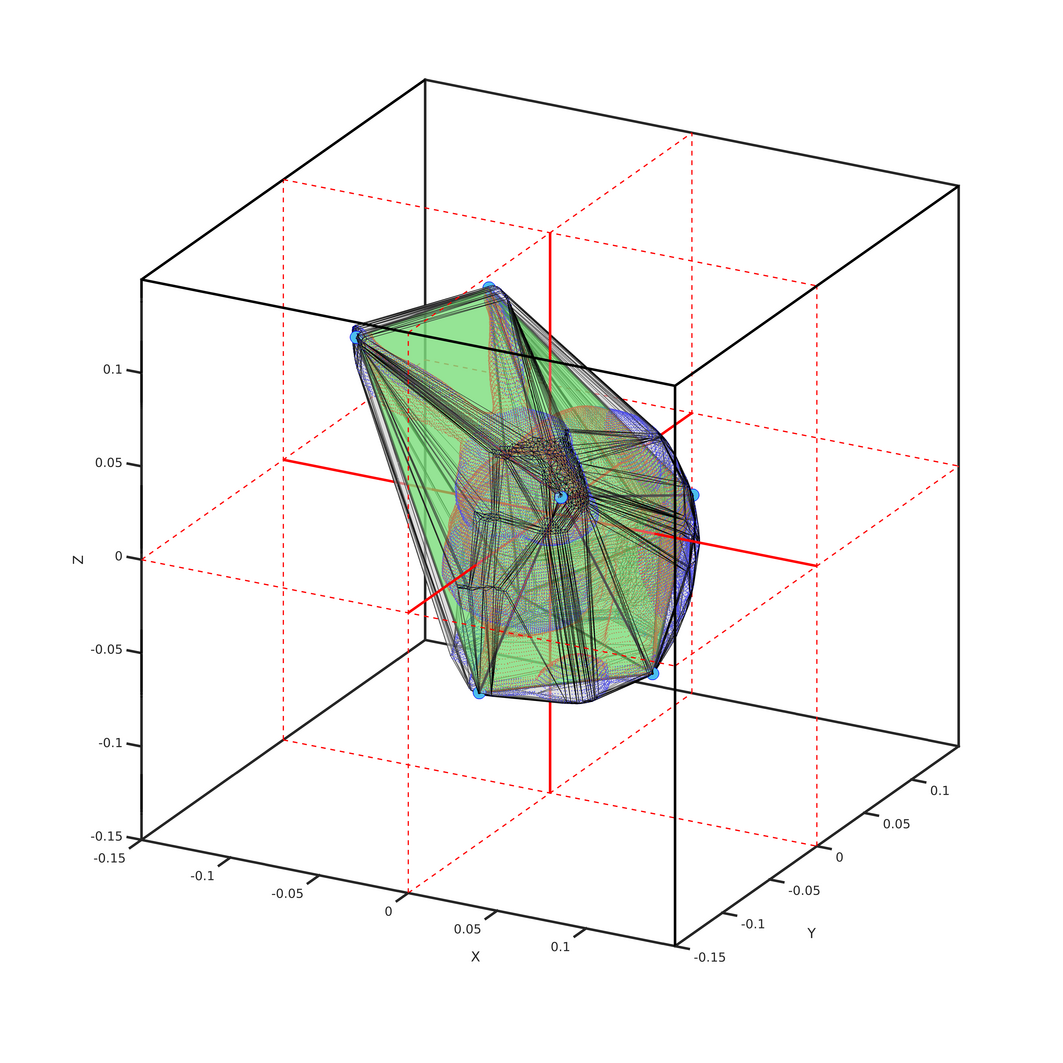}
    \caption{Cardinality 8}
  \end{subfigure}
  \begin{subfigure}[b]{0.49\textwidth}
    \centering  
    \includegraphics[width=\textwidth]{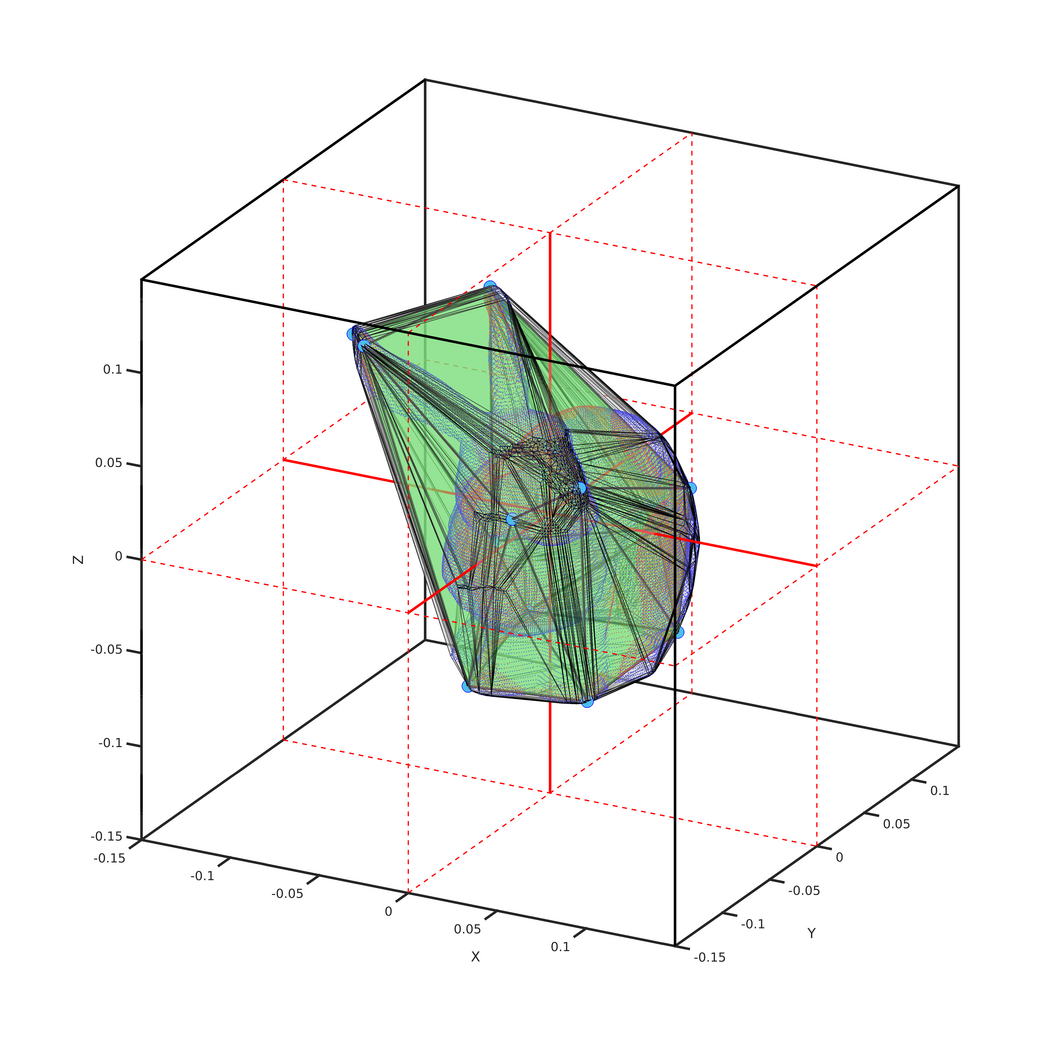}
    \caption{Cardinality 12}
  \end{subfigure}
  \caption{Convex hull and its AA approximation of the Stanford Bunny. The black lines forming the grey polyhedron connect the vertices of a convex hull of the Bunny. The green polyhedrons are the AA approximation of the convex hull, the grey polyhedron, for archetype cardinalities of 4, 6, 8 and 12. The cyan circles at the vertices of the green polyhedrons correspond to the locations of the archetypes of the Bunny.}
\label{fig:sb}
\end{figure}

To illustrate both the concept of a convex hull\footnote{\url{https://en.wikipedia.org/wiki/Convex_hull}, a convex polytop with the smallest volume encapsulating the entire pointset.} and its archetypal approximation, we present in Figure \ref{fig:sb} the Stanford Bunny convex hull and archetypes thereof. The black lines in Figure \ref{fig:sb} forming the grey polyhedron connect the vertices of a convex hull of the Bunny - imagine the grey polyhedron as a tight cellophane wrapper around a chocolate Easter Bunny. The green polyhedrons are the AA approximation of the convex hull, the tight cellophane wrapper, for archetype cardinalities of 4, 6, 8 and 12. The cyan circles at the vertices of the green polyhedrons correspond to the locations of the archetypes of the Bunny. In the case of the Bunny, `extreme configurations' of the dataset correspond approximately to the extremities of the Bunny's shape: ear tips, snout, tail, paws, etc,~...

The monthly records of detrended sea surface temperature anomalies (dSSTAs) over the 1982 to 2022 period is the pointset under consideration and each point of the set corresponds to an individual dSSTA monthly record over the region of interest. We decided to go `global' in our choice of region for the reasons exposed in the data section 3 of the main submission and we will come back to it in the following.

\begin{figure}[thbp]
    \centering\includegraphics[width=0.8\textwidth]{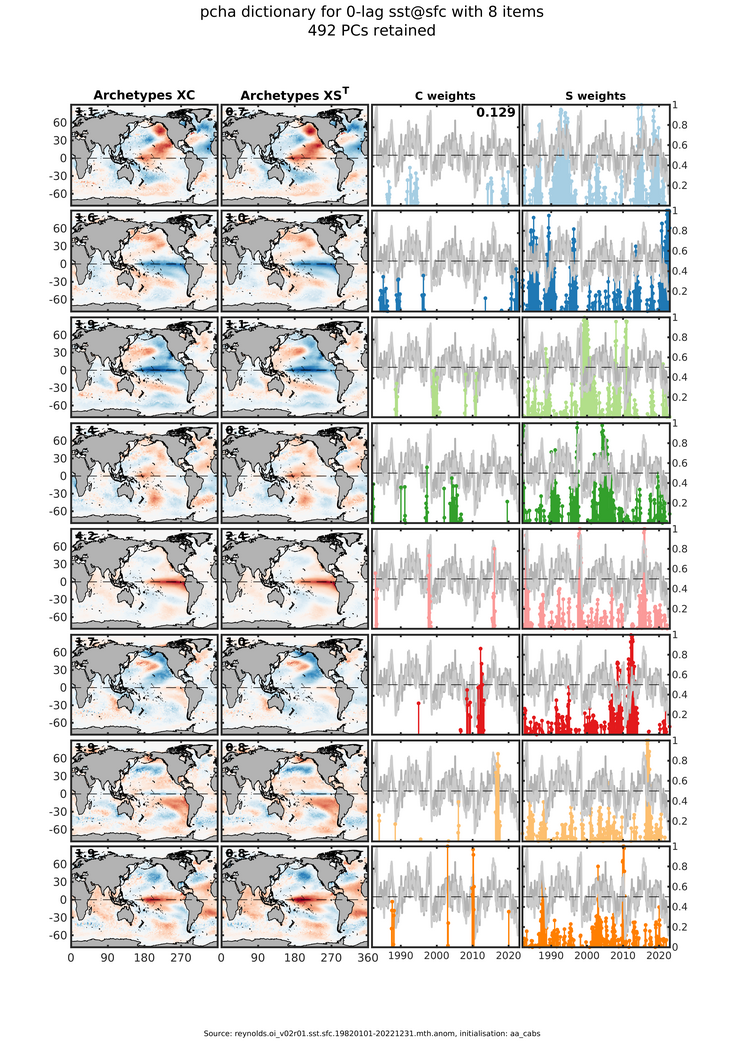}
    \caption{Adapted from \citeauthor{bla22:aies}~(\citeyear{bla22:aies}), archetypal analysis spatial pattern and time series results using detrended monthly SST anomalies over 1982-2022 for cardinality 8. The first and second columns on each subplot show archetypes constructed by $\textbf{XC}$ and $\textbf{X}\tilde{\textbf{S}}^T$, with $\tilde{\textbf{S}}_{p\times t}=\textbf{S}_{p\times t}/\sum^{t}_{i=1} S(p,i)$, respectively. The third and fourth columns show $\textbf{C}$ and $\textbf{S}$ matrix time series, respectively, with Multivariate ENSO Index (MEI) time series (grey) included on both. The maximum amplitude of the patterns is given in bold in $^{\circ}$C on top left corner of the each map. The maximum probability of 0.129 value across all archetypes (rows) for all times of the $\textbf{C}$ matrix is reported in bold on top right corner.}
    \label{fig:global}
\end{figure}

The main advantages of the AA representation, when compared to principal component analysis for example as discussed in \cite{ris21:mwr} and \cite{bla22:aies}, are 1) that the convexity characteristic of AA is crucial and leads to a probabilistic interpretation of both archetypes, $\textbf{XC}$, and the dataset representation, $\textbf{XCS}$, the matrices $\textbf{C}$ and $\textbf{S}$ being stochastic - whose columns or rows can be viewed as the probability of expression of data records - and 2) that archetypes are `closer' to individual data points or snapshots and are more representative than EOF patterns, as the $\textbf{C}$ matrix is sparse, meaning that archetypes are the weighted sum of only a few `(arche)typical' and similar data records, as illustrated in the third column of Figure \ref{fig:global} reproduced from \cite{bla22:aies}. In fact, for some datasets, the archetypes can correspond to \textbf{single} data record or snapshot. EOF patterns may never be observed, as the covariance of $\textbf{X}$ is the time-mean (or the statistical expectation) of the product of anomalies derived from the climatology computed over the entire data set.

\begin{figure}[thbp]
  \centering
  \noindent
  \begin{subfigure}[b]{\textwidth}
    \centering
    \includegraphics[width=0.8\linewidth]{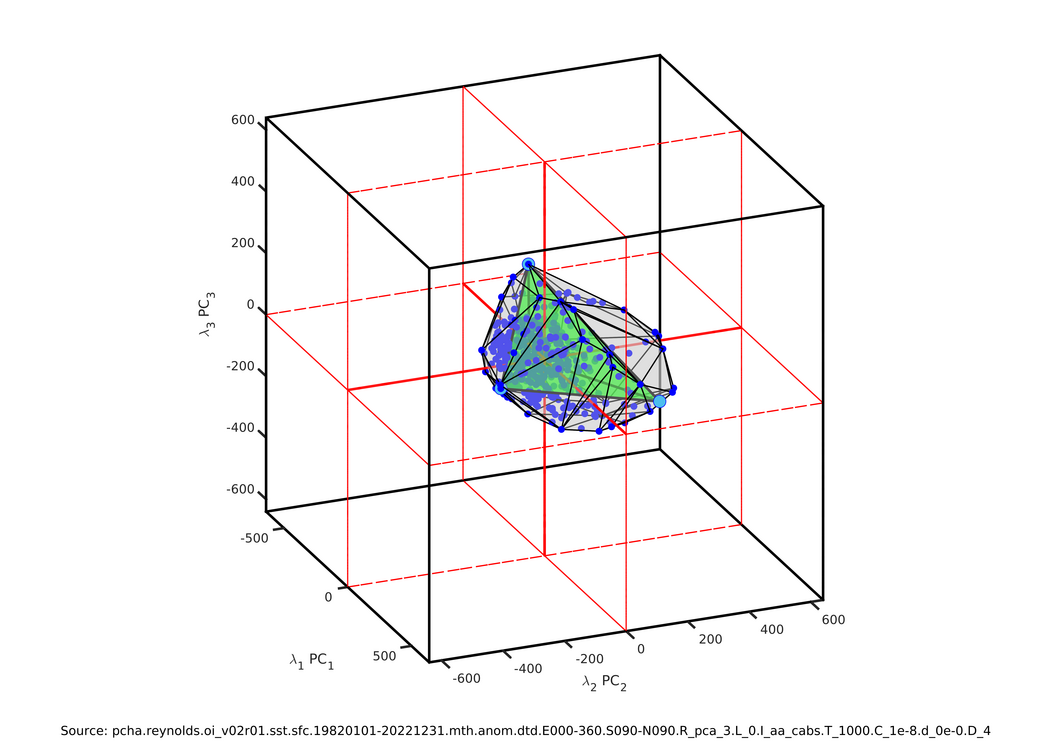}
    \caption{Cardinality 4}
  \end{subfigure} 
  \begin{subfigure}[b]{\textwidth}
    \centering
    \includegraphics[width=0.8\linewidth]{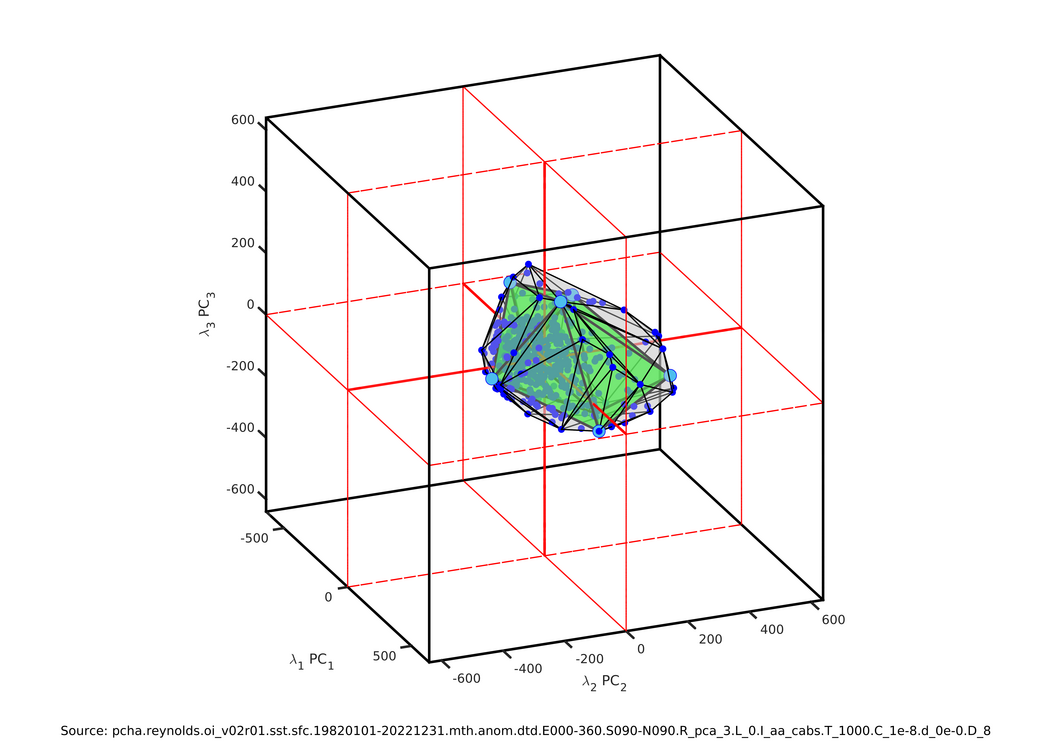}
    \caption{Cardinality 8}
  \end{subfigure}
  \caption{Adapted from \citeauthor{ris21:mwr}~(\citeyear{ris21:mwr}), the convex hull of dSSTAs truncated to the first three PCs. The cloud of points in blue represents monthly dSSTAs for all months over the 1982 to 2022 period, where each value is represented by the leading three PCs of the data. The black lines forming the polyhedron connect the vertices of a convex hull around all the blue points. The green tetrahedron is the AA approximation of the polyhedral convex hull for 4 a) and 8 b) archetypes. The cyan circles at the vertices of the tetrahedron are the locations of the archetypes.}
\label{fig:ch4_and_8}
\end{figure}

We analyse a monthly dataset, the pointset marked by dark blue dots in Figures \ref{fig:ch4_and_8}a and \ref{fig:ch4_and_8}b. These figures show the dSSTA convex hull - the grey polytop - and the archetypes or `extreme configurations' of dSSTAs - the vertices marked by cyan dots - when approximated by the first 3 PCs of dSSTAs for the whole globe. The green tetrahedron and octahedron are the AA approximation of the convex hull for 4 and 8 archetypes. They illustrate the `extremal' character of the archetypes and show that AA does uncover extreme configurations of the underlying dataset. We note that all 492 PCs, not just the first 3, are used in the analysis presented, representing 100\% of the dSSTAs variance.
It is important to make the distinction between `extreme configurations' of anomalies of any geophysical fields over the spatial domain under consideration - \textbf{a global property} - and extreme values of fields at a given spatial location - \textbf{a local property}. These do not necessarily correspond. For example, the archetype identified as EP-Ni\~no not only corresponds to an `extreme configuration' of the dSSTA pointset, but also displays the largest local dSSTA anomaly (+4.2$^{\circ}$C) in Figure \ref{fig:global}, row 5. Not all archetypes necessarily share this property as can be seen in Figure \ref{fig:global} for all the other rows; they correspond however to `extreme configurations' of dSSTAs over the geographical domain used for this study.

Finally, we have introduced our own measure of significance based on weighted covariance (Equations 4 and 5), to show that the archetypal patterns discovered are repeatable and reoccur over the period considered in a conditional time-mean sense by comparing pattern (conditional) mean-to-spread (or signal-to-noise) ratios. We note the `in-sample' signal-to-noise ratios reported in both the main submission and hereafter are a very conservative measure of reliability, especially for composites based of the affiliation weights, the stochastic matrix $\textbf{S}$. As mentioned in the core of the paper in Section 4, a less stringent measure of reliability or significance -- but intractable for spatial dimensions larger than 9 -- would be to sample the convex hull to generate pattern spread. An even less stringent approach as in \cite{cha22:ncom}, would be to employ a brute-force Monte-Carlo approach, 1) to generate an ensemble of random synthetic stochastic matrices to replicate the features of both the $\textbf{C}$ and $\textbf{S}$, 2) to compute the composites on these, and 3) to declare each point of the composites `significant' if not falling within a certain percentile range constructed from the synthetic ensemble. As archetypes correspond by construction to extreme configurations, the latter method mentioned is too permissive. The composites generated from the random stochastic matrices are not necessarily located on the convex hull of the dataset and cannot therefore be considered `legitimate' samples as archetypes are located on the convex hull by construction. We have now added stippling to all composited field figures to indicate the level significance as measured by the archetypal pattern (conditional) mean-to-spread ratios.

\section{The choice of SST domain and period of interest}

As in \cite{bla22:aies}, the dSSTA data matrix, $\textbf{X}_{s\times t}$, is weighted spatially by $\textbf{W}$ with elements $W_{ij} = \sqrt{cos(\theta_{i})}$, $\theta_{i}$ being the latitude of the grid cell $i=1,\ldots,s$ for all times $j=1,\ldots,t$ prior applying PCA, AA, KM and FCM to $\textbf{X}_{s \times t} \odot \textbf{W}_{s \times t}$\footnote{Element-wise or Hadamard matrix product, $(A \odot B)_{ij}=A_{ij}B_{ij}$.} rather than $\textbf{X}_{s\times t}$ in order to compensate for unequal grid spacing. As not to render our notation too cumbersome, we omit the weights, $W_{i}$. The tropical data points are therefore carrying more weight than mid- and high-latitude ones. However, when KM, FCM and AA composites are computed, unweighted snapshots based on the dSSTA cluster affiliation sequences alone are used.

The choice of the domain driving the analysis is motivated by minimizing the number of arbitrary choices with the explicit goal of capturing both the tropical character and remote teleconnection of the ENSO phenomenon at zero monthly lag. We have repeated the analysis for Pacific Ocean regions over a wide range of latitude bands ranging from $\pm 10^{\circ}$ to $\pm 90^{\circ}$ centered on the Equator with a fixed longitudinal sector from $120^{\circ}$E to $280^{\circ}$E and compared archetypes, $\textbf{XC}$, for cardinality 8 using pattern correlation between overlapping regions of the restricted and global domains in Table \ref{tab:match1}. From the latitudinal band $\pm 50^{\circ}$ and higher, individual regional archetypes can be uniquely matched to the global ones, with high correlation values. The correlation values given in Tables \ref{tab:match2} correspond to the archetype labels of Table \ref{tab:match1}. The labels $i=1,\ldots,8$ in each row indicate the archetype ranks based on the time mean of the AA stochastic matrix $\textbf{S}$ - the mean probability of expression of the archetypes across the whole period - in decreasing order. The time means are expressed as $\overline{S}_{i}=(\sum^{T}_{j=1} S_{ij})/T$ for archetypes $i=1,\ldots,8$ and $j=1,\ldots,T$ monthly records.
\begin{table}[htbp]
    \centering
    \begin{tabular}{c|llllllll}
        Global AA        & 1 & 2 & 3 & 4 & 5 & 6 & 7 & 8 \\
        Pacific &&&&&&&&\\
        \hline
        \hline
        $\pm 10^{\circ}$ & \textbf{4} & 3 & 2 & \textbf{4} & \textbf{5} & \textbf{5} & 7 & \textbf{5} \\
        $\pm 15^{\circ}$ & 1 & 3 & \textbf{2} & \textbf{2} & 8 & 7 & \textbf{5} & \textbf{5} \\
        $\pm 20^{\circ}$ & 1 & \textbf{2} & 8 & 6 & \textbf{2} & \textbf{7} & 5 & \textbf{7} \\
        $\pm 25^{\circ}$ & 1 & 3 & 8 & \textbf{2} & \textbf{2} & \textbf{7} & 5 & \textbf{7} \\ 
        $\pm 30^{\circ}$ & 1 & \textbf{3} & 8 & 2 & 7 & \textbf{5} & \textbf{5} & \textbf{3} \\
        $\pm 35^{\circ}$ & 8 & 1 & \textbf{2} & 3 & \textbf{2} & 5 & 6 & 7 \\
        $\pm 40^{\circ}$ & 8 & 1 & 2 & \textbf{7} & 5 & 3 & 6 & \textbf{7} \\
        $\pm 45^{\circ}$ & 4 & 2 & 3 & 1 & \textbf{7} & 5 & 8 & \textbf{7} \\
        \hline
        $\pm 50^{\circ}$ & 3 & 2 & 4 & 1 & 7 & 5 & 6 & 8 \\
        $\pm 55^{\circ}$ & 2 & 3 & 4 & 1 & 7 & 5 & 6 & 8 \\
        $\pm 60^{\circ}$ & 2 & 3 & 1 & 4 & 7 & 5 & 6 & 8 \\
        $\pm 65^{\circ}$ & 2 & 3 & 1 & 4 & 7 & 5 & 6 & 8 \\
        $\pm 70^{\circ}$ & 2 & 3 & 1 & 4 & 7 & 5 & 6 & 8 \\
        $\pm 75^{\circ}$ & 2 & 3 & 1 & 4 & 7 & 5 & 6 & 8 \\
        $\pm 80^{\circ}$ & 2 & 3 & 1 & 4 & 7 & 5 & 6 & 8 \\
        $\pm 85^{\circ}$ & 2 & 3 & 1 & 4 & 7 & 5 & 6 & 8 \\
        $\pm 90^{\circ}$ & 2 & 3 & 1 & 4 & 7 & 5 & 6 & 8 \\
        \hline
    \end{tabular}
    \caption{Archetypes, $\textbf{XC}$, matching between Global and Pacific Ocean over a wide range of latitude bands ranging from $\pm 10^{\circ}$ to $\pm 90^{\circ}$ centered on the Equator with a fixed longitudinal sector from $120^{\circ}$E to $280^{\circ}$E. The horizontal line indicates the latitudinal band where archetypes can be matched uniquely to the global ones. Bold faced labels correspond to archetypes which cannot be uniquely matched to the global ones.}
    \label{tab:match1}
\end{table}
\begin{table}[htbp]
    \centering
    \begin{tabular}{c|cccccccc}
        Global AA        & 1 & 2 & 3 & 4 & 5 & 6 & 7 & 8 \\
        Pacific &&&&&&&&\\
        \hline
        \hline
        $\pm 10^{\circ}$& 0.9135  &  0.9791  &  0.8855  &  0.7630  &  0.9080  &  0.9784  &  0.7779  &  0.7734 \\
        $\pm 15^{\circ}$& 0.9082  &  0.9665  &  0.8059  &  0.9012  &  0.9180  &  0.7882  &  0.9882  &  0.7807 \\
        $\pm 20^{\circ}$& 0.9161  &  0.9310  &  0.9349  &  0.6830  &  0.8491  &  0.9514  &  0.9939  &  0.4761 \\
        $\pm 25^{\circ}$& 0.9295  &  0.8315  &  0.9287  &  0.9238  &  0.8267  &  0.9487  &  0.9940  &  0.4211 \\
        $\pm 30^{\circ}$& 0.8911  &  0.9336  &  0.8944  &  0.9450  &  0.6683  &  0.9515  &  0.9644  &  0.3646 \\
        $\pm 35^{\circ}$& 0.8300  &  0.8842  &  0.9357  &  0.9710  &  0.6274  &  0.9923  &  0.8793  &  0.2720 \\
        $\pm 40^{\circ}$& 0.6142  &  0.8883  &  0.9283  &  0.5708  &  0.9926  &  0.9683  &  0.9222  &  0.2479 \\
        $\pm 45^{\circ}$& 0.5753  &  0.9263  &  0.9415  &  0.9248  &  0.7387  &  0.9883  &  0.9637  &  0.2457 \\
        \hline
        $\pm 50^{\circ}$& 0.9878  &  0.8353  &  0.5397  &  0.9406  &  0.9698  &  0.9940  &  0.9775  &  0.9533 \\
        $\pm 55^{\circ}$& 0.9580  &  0.9948  &  0.5853  &  0.9392  &  0.9740  &  0.9943  &  0.9858  &  0.9544 \\
        $\pm 60^{\circ}$& 0.9687  &  0.9956  &  0.9362  &  0.5765  &  0.9738  &  0.9948  &  0.9877  &  0.9531 \\
        $\pm 65^{\circ}$& 0.9683  &  0.9960  &  0.9367  &  0.5807  &  0.9766  &  0.9951  &  0.9899  &  0.9543 \\
        $\pm 70^{\circ}$& 0.9695  &  0.9963  &  0.9361  &  0.5894  &  0.9777  &  0.9953  &  0.9894  &  0.9548 \\
        $\pm 75^{\circ}$& 0.9701  &  0.9965  &  0.9344  &  0.5885  &  0.9770  &  0.9955  &  0.9888  &  0.9546 \\
        $\pm 80^{\circ}$& 0.9699  &  0.9965  &  0.9341  &  0.5873  &  0.9770  &  0.9955  &  0.9884  &  0.9548 \\
        $\pm 85^{\circ}$& 0.9699  &  0.9966  &  0.9342  &  0.5882  &  0.9771  &  0.9955  &  0.9885  &  0.9549 \\
        $\pm 90^{\circ}$& 0.9698  &  0.9966  &  0.9342  &  0.5894  &  0.9771  &  0.9955  &  0.9885  &  0.9547 \\
        \hline
    \end{tabular}
    \caption{Archetypes correlation values corresponding to matched Global and Pacific Ocean over a wide range of latitude bands ranging from $\pm 10^{\circ}$ to $\pm 90^{\circ}$ centered on the Equator with a fixed longitudinal sector from $120^{\circ}$E to $280^{\circ}$E given in Table S\ref{tab:match1}.}
    \label{tab:match2}
\end{table}

Alternatively, when comparing archetypes from one latitudinal band to the next in increments of $\pm 5^{\circ}$, starting with the Pacific region $\pm 10^{\circ}$ latitudinal band as in Table \ref{tab:match3}, individual regional archetypes can only be uniquely matched from the latitudinal band $\pm 55^{\circ}$ and higher, an indication that to discriminate between extreme dSSTAs conditions, the `latitudinal reach' has to be increased. Both matching strategies show that, to discriminate between archetypes across bands, it is necessary to select Pacific regions which latitudinal extent larger than about $\pm 50^{\circ}$ centered on the Equator. We note that the pattern correlation between consecutive latitudinal bands, in Table \ref{tab:match4}, are higher than those reported in Table \ref{tab:match2}, and remain high when the whole Pacific region archetypes are compared to the global archetypes. As we strive for parsimony, but also would like to capture the global teleconnections (at lag zero) across oceanic basins, we have therefore opted for a global domain.

An other compelling and dynamically driven argument for `going global' that the authors have put forward in the introduction is that processes associated to the ENSO phenomenon may impact a large range of latitudes and potentially tele-connect to other oceanic basins over the lifetime of an ENSO phase. For some El Ni\~no events for example, westerly wind events triggering tropically trapped oceanic Kelvin waves in the Pacific, followed by Eastern boundary trapped coastal Kelvin waves along the West coast of both the North and South American continents and associated oceanic Rossby waves radiating back westwards across the Pacific basin, are observed over wide range of latitudes in both hemispheres. Given the speed of atmospheric teleconnections, it can be argued that monthly mean SST records at latitudes other than the Tropics alone would have been influenced via the atmospheric bridge and its interaction with mid- to high-latitude atmospheric dynamics.

However, for applications targeting specific physical processes, it is necessary 1) to restrict the geographical domain as illustrated in \cite{cha22:ncom}, where AA has been applied to study the proximate and remote drivers of marine heat waves (MHWs) around the Australian continent coastline and 2) to use sampling rates commensurate with the typical time-scales of the mechanisms under study. AA has been shown to recover MHW events diagnosed by standard methods reviewed by \cite{oli21:arms}, for example. Although not in scope, it can be shown that AA can be applied to phenomena occurring in the Tropical wave guides. As mentioned earlier, these studies require 1) to adjust sampling rates to be commensurate with time-scale of the phenomena and 2) to choose the appropriate atmospheric and oceanic fields.

\begin{table}[htbp]
    \centering
    \begin{tabular}{c|llllllll}
        $\pm 10^{\circ}$  & 1 & 2 & 3 & 4 & 5 & 6 & 7 & 8 \\
        \hline
        \hline
        $\pm 15^{\circ}$ & \textbf{1}  &  2  &  3  &  \textbf{1}  &  5  &  7  &  6  &  8\\
        $\pm 20^{\circ}$ & 1  &  \textbf{2}  &  5  &  \textbf{2}  &  3  &  8  &  \textbf{7}  &  \textbf{7}\\
        $\pm 25^{\circ}$ & 1  &  4  &  3  &  2  &  5  &  6  &  7  &  8\\
        $\pm 30^{\circ}$ & 1  &  2  &  3  &  4  &  \textbf{5}  &  \textbf{5}  &  6  &  8\\
        $\pm 35^{\circ}$ & 2  &  4  &  1  &  3  &  5  &  \textbf{6}  &  \textbf{6}  &  8\\
        $\pm 40^{\circ}$ & 1  &  2  &  3  &  6  &  4  &  5  &  7  &  8\\
        $\pm 45^{\circ}$ & 1  &  4  &  2  &  5  &  \textbf{6}  &  3  &  \textbf{6}  &  8\\
        $\pm 50^{\circ}$ & 3  &  \textbf{1}  &  \textbf{1}  &  4  &  \textbf{5}  &  6  &  8  &  \textbf{5}\\
        \hline
        $\pm 55^{\circ}$ & 2  &  1  &  3  &  4  &  5  &  6  &  7  &  8\\
        $\pm 60^{\circ}$ & 1  &  2  &  4  &  3  &  5  &  6  &  7  &  8\\
        $\pm 65^{\circ}$ & 1  &  2  &  3  &  4  &  5  &  6  &  7  &  8\\
        $\pm 70^{\circ}$ & 1  &  2  &  3  &  4  &  5  &  6  &  7  &  8\\
        $\pm 75^{\circ}$ & 1  &  2  &  3  &  4  &  5  &  6  &  7  &  8\\
        $\pm 80^{\circ}$ & 1  &  2  &  3  &  4  &  5  &  6  &  7  &  8\\
        $\pm 85^{\circ}$ & 1  &  2  &  3  &  4  &  5  &  6  &  7  &  8\\
        $\pm 90^{\circ}$ & 1  &  2  &  3  &  4  &  5  &  6  &  7  &  8\\
        Global           & 2  &  3  &  1  &  4  &  7  &  5  &  6  &  8\\
        \hline
    \end{tabular}
    \caption{Archetypes matching between consecutive Pacific domains with increasing range of latitudes. The last row corresponds to the matching archetypes between the Pacific domain and the whole globe. The horizontal line indicates the latitudinal band where archetypes can be matched uniquely from one band to the next. Bold faced labels correspond to archetypes which cannot be uniquely matched to those of the next band.}
    \label{tab:match3}
\end{table}
\begin{table}[htbp]
    \centering
    \begin{tabular}{c|cccccccc}
        $\pm 10^{\circ}$  & 1 & 2 & 3 & 4 & 5 & 6 & 7 & 8 \\
        \hline
        \hline
        $\pm 15^{\circ}$ & 0.9435  &  0.9687  &  0.9901  &  0.8070  &  0.9598  &  0.9940  &  0.9848  &  0.9963\\
        $\pm 20^{\circ}$ & 0.9931  &  0.8755  &  0.9852  &  0.9330  &  0.9868  &  0.8422  &  0.9970  &  0.7690\\
        $\pm 25^{\circ}$ & 0.9911  &  0.9994  &  0.9939  &  0.9044  &  0.9883  &  0.9600  &  0.9993  &  0.9869\\
        $\pm 30^{\circ}$ & 0.9507  &  0.9718  &  0.9773  &  0.9839  &  0.8196  &  0.8239  &  0.9768  &  0.9984\\
        $\pm 35^{\circ}$ & 0.9717  &  0.9713  &  0.9279  &  0.9901  &  0.9734  &  0.9729  &  0.9384  &  0.9782\\
        $\pm 40^{\circ}$ & 0.9683  &  0.9888  &  0.9978  &  0.9991  &  0.9883  &  0.9994  &  0.9885  &  0.9958\\
        $\pm 45^{\circ}$ & 0.8795  &  0.9794  &  0.9946  &  0.9632  &  0.9969  &  0.9774  &  0.3373  &  0.9980\\
        $\pm 50^{\circ}$ & 0.9694  &  0.8601  &  0.9126  &  0.9415  &  0.7611  &  0.9914  &  0.9916  &  0.2509\\
        \hline
        $\pm 55^{\circ}$ & 0.9962  &  0.9472  &  0.9880  &  0.9962  &  0.9957  &  0.9999  &  0.9957  &  0.9982\\
        $\pm 60^{\circ}$ & 0.9973  &  0.9994  &  0.9992  &  0.9983  &  0.9983  &  0.9999  &  0.9990  &  0.9994\\
        $\pm 65^{\circ}$ & 0.9998  &  0.9999  &  0.9997  &  0.9997  &  0.9996  &  1.0000  &  0.9991  &  0.9995\\
        $\pm 70^{\circ}$ & 0.9999  &  1.0000  &  0.9999  &  0.9999  &  0.9998  &  1.0000  &  0.9998  &  0.9999\\
        $\pm 75^{\circ}$ & 0.9999  &  1.0000  &  1.0000  &  0.9998  &  0.9999  &  1.0000  &  0.9998  &  1.0000\\
        $\pm 80^{\circ}$ & 1.0000  &  1.0000  &  1.0000  &  1.0000  &  1.0000  &  1.0000  &  1.0000  &  1.0000\\
        $\pm 85^{\circ}$ & 1.0000  &  1.0000  &  1.0000  &  1.0000  &  1.0000  &  1.0000  &  1.0000  &  1.0000\\
        $\pm 90^{\circ}$ & 1.0000  &  1.0000  &  1.0000  &  1.0000  &  1.0000  &  1.0000  &  1.0000  &  1.0000\\
        Global           & 0.9698  &  0.9966  &  0.9342  &  0.5894  &  0.9771  &  0.9955  &  0.9885  &  0.9547\\
        \hline
    \end{tabular}
    \caption{Archetypes correlation values corresponding consecutive Pacific regions with increasing range of latitudes for matched labels given in Table S\ref{tab:match3}.}
    \label{tab:match4}
\end{table}

The justification for choosing the recent satellite 1982-2022 period for our analysis is mainly driven by the fact that gridded SST reconstructions\footnote{We differentiate here between observations and reconstructions, as reconstructions are at times conflated with observations. Reconstructions are not observations and are only an approximation of the true historical dSSTAs.} for periods prior the advent of satellite observations suffer from a raft of problems reviewed for example by \cite{ken19:fms}, \cite{ken14:rog} and \cite{ken19:jgra}. These correspond to unresolved instrumentation biases and incomplete spatiotemporal observation coverage leading to inhomogeneous spatio-temporal climatology, trend and variance characterisation as illustrated for the Earth's surface temperature and sea level reconstructions in \cite{chr2009:jcli} and \cite{chr2010:jcli}, respectively. Some reconstruction methods using principal components or EOFs patterns derived from the recent and well-sampled SST data to reconstruct the sparsely observed historical records suffer potentially from the more pernicious, if not insurmountable, problem of assuming that the SST variability across time has not changed from the historical to the more recent period, assuming `de facto' covariance stationarity. To date, the authors are not aware of any publication where the effects of insufficient versus adequate spatio-temporal sampling on reconstructions have been rigorously assessed. One may note that this pernicious problem may affect both ocean and atmospheric reanalyses as gridded global SST records are generally used as boundary conditions or forcing fields for both. Historical reconstructions are nevertheless useful in providing `qualitative' assessment of large spatial- and long time-scale variability of the climate system with appropriate caveats. In contrast to principal components analysis on which some ENSO phase characterisations are based, clustering algorithms in general and AA in particular have the potential to reveal dataset anomalies, either real - due to regime shifts - or spurious - due to change in the observing platforms, for example, as these algorithms try to match individual data records across time.

To illustrate these concerns, we report hereafter our analysis for COBE-SST \citep{ish05:ijc} and ERSST version 5 \citep{hua17:jcli} dSSTAs over the 1900 to 2022 period for cardinalities 4 and 8. At each grid point, we use a quadratic trend to fit and approximately remove the non-linear trend in anthropogenic warming mainly observed in the second half of the record. We note, however, that using a linear trend instead does not substantially change the results.

\begin{figure}[htbp]
  \centering
  \noindent
  \begin{subfigure}[b]{0.49\textwidth}
    \centering  
    \includegraphics[width=\textwidth]{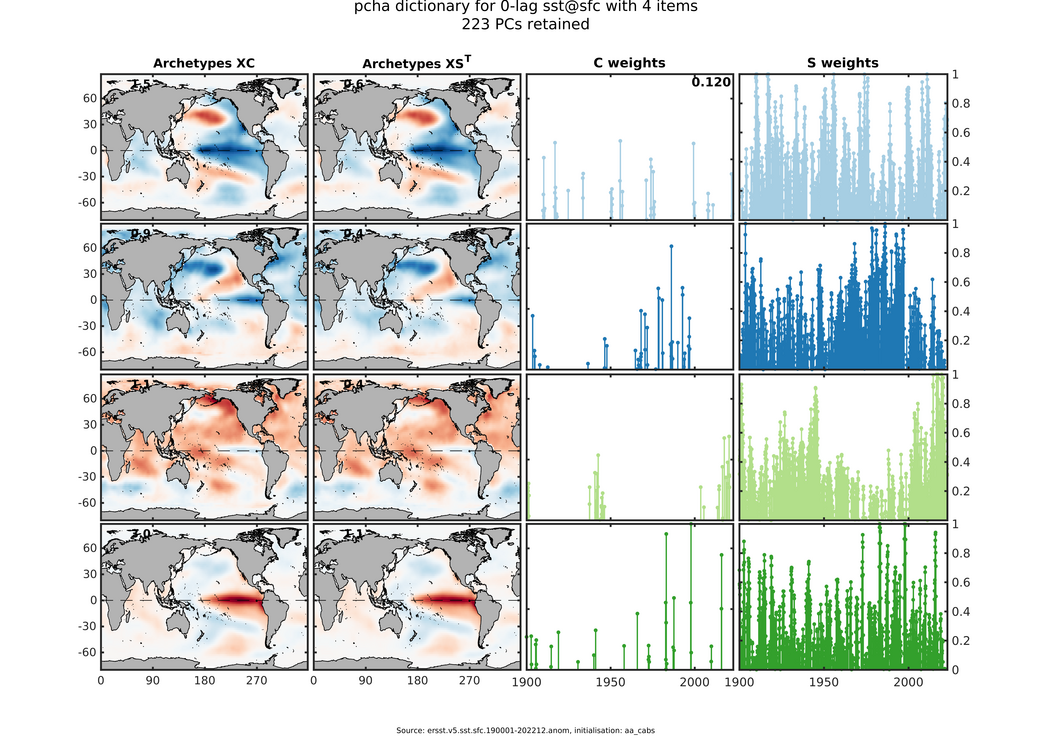}\\
  \end{subfigure}
  \begin{subfigure}[b]{0.49\textwidth}
    \centering  
    \includegraphics[width=\textwidth]{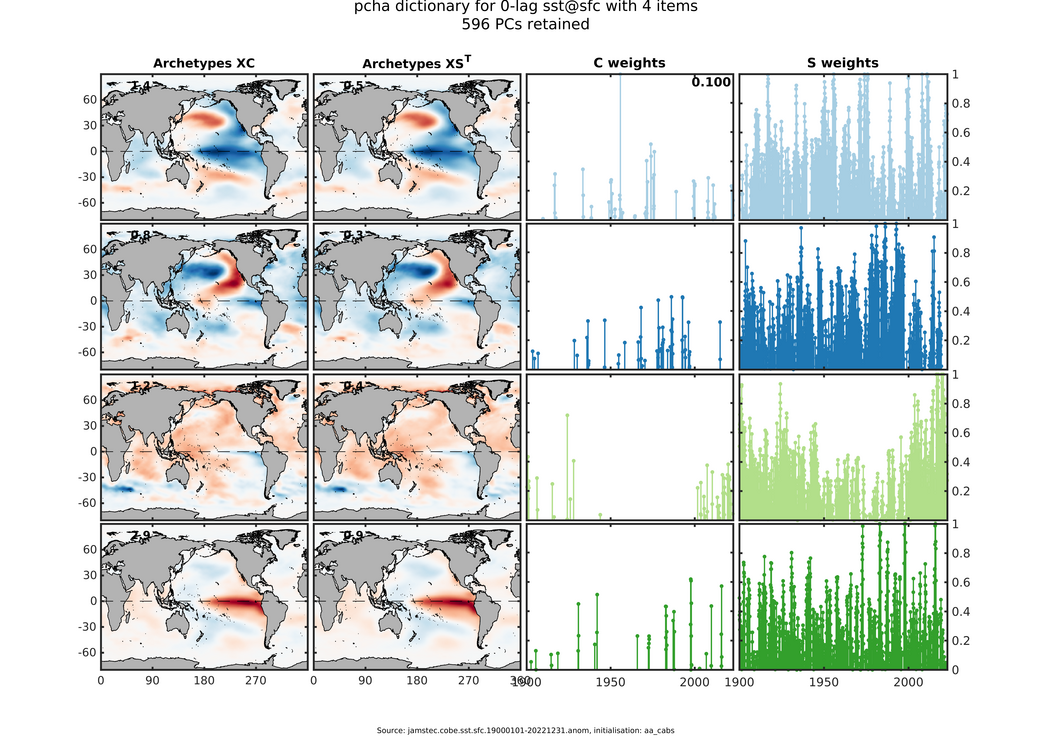}\\
  \end{subfigure}
\caption{AA spatial pattern and time series results using ERSST (left) and COBE (right) quadratically detrended monthly SST anomalies over 1900-2022 for cardinality 4. The two left columns on each subplot show archetypes and composites constructed by $\textbf{XC}$ and $\textbf{X}\Tilde{\textbf{S}}^T$, with $\Tilde{\textbf{S}}_{p\times t}=\textbf{S}_{p\times t}/\sum^{t}_{i=1} S(p,i)$, respectively. The two right columns show $C$ and $S$ matrix time series, respectively, with SOI time series (grey) included on both.}
\label{fig:aa4}
\end{figure}

\begin{figure}[htbp]
  \centering
  \noindent
  \begin{subfigure}[b]{0.49\textwidth}
    \centering  
    \includegraphics[width=\textwidth]{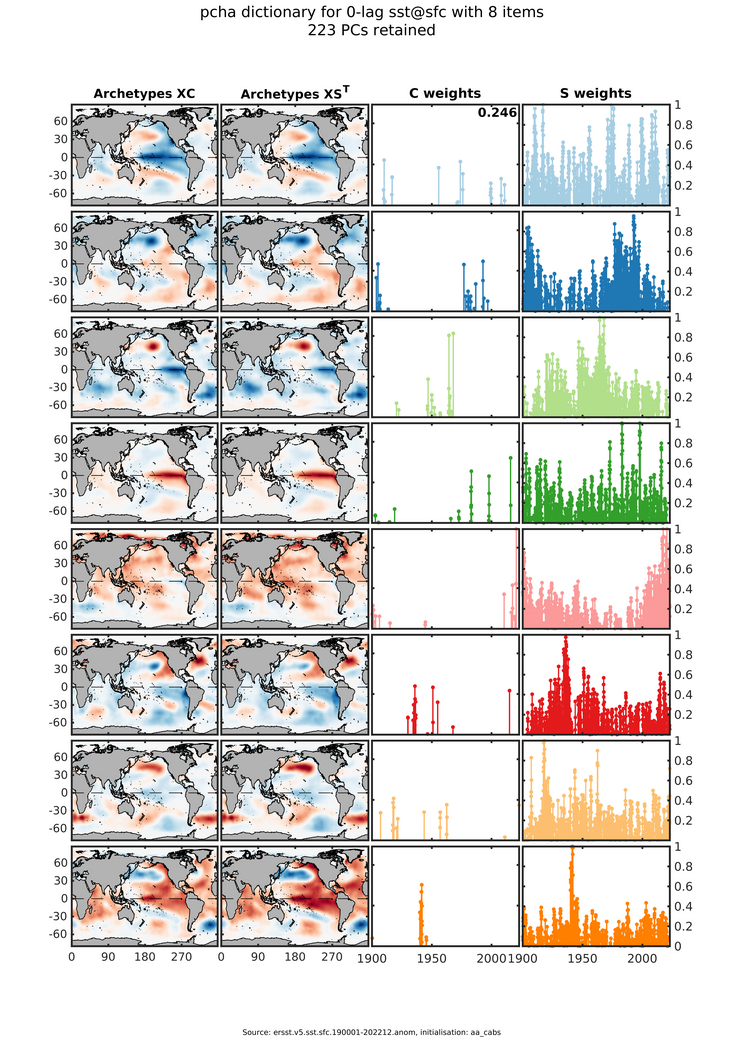}\\
  \end{subfigure}
  \begin{subfigure}[b]{0.49\textwidth}
    \centering  
    \includegraphics[width=\textwidth]{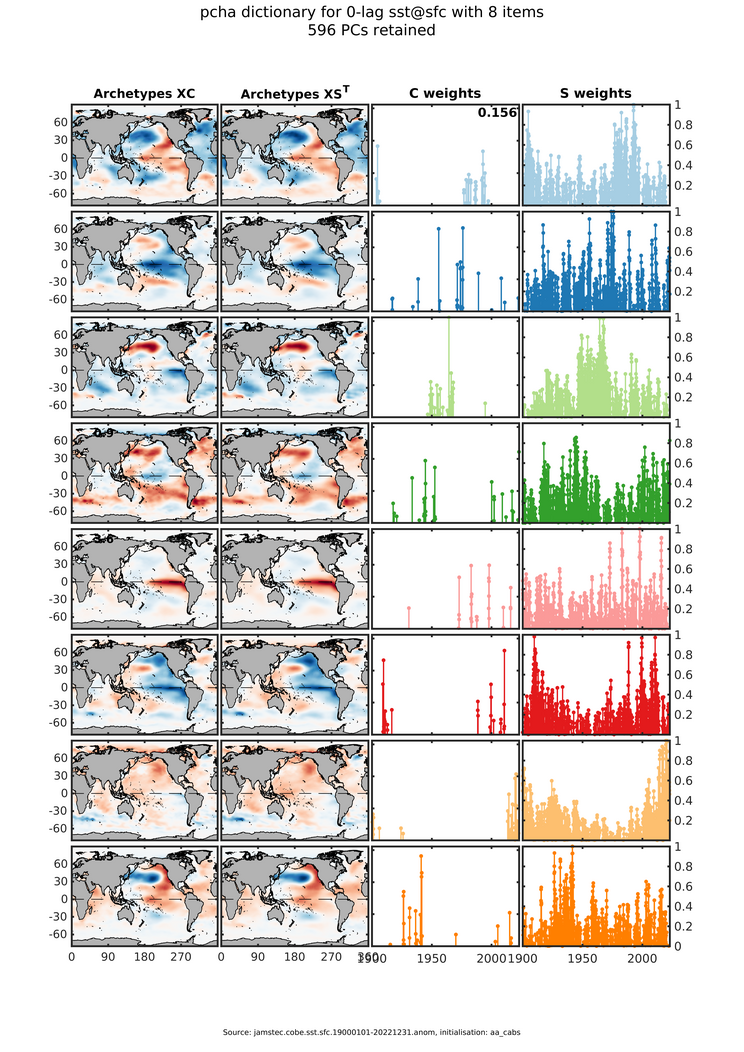}\\
  \end{subfigure}
\caption{Same as in Figure \ref{fig:aa4}, but for cardinality 8. The archetype labels (rows) are based on the time mean of the AA stochastic matrix $\textbf{S}$ - the mean probability of expression of the archetypes across the whole period - ranked in decreasing order. The time means are expressed as $\overline{S}_{i}=(\sum^{T}_{j=1} S_{ij})/T$ for archetypes $i=1,\ldots,8$ and $j=1,\ldots,T$ monthly records and vary across reconstructions.}
\label{fig:aa8}  
\end{figure}

By comparing Figures \ref{fig:aa4} and \ref{fig:aa8}, we first note that the `nestedness' property of AA for these historical reconstruction dSSTAs holds for each individual reconstruction -- the AA nestedness being the property of patterns being uncovered for a lower cardinality can be matched to those for a higher cardinality. The archetypal patterns can be also approximately matched across reconstructions for cardinality 4.

The matching across reconstructions for cardinality 8 is more challenging, probably the result of the different methodologies used to in-fill data poor regions. However, a more concerning aspect of these results can be seen in the expression of the $\textbf{S}$ weights, the last columns of each subplots in Figures \ref{fig:aa4} and \ref{fig:aa8}, which correspond to the probability of expression dSSTA archetypes in each and every data records. In both reconstructions, the time evolution of $S_{pt}$, for any given $p=1,\ldots,4$ or $8$, displays elevated or reduced values over multiple years, at time lasting decades, for no apparent reason. For example on Figure \ref{fig:aa8}, ERSST archetype 2 (left, 2nd row) and the corresponding COBE archetype 1 (right, 1st row) display high probabilities of expression from 1975 to 2000. These are not replicated in AA results stemming from OISST shown in Figure \ref{fig:global}. The behaviour is apparent in both extended reconstructions for multiple intervals and archetypes and possibly indicates that the paucity of observations in the earlier part of the record leads to poorly constrained the gridded datasets. Our cautionary approach in not including these long reconstructions for a quantitative study is therefore warranted.

\begin{figure}[htbp]
        \centering
            \begin{subfigure}[t]{0.32\textwidth}
            \includegraphics[width=\textwidth]{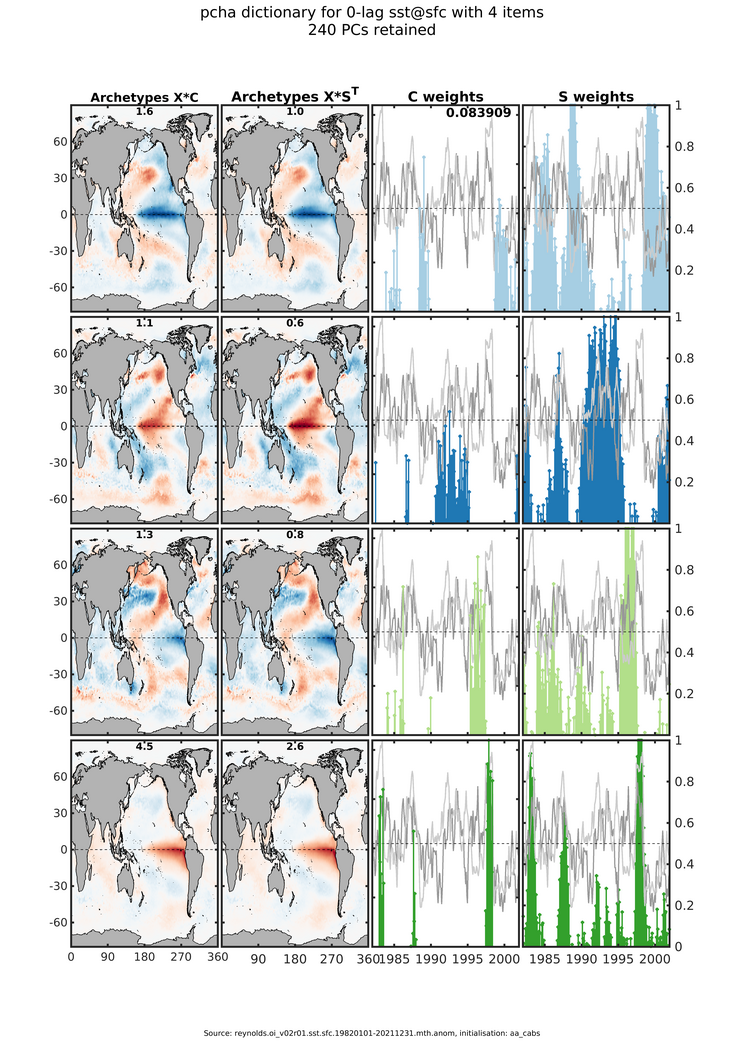}
            \label{fig:full_ssta_aa}
            \caption{1982-2001}
            \end{subfigure}
            \begin{subfigure}[t]{0.32\textwidth}
            \includegraphics[width=\textwidth]{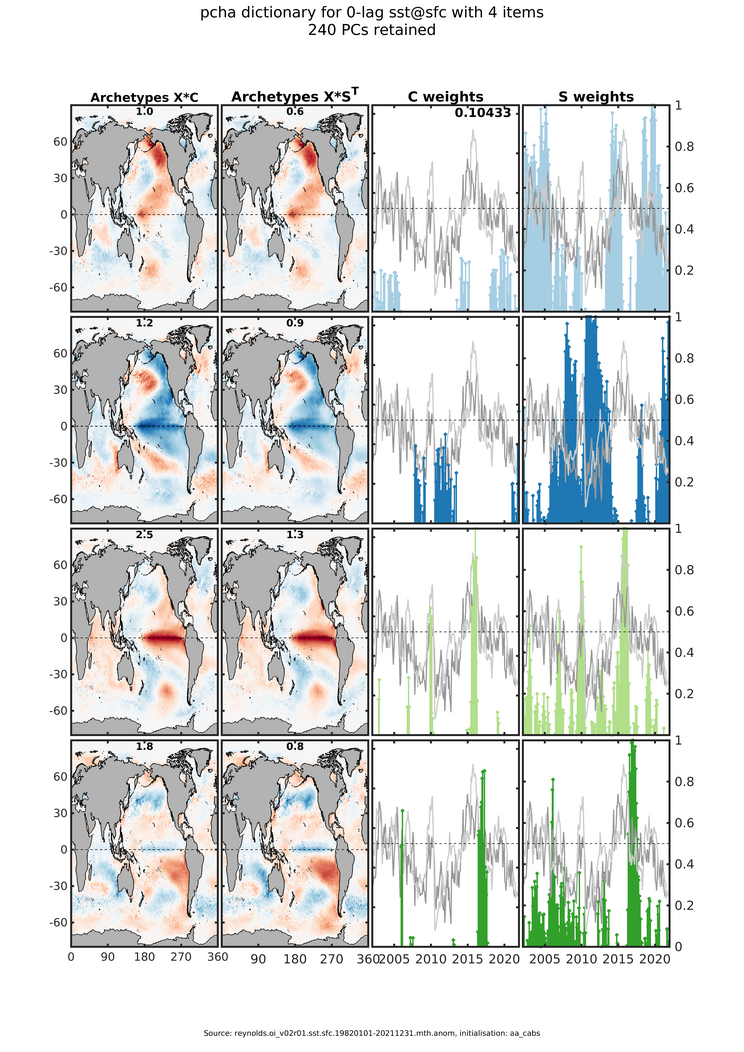}
            \label{fig:dtd_ssta_aa}
            \caption{2002-2021}
            \end{subfigure}
            \begin{subfigure}[t]{0.32\textwidth}
            \includegraphics[width=\textwidth]{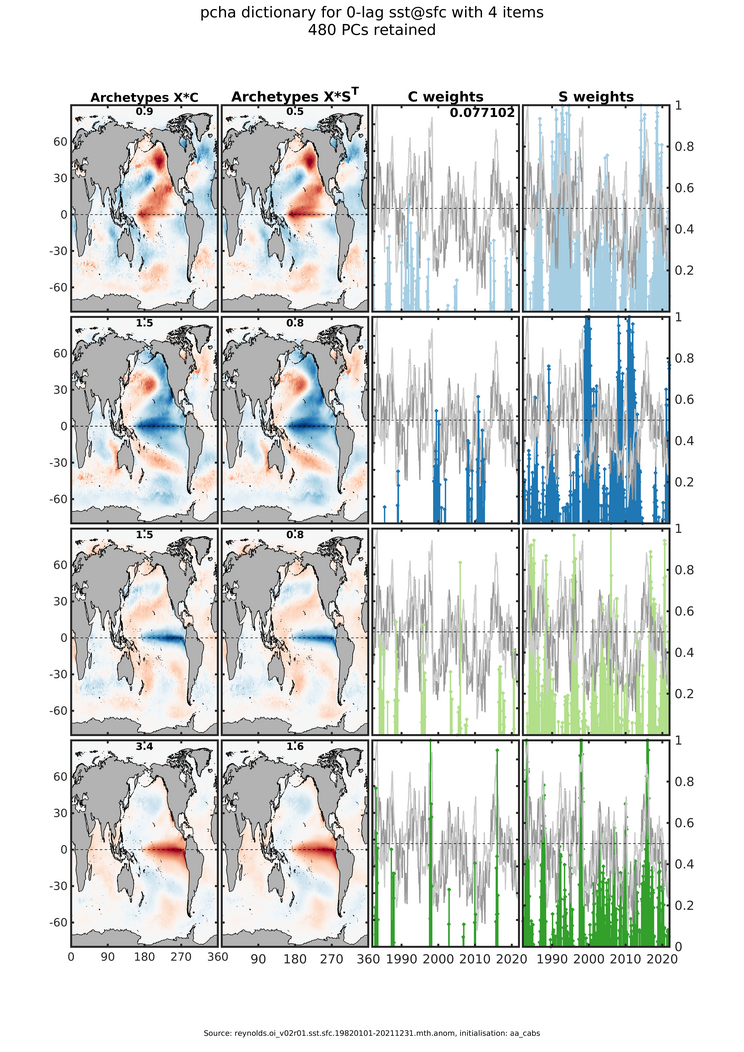}
            \label{fig:full_ssta_aa}
            \caption{1982-2021}
            \end{subfigure}
        \caption{Comparison of archetypes and associated affiliation sequences for cardinality 4 over a) 1982-2001, b) 2002-2021 and c) 1982-2021 periods. For each period, the archetypes have been ranked from the most to the least probable as measured by the mean probability of expression of a given archetype over time as in Figures \ref{fig:aa4} and \ref{fig:aa8}.}
        \label{fig:period}
\end{figure}

Finally, to answer concerns regarding the robustness of AA results with respect to the period chosen for the analysis, Figure \ref{fig:period} compares archetypes and associated affiliation sequences for cardinality 4 over a) 1982-2001, b) 2002-2021 and c) 1982-2021 periods. The AA results show reasonable agreement across periods. For each period, the archetypes have been ranked from the most probable to least probable as measured by the mean probability of expression of a given archetype over time. As one does not expect the relative frequency of expression to stay same across the 3 periods, the archetype ranking in Figure \ref{fig:period} from the most to the least probable has changed.

Although reassuring, it is important to realise that these results cannot and should not be necessarily expected \textsl{a priori}, unless one assumes that the SST expression of the ENSO phenomenon does not vary over time (imputed stationarity), a fact not supported by observations and originally reported by \cite{wyr75:jpo}.

\section{The choice of cardinality and AA nestedness}
The problem of cardinality in clustering algorithms is a common issue as discussed in \cite{chr07:jcli}. How many clusters are necessary to capture the `essence' of any given process? This question is similar to how many EOF/PC pairs are needed to capture the variability and associated phases of a phenomenon. For the later, researchers focus typically on significant gaps between eigenvalues (or significant jump in the explained variance of the principal component analysis modes) of the spectral characteristics of representative geophysical fields. For the former, they may rely on metrics such as the gap statistics introduced by \cite{tib01:jrss}, applied to both K-means and hierarchical clustering in this seminal publication, which compares the change in within-cluster dispersion with that expected under an appropriate reference null distribution. 

For AA, the original publication of \cite{cut94:tec} did not introduce of a metric or parsimony criterion to estimate the appropriate number of clusters required for the characterisation of the dataset under investigation. However, our submission relies on the following features of the AA representation, these are 1) the archetype nestedness \citep{bla22:aies} , 2) the discrimination score \citep{ris21:mwr} and 3) the in-sample AA variance and associated signal-to-noise ratios introduced for the first time for both archetypes and AA composites in Section 4 of our submission.

\begin{landscape}
\begin{figure}[htbp]
\centering
    \begin{subfigure}[b]{0.65\textwidth}
        \includegraphics[height=0.85\textheight]{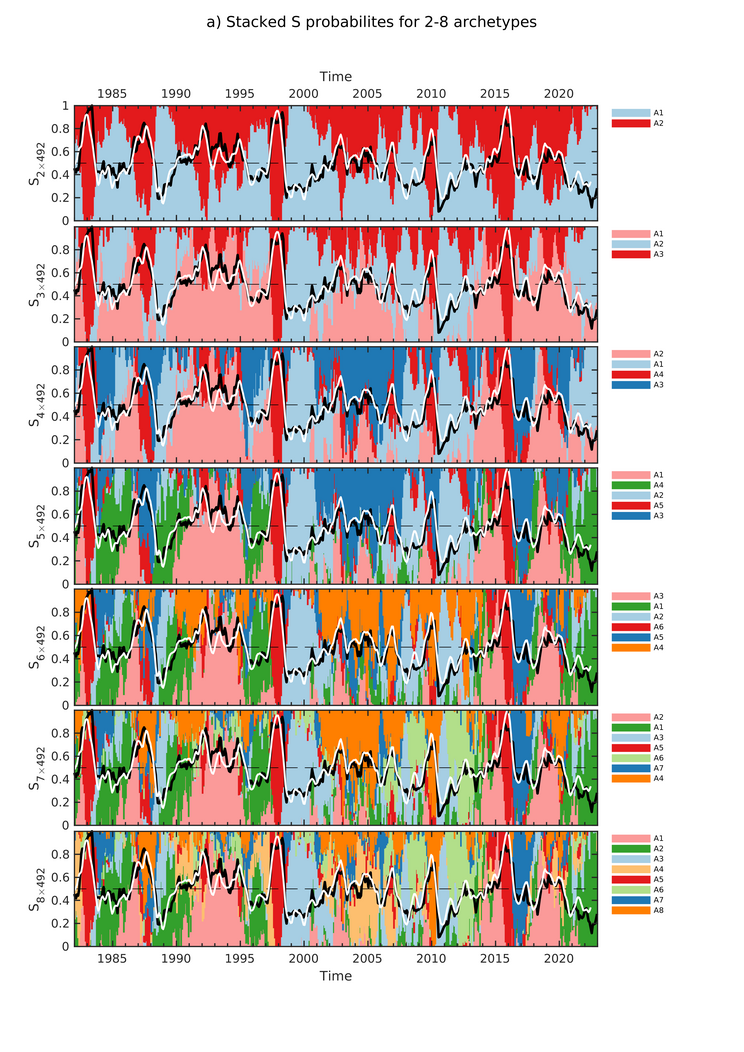}
    \end{subfigure}
    \begin{subfigure}[b]{0.65\textwidth}
        \includegraphics[height=0.85\textheight]{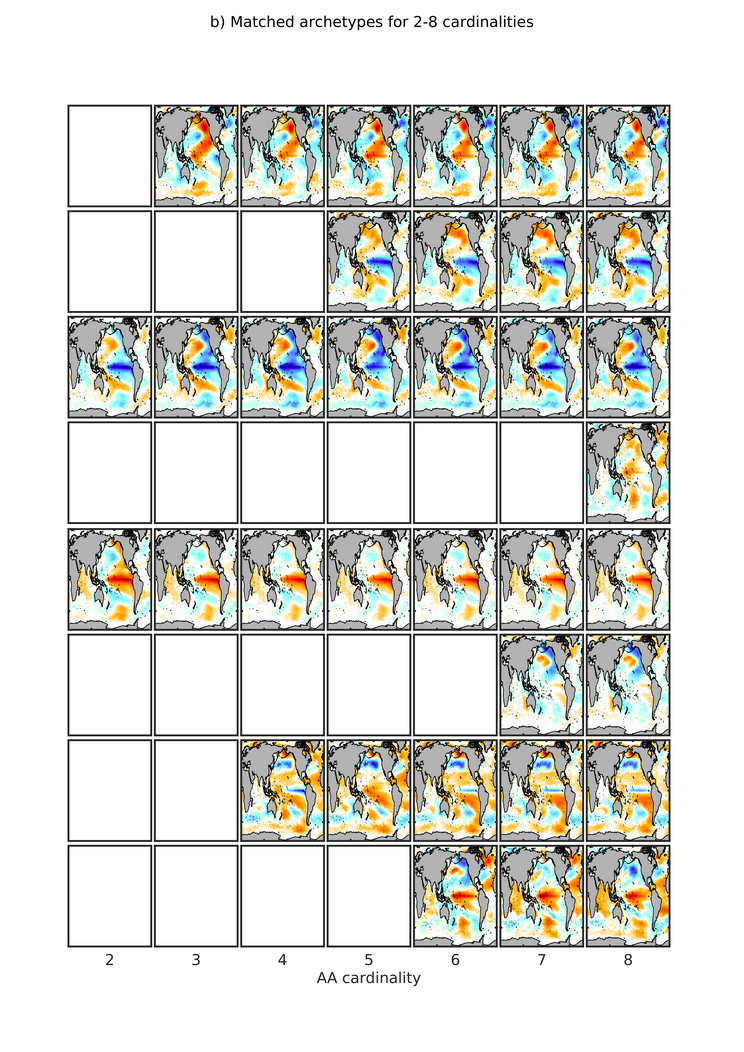}
    \end{subfigure}
    \caption{a) Stacked bar-plots of $S$-matrix probabilities for detrended monthly SST anomalies over 1982-2022 and b) corresponding matched archetypes using pattern correlation for cardinalities ranging from 2 to 8. Each row in a) corresponds to AA results ranging from 2 (top row) to 8 (bottom row), with MEI on ONI time series (black and white) included on all. The bar color codes correspond to matched archetypes referenced to AA results for a cardinality of 8, whereas the labels $A_{i}$ in each row indicate the archetype ranks based on the time mean of the AA stochastic matrix $S_{n_{AA}}$ with $n_{AA} = 2,\ldots,8$ and $i=1,\ldots,n_{AA}$ in decreasing order of $\overline{S}_{n_{AA}}(i)$. The 7 columns by 8 rows AA patterns in b) correspond to matched archetypes (rows) referenced to the AA results for cardinality of 8 (last column) across cardinalities ranging from 2 (first column) to 8 (last column).}
    \label{fig:dsst_nest}
\end{figure}
\end{landscape}

\begin{landscape}
\begin{figure}[htbp]
\centering
    \begin{subfigure}[b]{0.65\textwidth}
        \includegraphics[height=0.85\textheight]{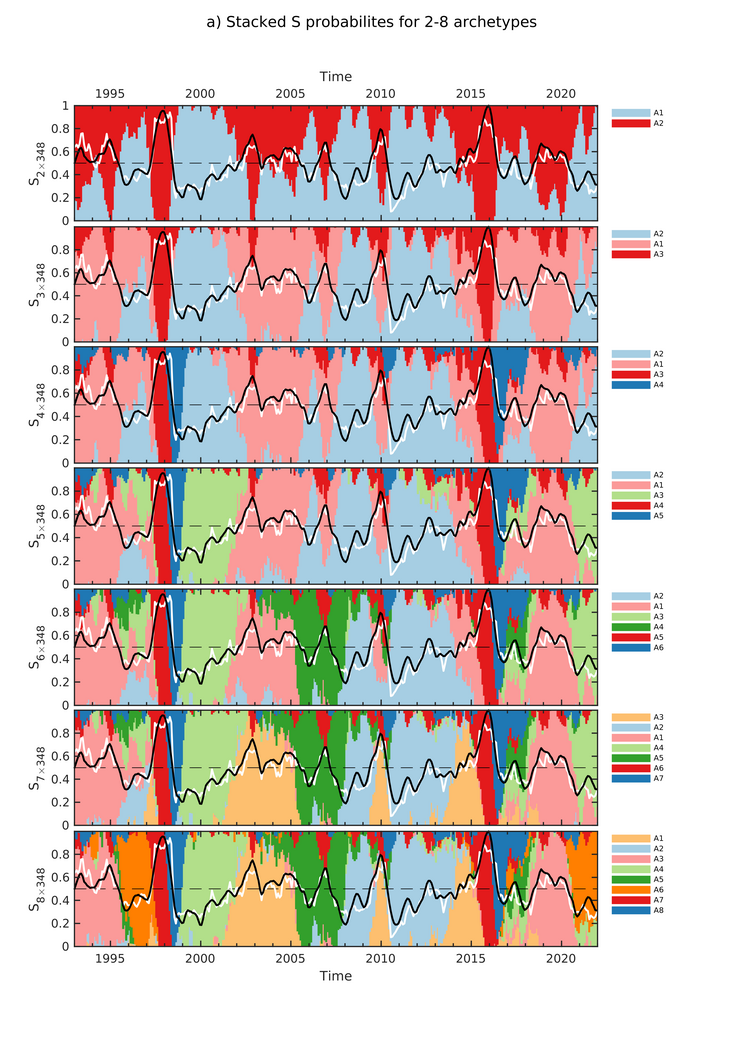}
    \end{subfigure}
    \begin{subfigure}[b]{0.65\textwidth}
        \includegraphics[height=0.85\textheight]{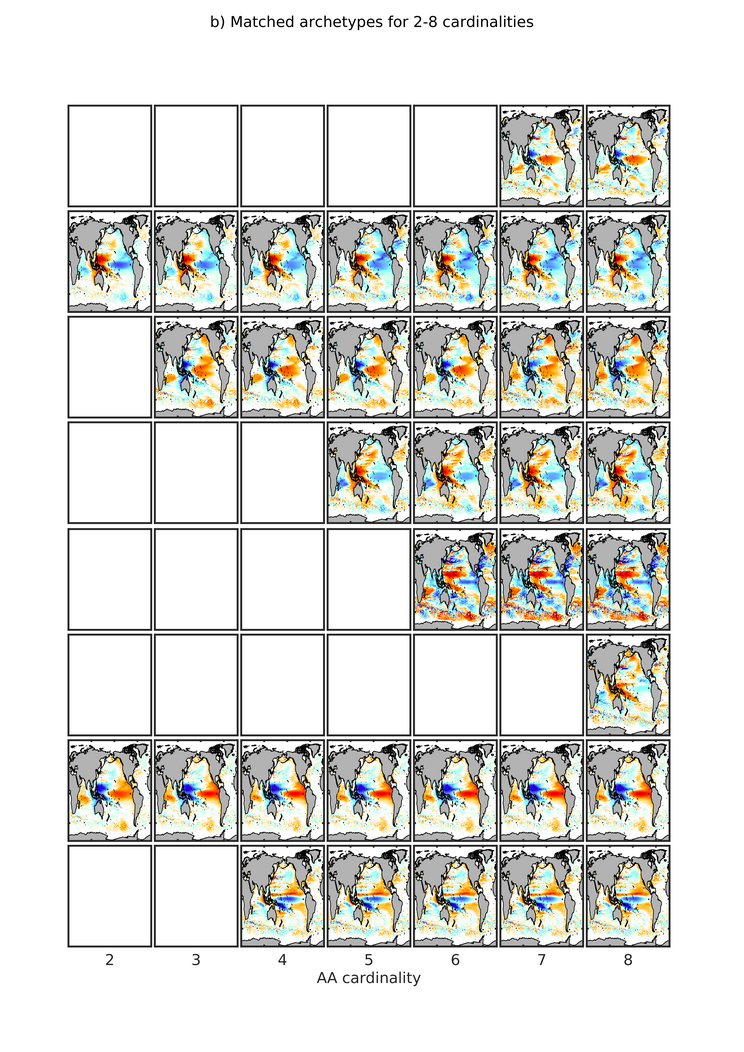}
    \end{subfigure}
    \caption{a) Stacked bar-plots of $S$-matrix probabilities for detrended monthly SLA anomalies over 1993-2021 and b) corresponding matched archetypes using pattern correlation for cardinalities ranging from 2 to 8. Each row in a) corresponds to AA results ranging from 2 (top row) to 8 (bottom row), with MEI on ONI time series (black and white) included on all. The bar color codes correspond to matched archetypes referenced to AA results for a cardinality of 8, whereas the labels $A_{i}$ in each row indicate the archetype ranks based on the time mean of the AA stochastic matrix $S_{n_{AA}}$ with $n_{AA} = 2,\ldots,8$ and $i=1,\ldots,n_{AA}$ in decreasing order of $\overline{S}_{n_{AA}}(i)$. The 7 columns by 8 rows AA patterns in b) correspond to matched archetypes (rows) referenced to the AA results for cardinality of 8 (last column) across cardinalities ranging from 2 (first column) to 8 (last column).}
    \label{fig:dsl_nest}
\end{figure}
\end{landscape}

Although unexpected \textsl{a priori} by the AA methodology as discussed by \cite{cut94:tec}, the archetypes approximately nest in Figures \ref{fig:dsst_nest} and  \ref{fig:dsl_nest}, a) and b), adapted from \cite{bla22:aies} Figure 9, where a) the stacked bar-plots of $S$-matrix probabilities for monthly dSSTAs over 1982-2022 and b) corresponding matched archetypes using pattern correlation are represented. Each row in Figures \ref{fig:dsst_nest} and \ref{fig:dsl_nest} a) corresponds to AA results for cardinalities ranging from 2 (top row) to 8 (bottom row), with MEI on ONI time series (black and white) included on all and scale to fit within the $\textbf{S}$-matrix affiliation bounds. The bar color codes correspond to matched archetypes referenced to AA results for a cardinality of 8, whereas the labels $A_{i}$ in each row indicate the archetype ranks based on the time mean of the AA stochastic matrix $S_{n_{AA}}$ with $n_{AA} = 2,\ldots,8$ and $i=1,\ldots,n_{AA}$ in decreasing order of $\overline{S}_{n_{AA}}(i)$. The 7 columns by 8 rows AA patterns in Figures \ref{fig:dsst_nest} and \ref{fig:dsl_nest} b) correspond to matched archetypes (rows) referenced to the AA results for cardinality of 8 (last column) across cardinalities ranging from 2 (first column) to 8 (last column).

Firstly, contrary to K-means and Fuzzy C-means, which do not nest as illustrated in Section 5 of the manuscript when comparing results for cardinalities 4 and 8, the archetypes reliably do. This `remarkable' AA property remains valid across geophysical realms as illustrated for example by our sea level anomaly results reported in Section 8. The phases or flavours identified for lower cardinality persist across a wide cardinality range as can be seen for dSSTAs and dSLAs in Figures \ref{fig:dsst_nest} and \ref{fig:dsl_nest}.

\begin{figure}[hbtp]
    \centering
    \includegraphics[width=1.0\textwidth]{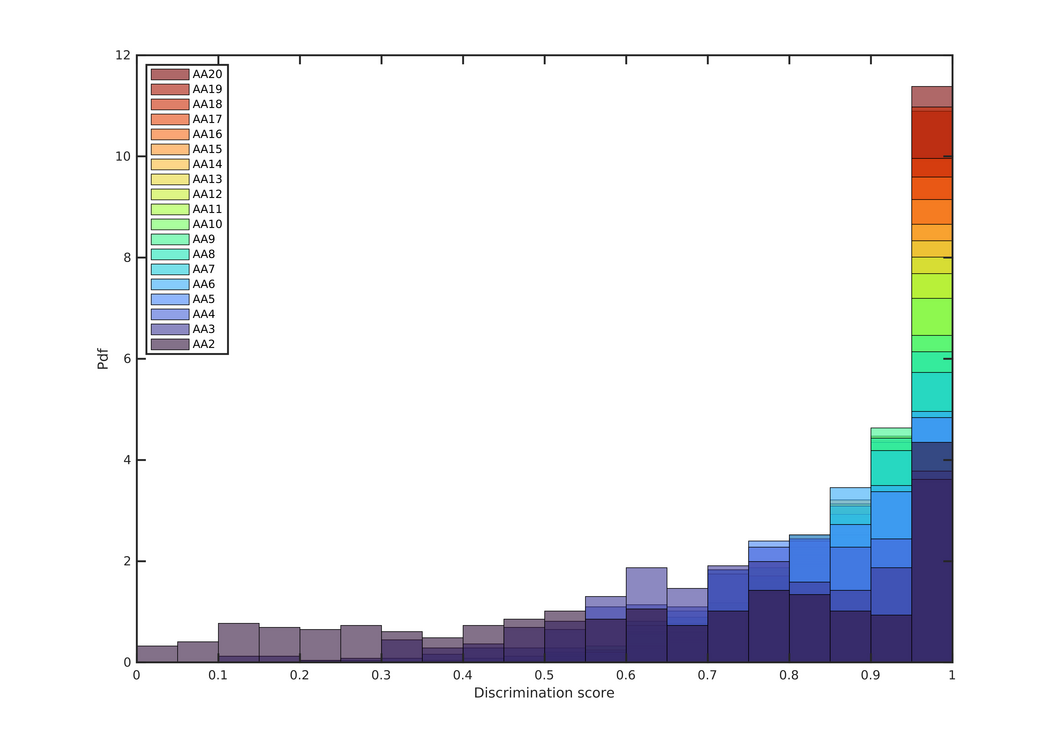}
    \caption{Discrimination score $\Delta^{p}_{i_{max}}(t)$ distribution as a function of archetype cardinality $p$, ranging from 2 to 20.}
\label{fig:dsst_disc}
\end{figure}

Secondly, the power of discrimination of AA between identified clusters is revealed by the discrimination score $\Delta^{p}_{i_{max}}(t)$ distribution as a function of archetype cardinality $p$, ranging from 2 to 20, reported in Figure \ref{fig:dsst_disc}: the higher the cardinality, the sharper the discrimination based the highest probability of expression of a given archetype, $i_{max}$, for a given snapshot of global dSSTA at time $t$. In other words, the archetypes and data snapshots resemblance increases with increasing cardinality.

Although we present the results for 4 archetypes which show similarities and differences with phases detected by traditional methods, we stress that, in order to adequately characterise the ENSO phenomenon and its diversity, higher cardinalities have to be invoked as these will help to further differentiate not only ENSO phases and their remote impacts, but also their onset and decay.

\section{Physical interpretation of AA results}
Based on 1) the `extremal' properties of AA, 2) the potentially wide latitudinal and longitudinal reach of ENSO teleconnections, 3) the AA nestedness and power of discrimination of the method, we posit that AA, especially our results for cardinality 8, provide sharper characterisation of both the tropical and global teleconnection of ENSO. The archetypal flavours of ENSO are closer to individual global dSSTA snapshots as they are `not' built on the expectation or the time-mean of dSSTA record squares as it is the case for principal component analysis where all dSSTA records are included with the same weight. This property has an immediate implication: the archetypes are dynamically more pertinent than time-mean statistics.

\begin{figure}[htbp]
    \centering
    \includegraphics[width=0.85\textwidth]{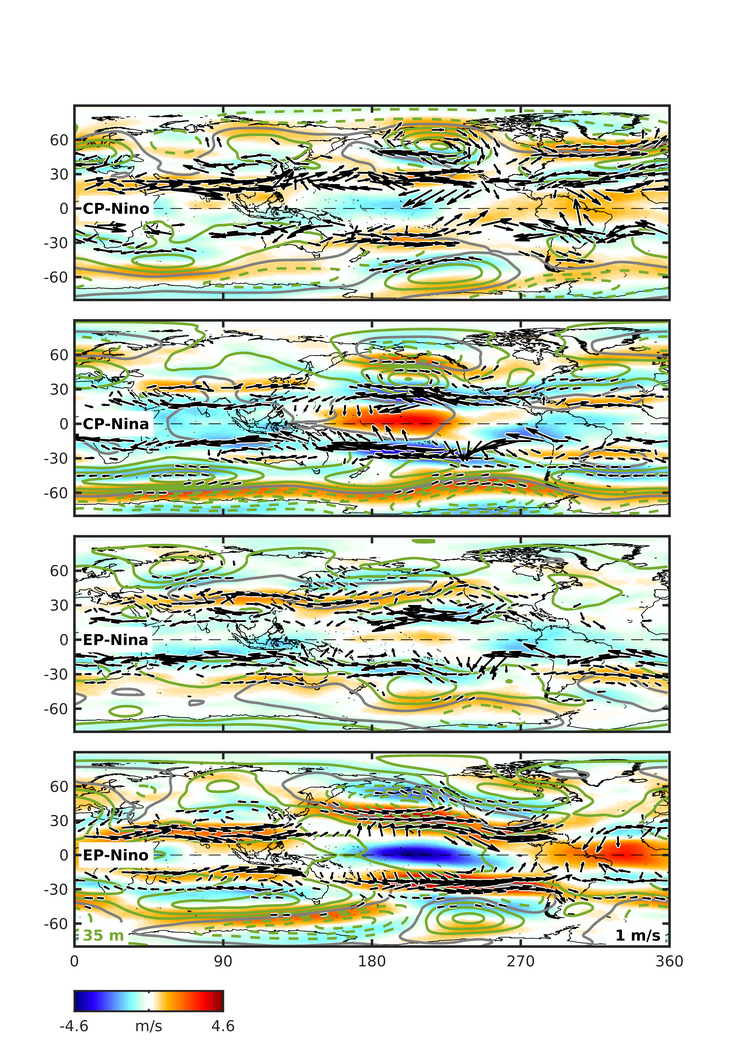}
    \caption{Adapted from \cite[Figure 10]{bla22:aies}, AA atmospheric field composites constructed from AA results of detrended monthly SST anomalies over 1982-2022 for cardinality 4 for the 300 hPa zonal wind component (shading) with superimposed 500 hPa geopotential height anomalies (green and grey contours) and thermal wind anomalies components (vectors). The corresponding ENSO flavours are reported in bold on the Equator on the left of each panel. The maximum absolute anomaly amplitudes for geopotential and thermal wind are reported on the bottom right and left of the last panel. The panels are ranked from the highest (top, more frequent) and to lowest (bottom, less frequent) mean probability of expression computed over the 1982-2022 monthly records $T$ using the affiliation sequence $\overline{S}_{p} = \frac{1}{T}\sum^{T}_{t=1} S_{pt},\, p=1,\ldots,4$.}
    \label{fig:aies10}
\end{figure}

\begin{figure}[htbp]
  \centering
  \noindent
  \begin{subfigure}[b]{0.45\textwidth}
    \centering  
    \includegraphics[width=\textwidth]{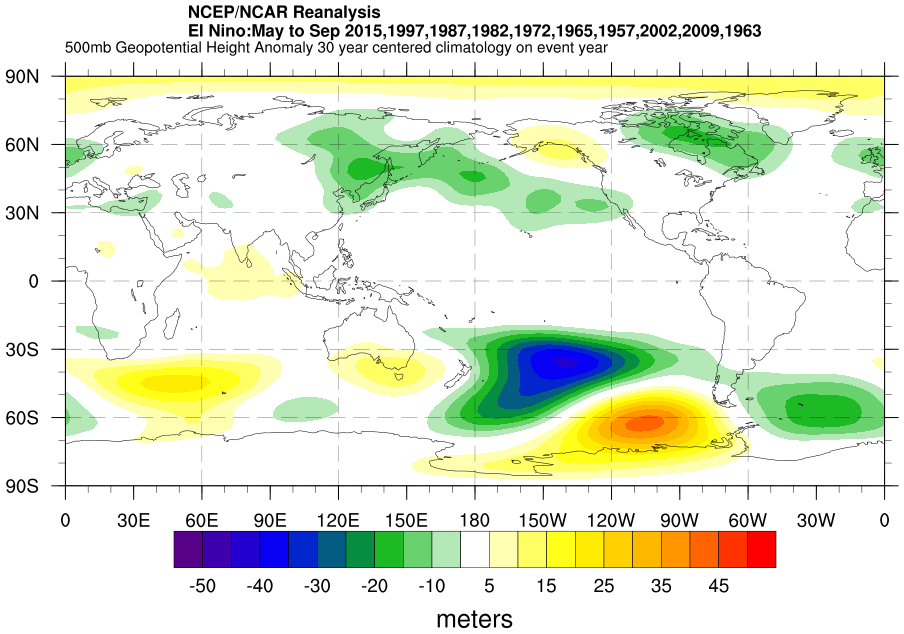}\\
  \end{subfigure}
  \begin{subfigure}[b]{0.45\textwidth}
    \centering  
    \includegraphics[width=\textwidth]{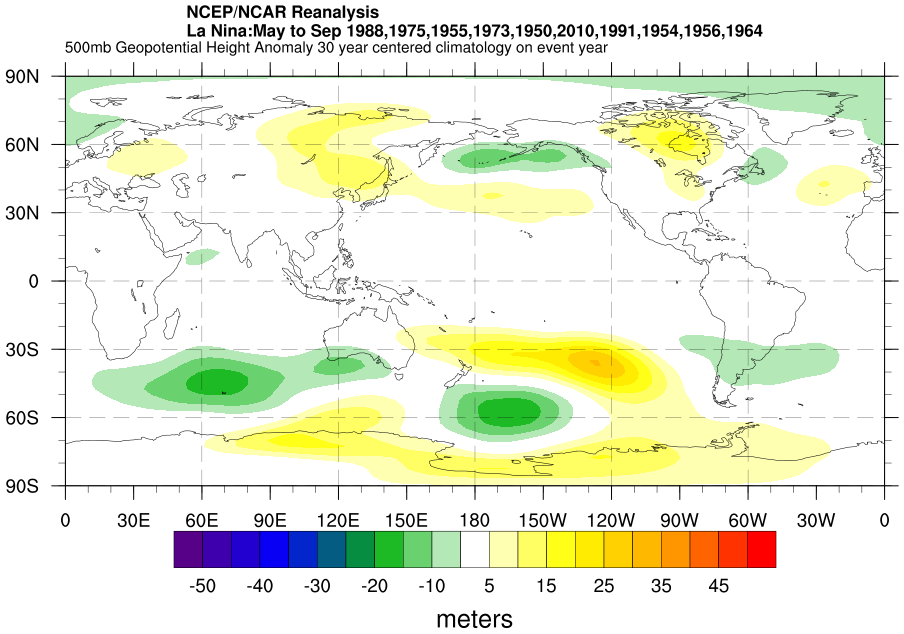}\\
  \end{subfigure}
  \centering
  \noindent
  \begin{subfigure}[b]{0.45\textwidth}
    \centering  
    \includegraphics[width=\textwidth]{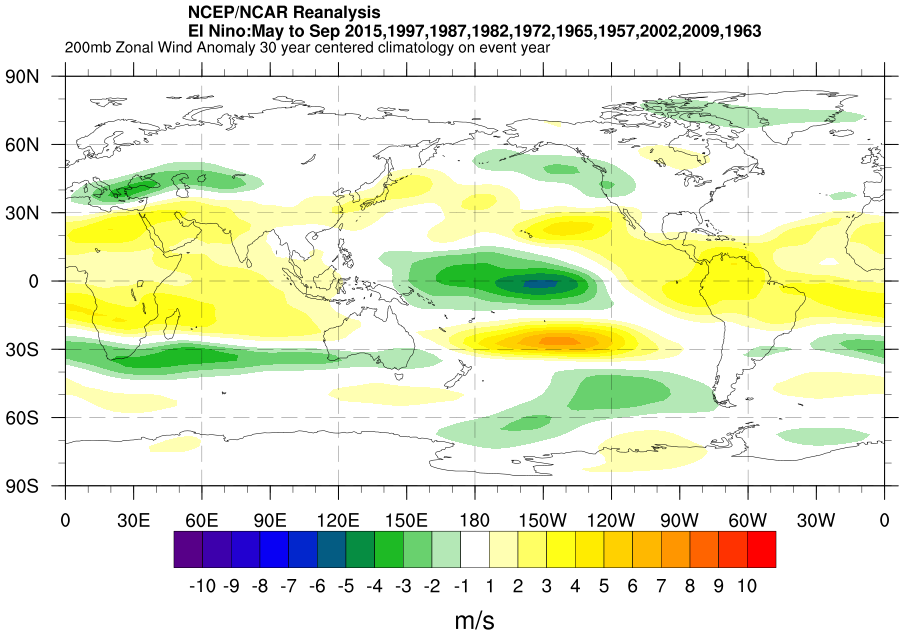}\\
  \end{subfigure}
  \begin{subfigure}[b]{0.45\textwidth}
    \centering  
    \includegraphics[width=\textwidth]{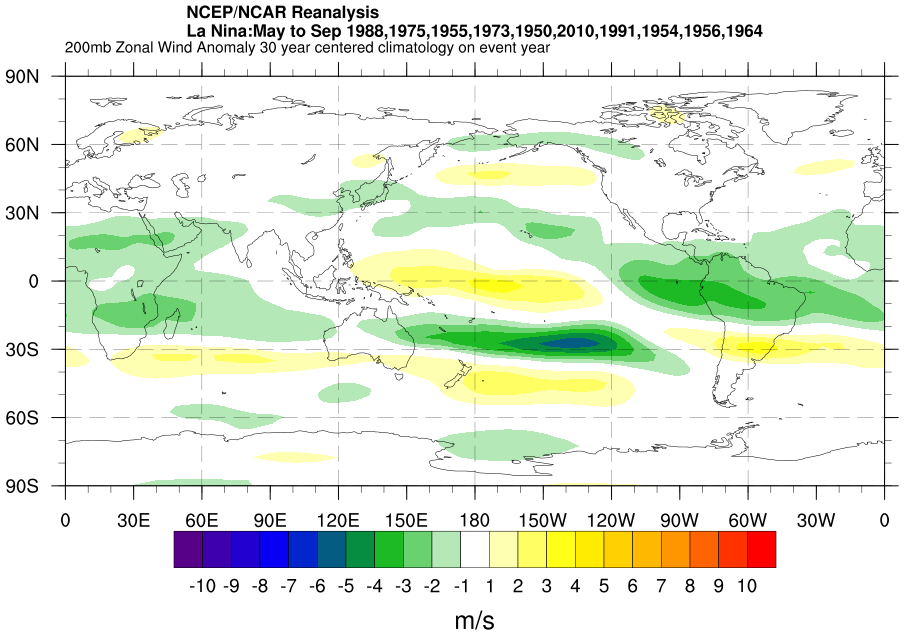}\\
  \end{subfigure}
  \caption{Geopotential anomalies and zonal wind anomalies at 500~hPa (top) and 200~hPa (bottom) Austral Winter composites associated with 10 El Ni\~no (left) and 10 La Ni\~na (right) events from 1948 to the present. These figures are provided by the NOAA Physical Sciences Laboratory, Boulder, Colorado, USA.}
\label{fig:noaa}
\end{figure}

Regarding the physical interpretation of AA results, the paper presents lag-0 composites illustrating the concurrent impact of ENSO AA flavour on rainfall, atmospheric convection and ocean subsurface fields. We argue that these composites, conditioned on the archetypes probability of expression in each snapshots, are dynamically more consistent than correlations or linear regressions (time-mean metrics) patterns derived from dSSTA principal components or ENSO indices. We show in Figure \ref{fig:aies10} updated from \cite[Figure 10]{bla22:aies}, that AA composites based on global dSSTAs for cardinality 4 recovers the ENSO Response Comparison Plots computed on-line from NOAA\footnote{\url{https://psl.noaa.gov/enso/compare.}}, reproduced in Figure \ref{fig:noaa}, for geopotential anomalies and zonal wind anomalies at 500~hPa and 200~hPa Austral Winter composites for selected El Ni\~no and La Ni\~na events from 1948 to the present. We note that the imprint of ENSO is global for both geopotential and zonal wind anomalies, reaching 80$^{\circ}$ in latitude, especially in the South Pacific sector. In this submission, the ENSO event selection is directly based on the AA factorisation results alone - the stochastic matrices $\textbf{C}$ or $\textbf{S}$ - the probabilities of expression of individual dSSTA records in dSSTA archetypes or the probabilities of expression of individual dSSTA archetypes in dSSTA records. Understanding this duality is important. Put in other words, both archetypes and (approximated) data records can be understood probabilistically.

For high temporal resolution, from 1 to 1/5 days, \cite{bac19:jcli} (Fig.~8) show that the local Atmosphere-Ocean predictability, based on a spectral (Granger) causality argument, is mainly driven by atmospheric processes apart from a narrow tropical band, where the Ocean drives the Atmosphere. However, for frequencies lower than 1/month, the regions where the Ocean (SST in the \cite{bac19:jcli} study) drives predictability expand to the sub-tropics and higher latitudes. These findings are based on local interactions between Atmosphere and Ocean only. The picture is likely to change when remote teleconnections are taken into account. A multivariate approach to assess non-local interactions between Ocean and Atmosphere is challenging. Lagged correlation and linear response theoretical arguments have been put forward to explain this relationship \citep{cap17:GRL, peg20:cd, cap21:scirep}. Lagged-regression, -correlation or -covariance analyses are often used to assess causality in weather and climate studies to investigate relationships between variables. However, these methods often suffer from drawbacks where variables auto-correlation and their non-linear interactions have a confounding effect as discussed recently by \cite{mcg18:jcli}. They advocate for the implementation of dedicated causality-probing methods \citep{gra69:eco, bar14:jnm} as in \cite{bac19:jcli}. Also, these approaches are mainly based on time-mean (expectation) statistics and are difficult to compare directly with AA results which are not. As consequence of the monthly temporal resolution used in this study and the potential causal and non-local relationships between tropical, extra-tropical and high-latitude SSTs, enlarging our domain of study can be justified as this choice can help to link the ENSO tropical signature - its location, extent and intensity - to concurrent extra-tropical conditions and to inform on its potential remote impacts.

Although enormous progress has been made in the understanding of ENSO, a detailed and comprehensive theory of ENSO in all its facets is still lacking and, given its global imprint and impact on the weather and climate, remains an active area of research. Our submission does not claim to add to the theoretical understanding of the phenomenon, but proposes an alternative, informative, and possibly `disruptive' \citep{par23:nat} way to look at both its tropical and global impacts. The results presented recover important characteristics of ENSO such as its irregularity, the asymmetry between opposite phases in terms of amplitudes, patterns and temporal evolution as recently reviewed by \cite{an21:book}.

\newpage
\section{AA results and composites for cardinality 4}\label{apx:aa4}
\begin{figure}[h]
    \centering
    \includegraphics[width=1.0\textwidth]{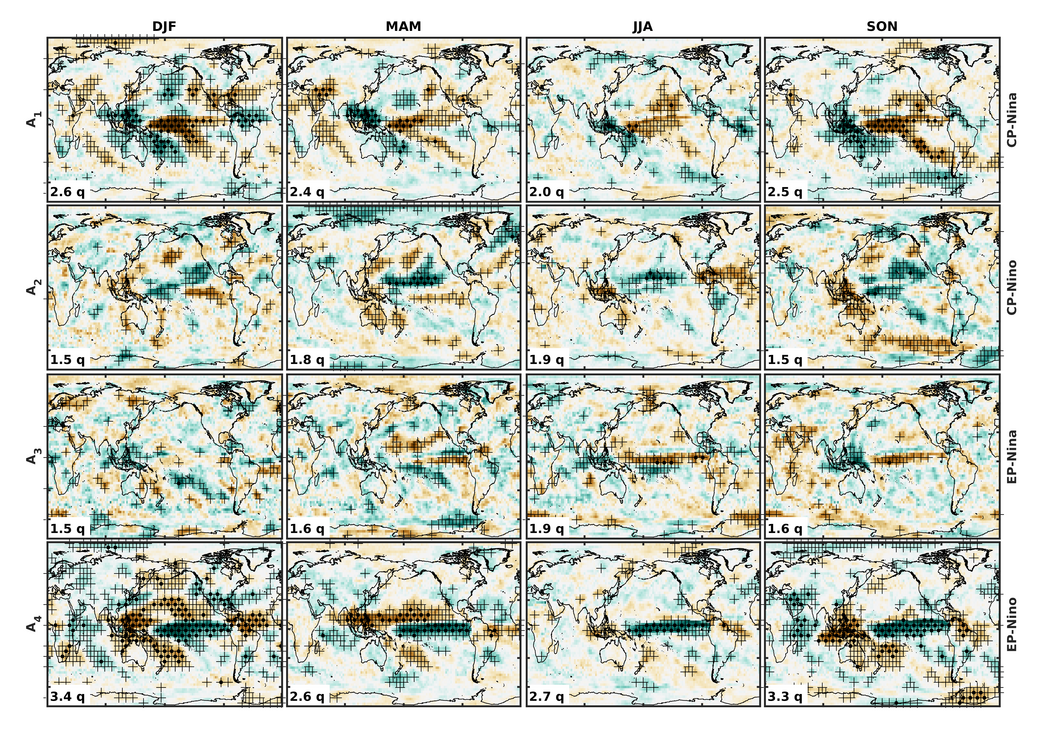}
    \caption{Complement Fig.~9 for cardinality 4, with centered rainfall undeciles composites and signal-to-noise ratios ($\textbf{+} > 0.15, \bullet > 0.3$, $(F^{2}_{S}/C^{S}_{F})^{1/2}$) based on time averages of the $\textbf{S}$-affiliation matrix over the 1982-2022 period for a cardinality of 4. As each individual subplot is scaled to increase contrast across archetypal patterns (rows), the boldfaced numbers (left bottom corner) label the maximum absolute SSTA for each archetypal pattern $\textbf{A}_{i}, i = 1,...,4$ and seasons (DJF, MAM, JJA and SON).}
    \label{fig:aa4snr_gpcp}
\end{figure}

\begin{figure}[h]
    \centering
    \includegraphics[width=1.0\textwidth]{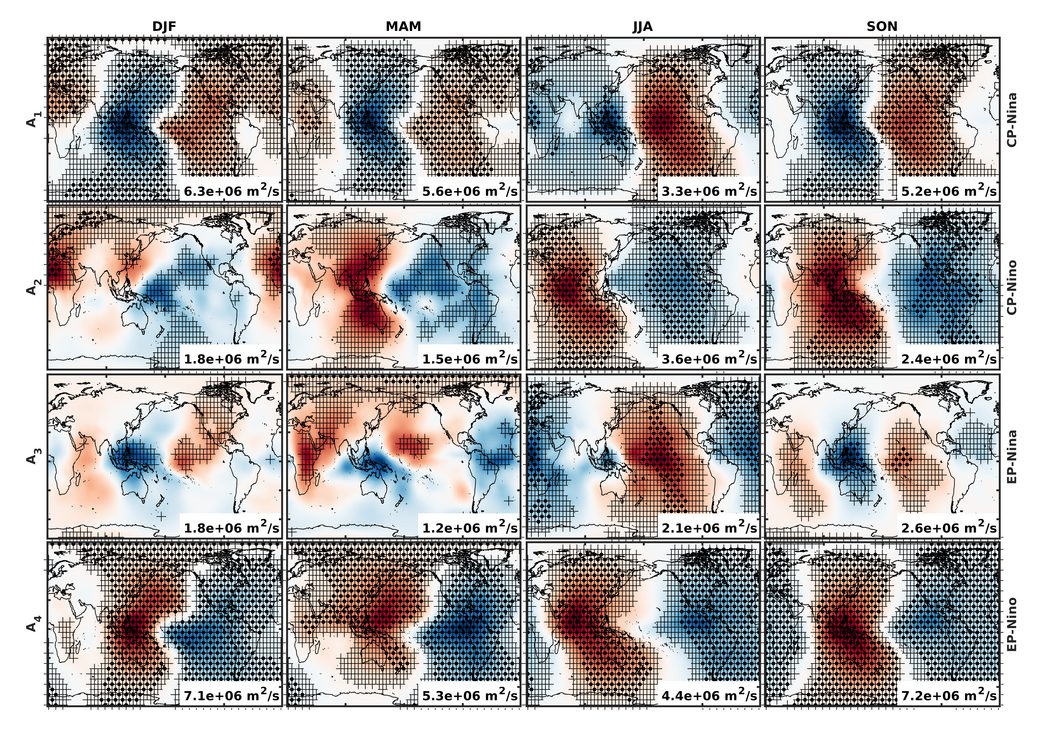}
    \caption{Complement Fig.~9 for cardinality 4, with $\Delta\chi_{150-850}$ anomalies composites and signal-to-noise ratios ($\textbf{+} > 0.15, \bullet > 0.3$, $(F^{2}_{S}/C^{S}_{F})^{1/2}$) based on time averages of the $\textbf{S}$-affiliation matrix over the 1982-2022 period for a cardinality of 4. As each individual subplot is scaled to increase contrast across archetypal patterns (rows), the boldfaced numbers (right bottom corner) label the maximum absolute SSTA for each archetypal pattern $\textbf{A}_{i}, i = 1,...,4$ and seasons (DJF, MAM, JJA and SON)}
    \label{fig:aa4snr_vpot}
\end{figure}

Figures \ref{fig:aa4snr_gpcp} and \ref{fig:aa4snr_vpot} show teleconnections composites corresponding to global precipitation and vertical convection proxy fields with corresponding signal-to-noise ratios as a measure of significance superimposed computed from $(F^{2}_{S}/C^{S}_{F})^{1/2}$ based on time averages of the $\textbf{S}$-affiliation matrix, Equations 3 and 4 of the main submission.
%%  M-file
%   JCLI_fig14S.m for GPCP
%
%%  M-file
%   JCLI_fig14S.m for VPOT
%

\section{AA results and composites for cardinality 8}\label{apx:aa8}
\begin{figure}[htbp]
    \centering
    \includegraphics[height=1.0\textheight]{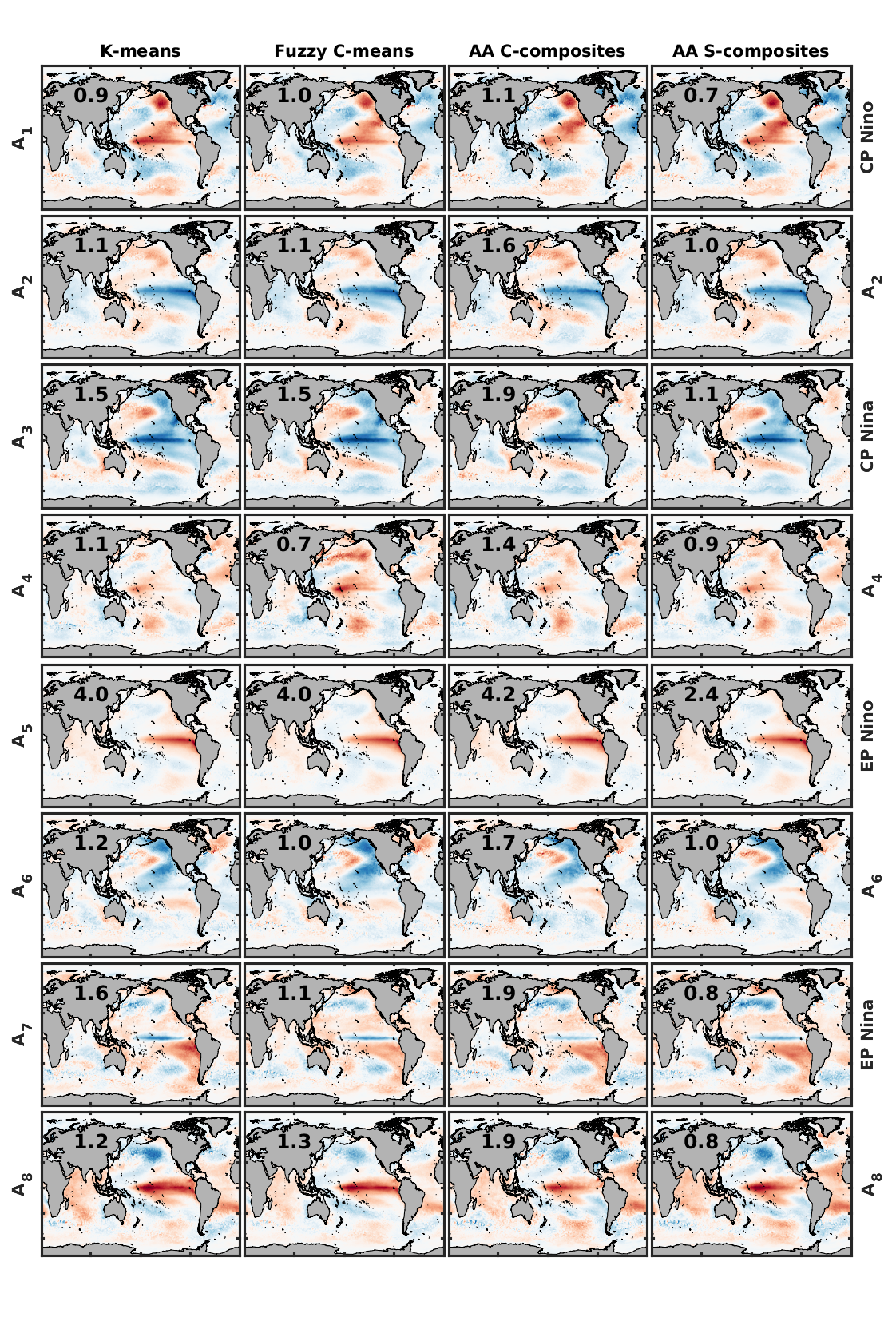}
    \caption{As Fig.~2 for cardinality 8 and dSSTAs}
    \label{fig:kca_8}
\end{figure}

This section summarises results obtained for a cardinality of 8. Figure \ref{fig:kca_8} shows reduced space (RS) SSTA cluster centers for K-means, fuzzy C-means, AA C-composites $\textbf{X}\textbf{C}$ and SSTA AA S-composites $\textbf{X}\bar{\textbf{S}}^{T}$ (columns). The cluster centers for each method (rows) are referenced to the AA results for a cardinality of 8 (last column) and assigned using the Euclidean distance to match patterns across methods. Each pattern has been normalised by its maximum absolute value in degree C, reported in bold on the Eurasian continent, for increased contrast across clusters. As for cardinality 4, the `fuzzyfier' exponent $m$ was set to 1.05 for C-means.

\begin{figure}[htbp]
    \centering
    \includegraphics[width=1.0\textwidth]{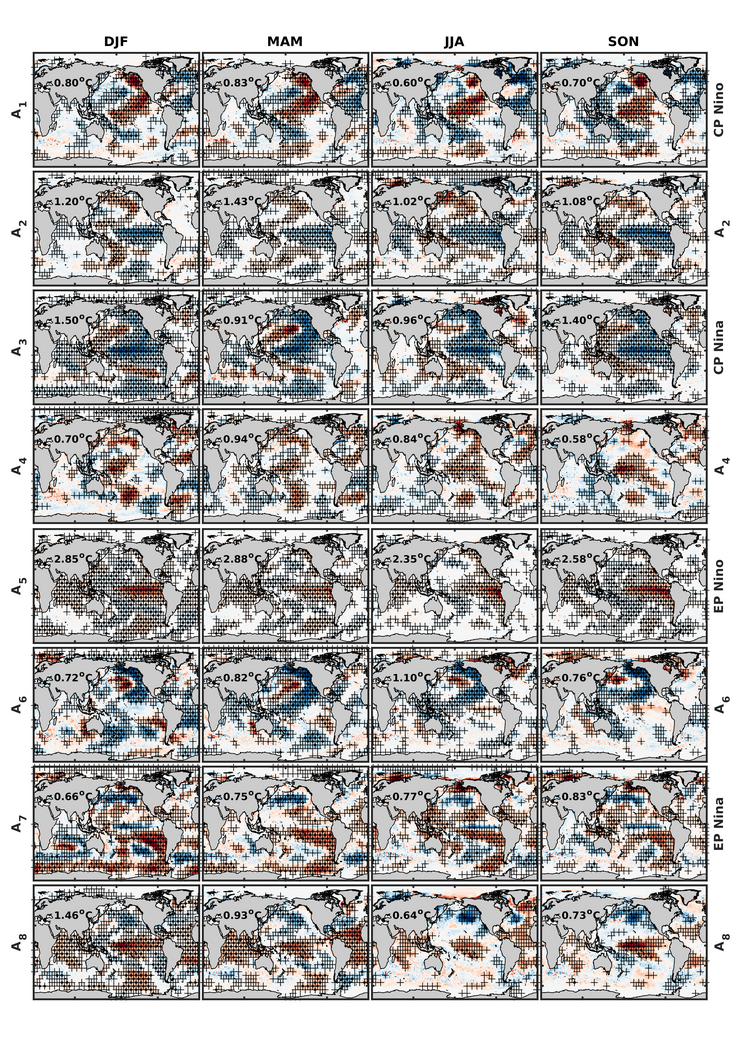}
    \caption{As Fig.~8 for cardinality 8 and dSSTAs}
    \label{fig:aasc8}
\end{figure}

\begin{figure}[htbp]
    \centering
    \includegraphics[width=1.0\textwidth]{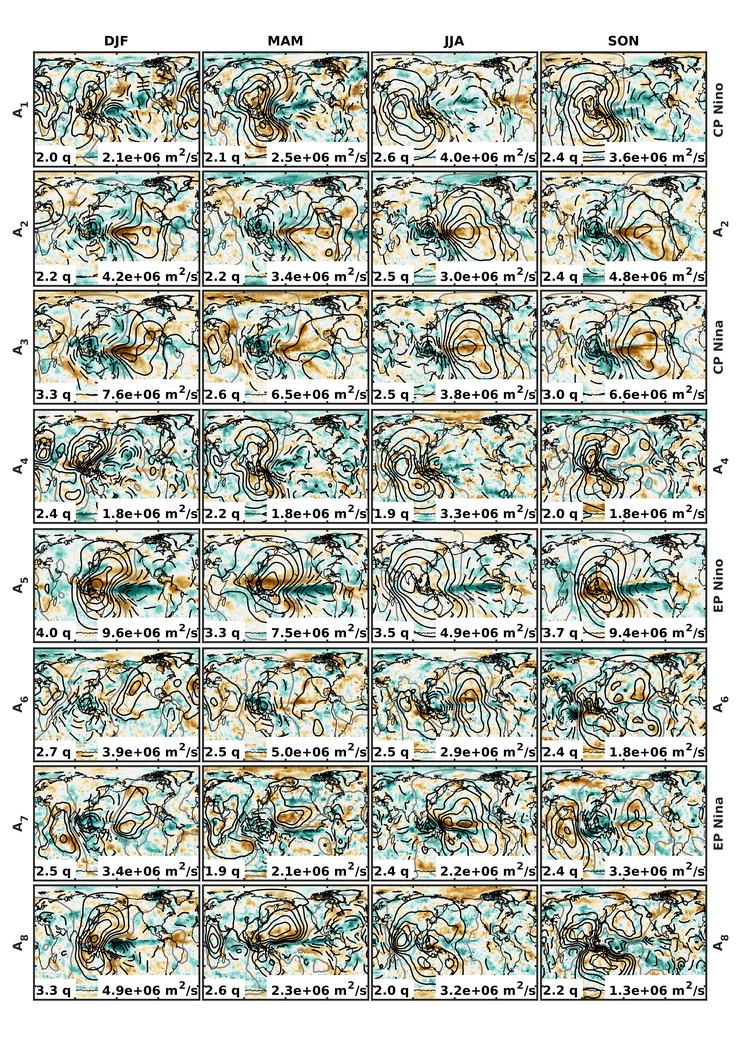}
    \caption{As Fig.~9 for cardinality 8, rainfall undeciles and $\Delta\chi_{150-850}$ anomalies.}
    \label{fig:vpot_xsaa8}
\end{figure}

\begin{figure}[htbp]
    \centering
    \includegraphics[width=0.9\textwidth]{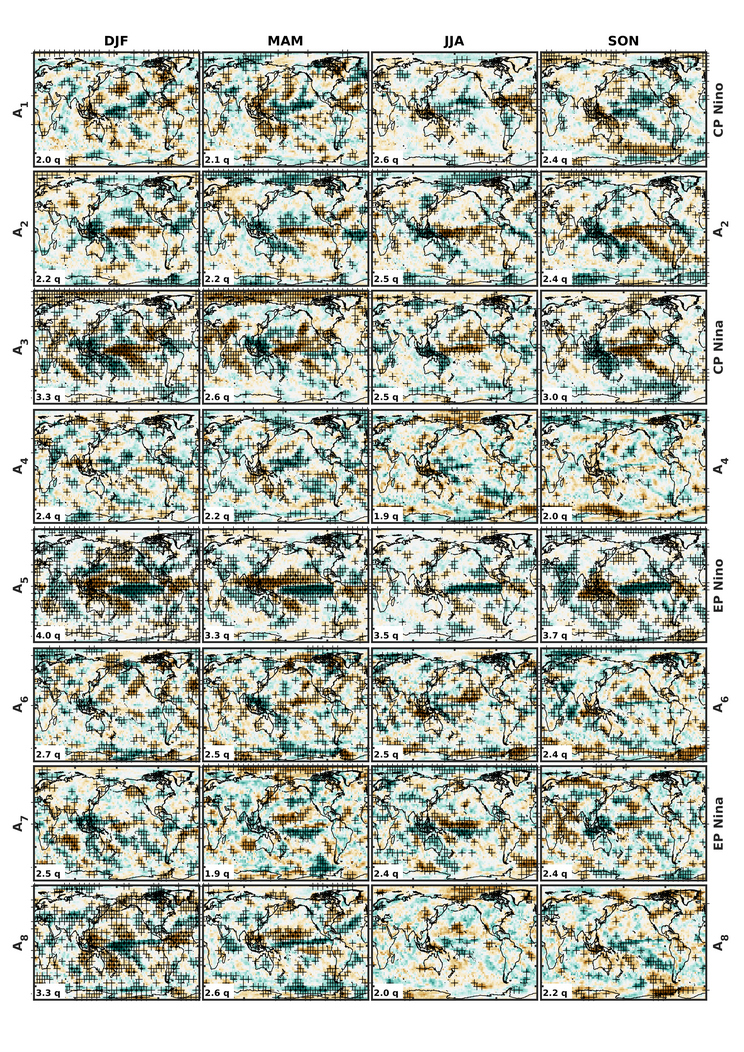}
    \caption{Complement to Fig.~\ref{fig:vpot_xsaa8} with rainfall undeciles and signal-to-noise ratios ($\textbf{+} > 0.15, \bullet > 0.3$, $(F^{2}_{S}/C^{S}_{F})^{1/2}$) based on time averages of the $\textbf{S}$-affiliation matrix.}
    \label{fig:aa8snr_gpcp}
\end{figure}

\begin{figure}[htbp]
    \centering
    \includegraphics[width=0.9\textwidth]{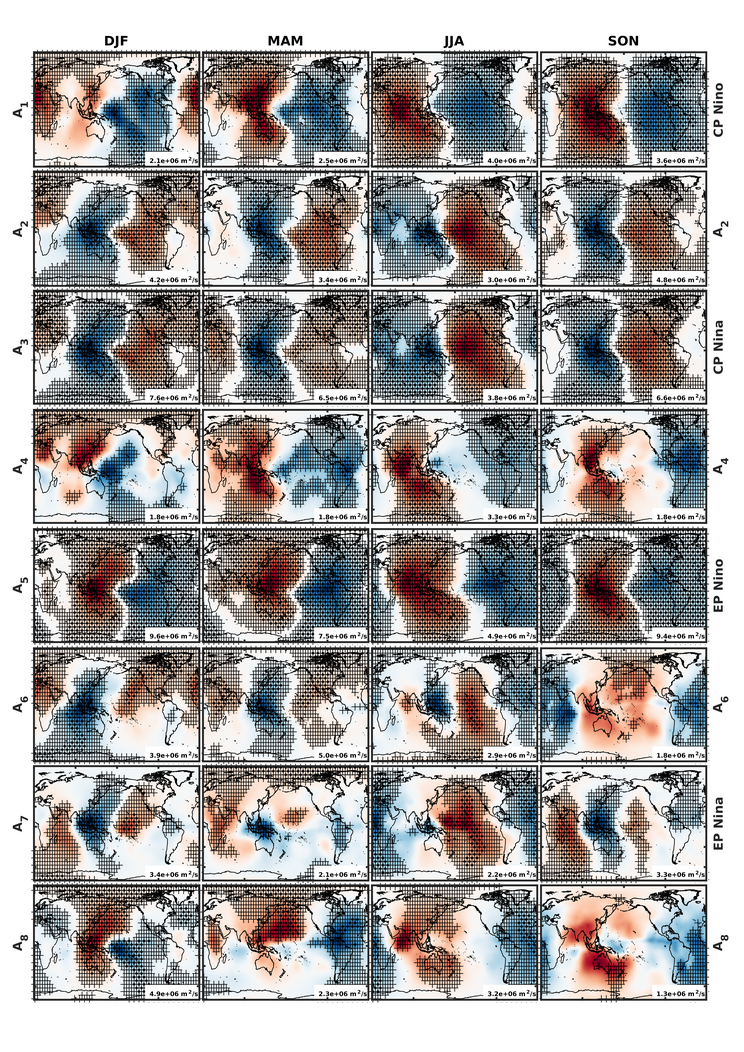}
    \caption{Complement to Fig.~\ref{fig:vpot_xsaa8} with $\Delta\chi_{150-850}$ anomalies and signal-to-noise ratios($\textbf{+} > 0.15, \bullet > 0.3$, $(F^{2}_{S}/C^{S}_{F})^{1/2}$) based on time averages of the $\textbf{S}$-affiliation matrix.}
    \label{fig:aa8snr_vpot}
\end{figure}

\begin{figure}[htbp]
    \centering
    \includegraphics[width=1.0\textwidth]{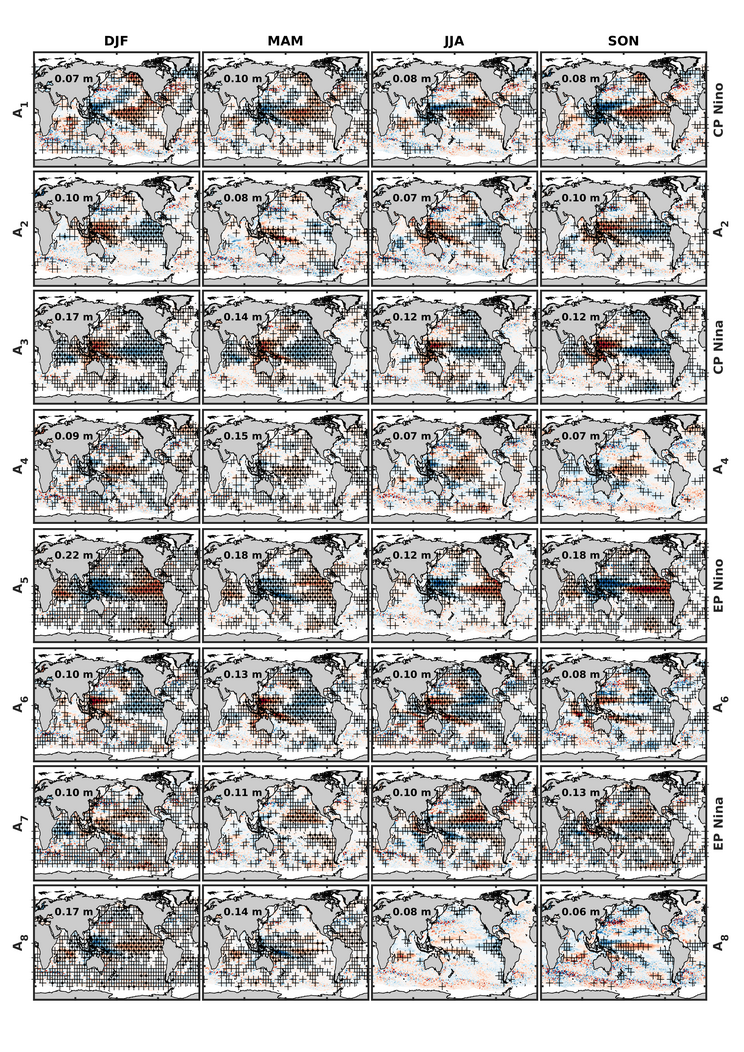}
    \caption{As Fig.~11 for cardinality 8 and dSLAs.}
    \label{fig:sla_xsaa8}
\end{figure}

\begin{figure}[htbp]
    \centering
    \includegraphics[width=1.0\textwidth]{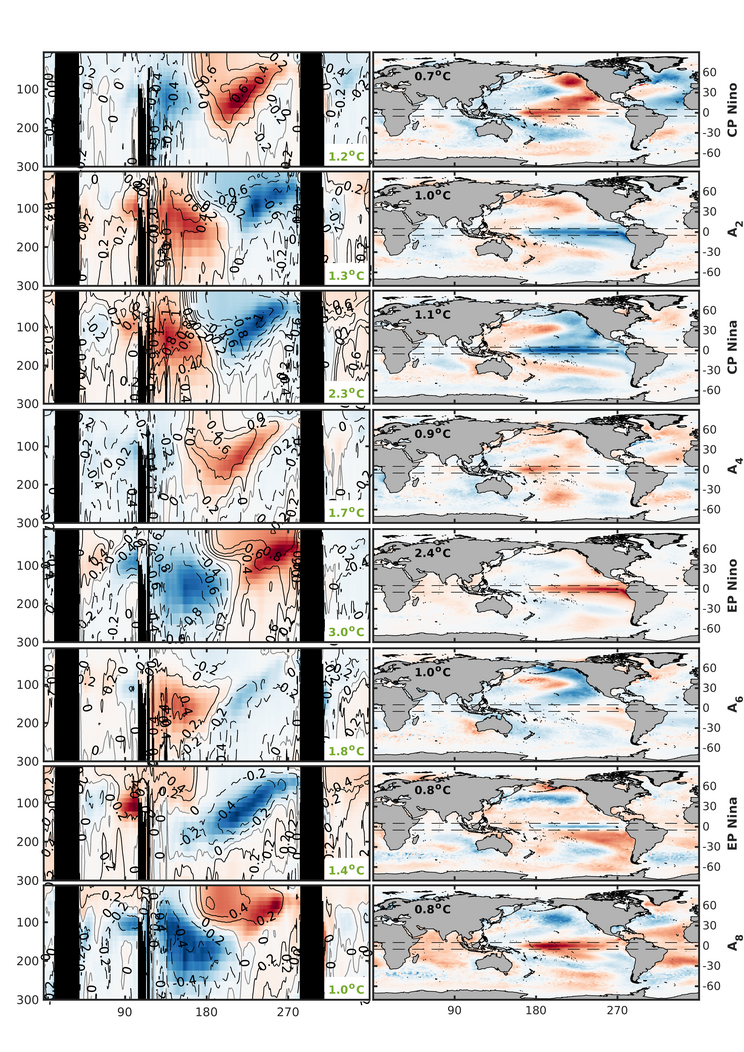}
    \caption{As Fig.~12 for cardinality 8 and potential temperature anomalies}
    \label{fig:sodalon_xsaa8}
\end{figure}

\begin{figure}[htbp]
    \centering
    \includegraphics[width=1.0\textwidth]{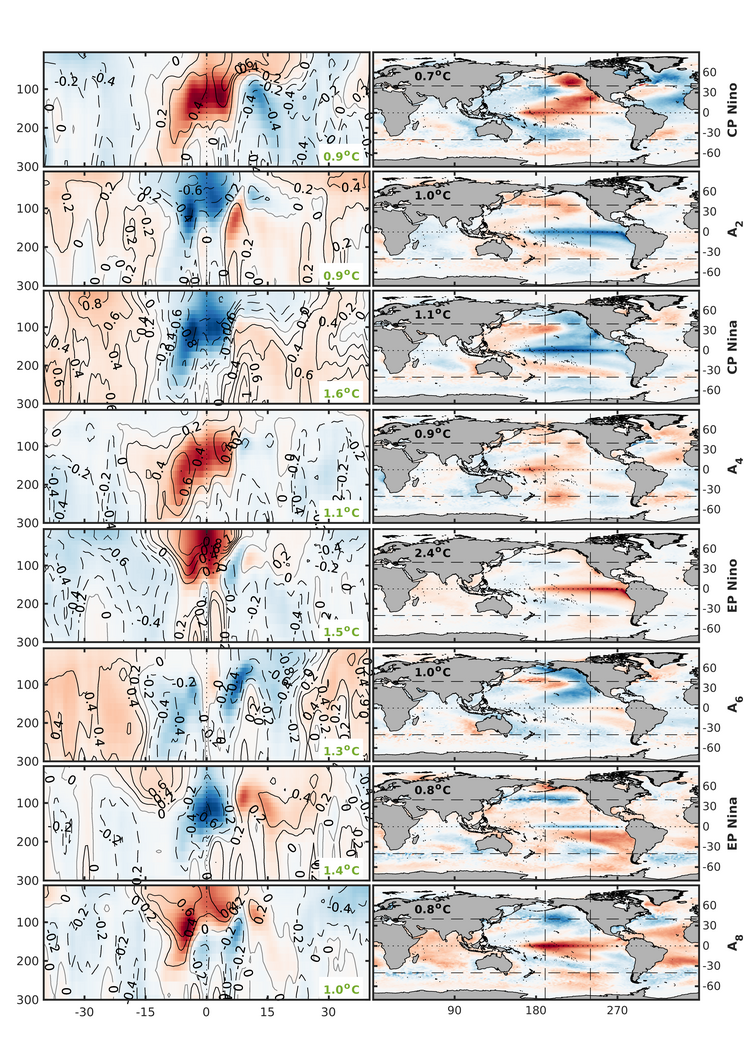}
    \caption{As Fig.~13 for cardinality 8 and potential temperature anomalies.}
    \label{fig:sodalat_xsaa8}
\end{figure}

\begin{figure}[htbp]
    \centering
    \includegraphics[width=1.0\textwidth]{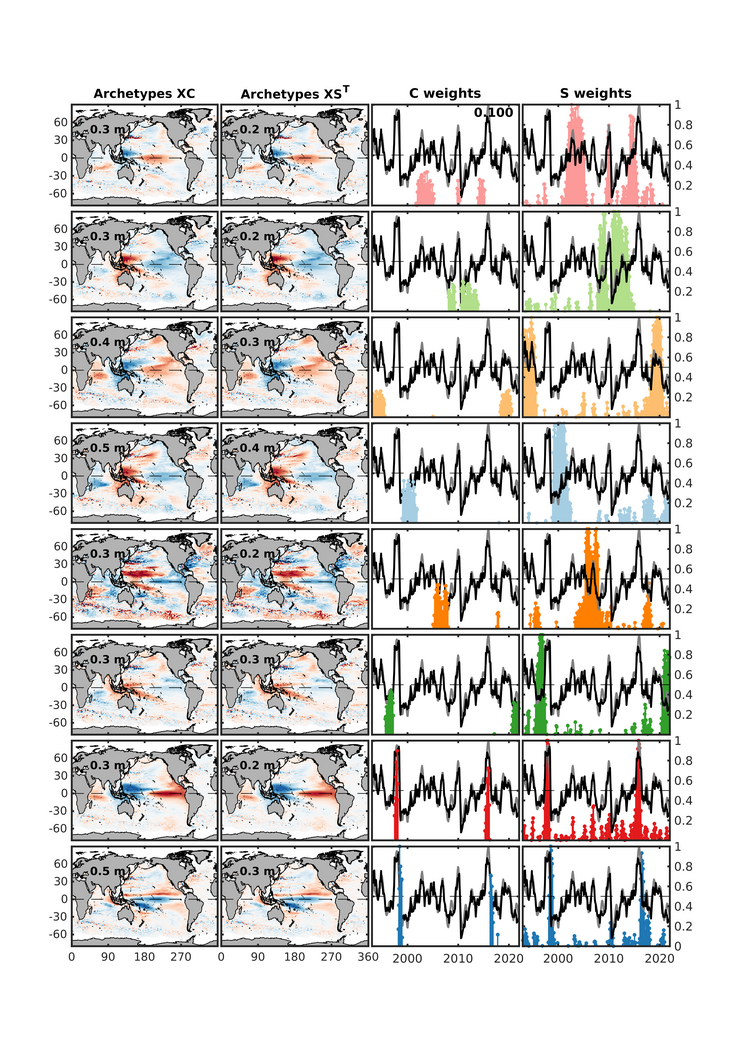}
    \caption{As Fig.~14 for cardinality 8 and dSLAs.}
    \label{fig:dsla_xcxsaa8}
\end{figure}

\begin{figure}[htbp]
    \centering
    \includegraphics[width=1.0\textwidth]{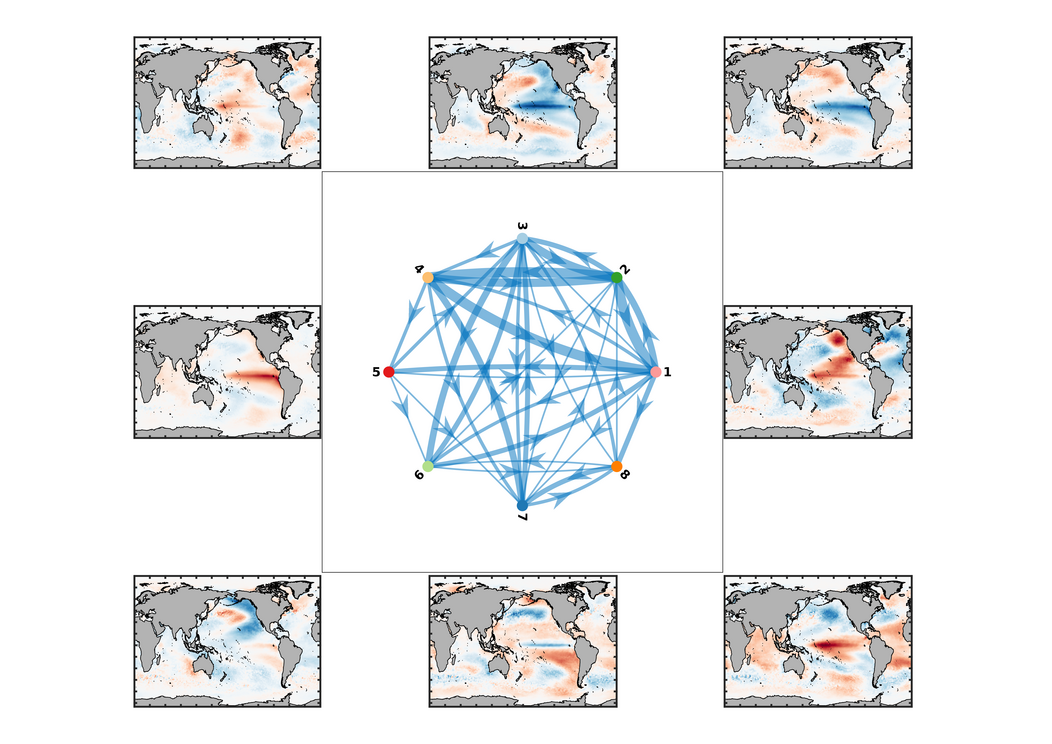}
    \caption{As Fig.~16 for cardinality 8 with the colour-coded graph nodes corresponding to the ENSO phase colours in Fig. 5 and 6.}
    \label{fig:trans_aa8}
\end{figure}

Seasonal dSSTA composites are given Figure S8 with the ENSO phases detected for a cardinality of 4 given on the right. Similarly, teleconnections composites corresponding to 1) global precipitation and vertical convection proxy fields, 2) detrended sea-level anomalies, 3) zonal and meridional tropical potential temperature profiles and 4) transition probabilities are reported in the Figures \ref{fig:aasc8}, \ref{fig:vpot_xsaa8}, \ref{fig:aa8snr_gpcp}, \ref{fig:aa8snr_vpot}, \ref{fig:sla_xsaa8}, \ref{fig:sodalon_xsaa8}, \ref{fig:sodalat_xsaa8}, \ref{fig:dsla_xcxsaa8} and \ref{fig:trans_aa8}.
%%  M-file
%   JCLI_fig5.m
%
%%  M-file
%   JCLI_fig11.m
%
%%  M-file
%   JCLI_fig15.m
%
%%  M-file
%   JCLI_figS15.m for GPCP
%
%%  M-file
%   JCLI_figS15.m for VPOT
%
%%  M-file
%   JCLI_fig17.m
%
%%  M-file
%   JCLI_fig19.m
%
%%  M-file
%   JCLI_fig21.m
%

\clearpage
\bibliographystyle{ametsocV6.bst}
\bibliography{supplement}